# Kinematics of dense gas in the L1495 filament*

A. Punanova[1,2], P. Caselli[1], J. E. Pineda[1], A. Pon[3], M. Tafalla[4], A. Hacar[5], and L. Bizzocchi[1]

[1] Max-Planck-Institut für extraterrestrische Physik, Giessenbachstrasse 1, 85748 Garching, Germany
[2] Ural Federal University, 620002, 19 Mira street, Yekaterinburg, Russia
e-mail: anna.punanova@urfu.ru
[3] Department of Physics and Astronomy, The University of Western Ontario, 1151 Richmond Street, London, ON, N6A 3K7, Canada
[4] Observatorio Astronómico Nacional (IGN), Alfonso XII 3, 28014 Madrid, Spain
[5] Leiden Observatory, University of Leiden, Niels Bohrweg 2, 2333 CA Leiden, The Netherlands

**ABSTRACT**

*Context.* Nitrogen bearing species, such as $NH_3$, $N_2H^+$, and their deuterated isotopologues, show enhanced abundances in CO-depleted gas, and thus are perfect tracers of dense and cold gas in star forming regions. The Taurus molecular cloud contains the long L1495 filament providing an excellent opportunity to study the process of star formation in filamentary environments.

*Aims.* We study the kinematics of the dense gas of starless and protostellar cores traced by the $N_2D^+$(2–1), $N_2H^+$(1–0), $DCO^+$(2–1), and $H^{13}CO^+$(1–0) transitions along the L1495 filament and the kinematic links between the cores and the surrounding molecular cloud.

*Methods.* We measure velocity dispersions, local and total velocity gradients and estimate the specific angular momenta of 13 dense cores in the four transitions using the on-the-fly observations with the IRAM 30 m antenna. To study a possible connection to the filament gas, we use the fit results of the $C^{18}O$(1–0) survey performed by Hacar et al. (2013).

*Results.* The velocity dispersions of all studied cores are mostly subsonic in all four transitions, with similar and almost constant dispersion across the cores in $N_2D^+$(2–1) and $N_2H^+$(1–0). A small fraction of the $DCO^+$(2–1) and $H^{13}CO^+$(1–0) lines show transonic dispersion and exhibit a general increase in velocity dispersion with line intensity. All cores have velocity gradients (0.6–6.1 km s$^{-1}$ pc$^{-1}$), typical of dense cores in low-mass star forming regions. All cores show similar velocity patterns in the different transitions, simple in isolated starless cores, and complex in protostellar cores and starless cores close to young stellar objects where the gas motions can be affected by outflows. The large-scale velocity field traced by $C^{18}O$(1–0) does not show any perturbation due to protostellar feedback and does not mimic the local variations seen in the other four tracers. Specific angular momentum $J/M$ varies in a range (0.6–21.0)$\times 10^{20}$ cm$^2$ s$^{-1}$ which is similar to the results previously obtained for dense cores. $J/M$ measured in $N_2D^+$(2–1) is systematically lower than $J/M$ measured in $DCO^+$(2–1) and $H^{13}CO^+$(1–0).

*Conclusions.* All cores show similar properties along the 10 pc-long filament. $N_2D^+$(2–1) shows the most centrally concentrated structure, followed by $N_2H^+$(1–0) and $DCO^+$(2–1), which show similar spatial extent, and $H^{13}CO^+$(1–0). The non-thermal contribution to the velocity dispersion increases from higher to lower density tracers. The change of magnitude and direction of the total velocity gradients depending on the tracer used indicates that internal motions change at different depths within the cloud. $N_2D^+$ and $N_2H^+$ show smaller gradients than the lower density tracers $DCO^+$ and $H^{13}CO^+$, implying a loss of specific angular momentum at small scales. At the level of cloud-core transition, the core's external envelope traced by $DCO^+$ and $H^{13}CO^+$ is spinning up, consistent with conservation of angular momentum during core contraction. $C^{18}O$ traces the more extended cloud material whose kinematics is not affected by the presence of dense cores. The decrease in specific angular momentum towards the centres of the cores shows the importance of local magnetic fields to the small scale dynamics of the cores. The random distributions of angles between the total velocity gradient and large scale magnetic field suggests that the magnetic fields may become important only in the high density gas within dense cores.

## 1. Introduction

Recent submillimetre studies of the nearest star-forming clouds with the *Herschel Space Observatory* show that interstellar filaments are common structures in molecular clouds and play an important role in the star-forming process (e.g. Men'shchikov et al. 2010; André et al. 2014). The filaments host chains of dense cores (e.g. Hacar et al. 2013; Könyves et al. 2014); some of the cores are pre-stellar – on the verge of star formation. Pre-stellar cores are cold (∼10 K), dense ($10^4$–$10^7$ cm$^{-3}$), and quiescent (thermal pressure dominates turbulent motions; e.g. Benson & Myers 1989; Fuller & Myers 1992; Lada et al. 2008; Caselli et al. 2008) self-gravitating structures (Ward-Thompson et al. 1999; Keto & Caselli 2008), characterised by high deuterium fractions (>10%, Crapsi et al. 2005). Pre-stellar cores represent the initial conditions in the process of star formation, thus their study is crucial to understand how stars and stellar systems form.

The target of our study, L1495 (Lynds 1962), is an extended filamentary structure in the Taurus molecular cloud, a nearby (140 pc distance, Elias 1978; Torres et al. 2012), relatively quiescent, low-mass star forming region. The selected filament con-

* This work is based on observations carried out under the projects 032-14 and 156-14 with the IRAM 30m Telescope. IRAM is supported by INSU/CNRS (France), MPG (Germany) and IGN (Spain).

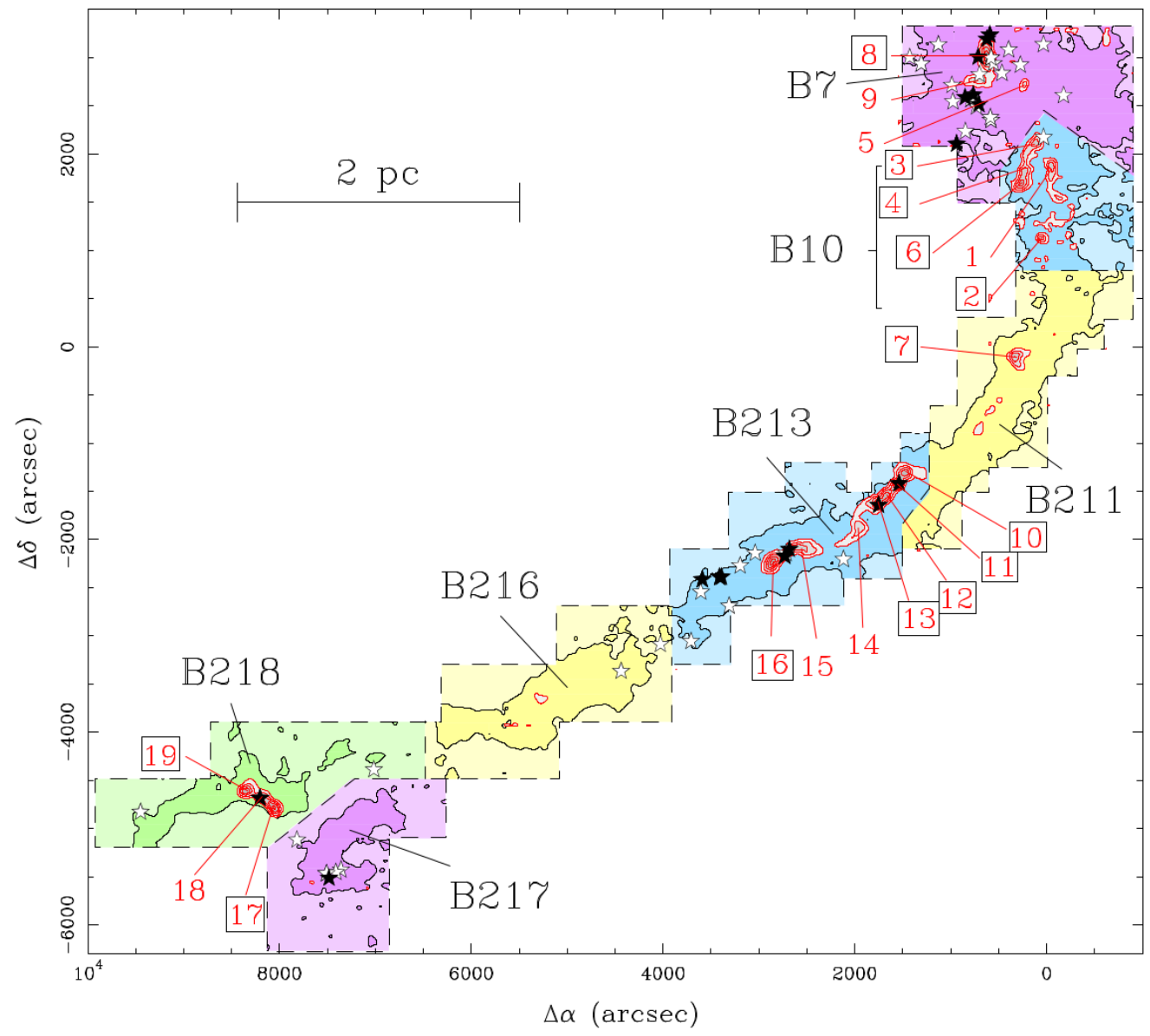

**Fig. 1.** Schematic view of the selected L1495 filament within the Taurus Molecular Cloud Complex (adapted from Hacar et al. 2013). The offsets refer to the centre at $\alpha = 4^h17^m47.1^s$, $\delta = +27°37'18''$ (J2000). The black solid line represents the lowest $C^{18}O$(1–0) contour (0.5 K km s$^{-1}$). The red lines represent the $N_2H^+$(1–0) emission (first contour and contour interval are 0.5 K km s$^{-1}$), which traces the dense cores. The $C^{18}O$(1–0) and $N_2H^+$(1–0) data are from Hacar et al. (2013). The red labels identify the cores found by Hacar et al. (2013), and black squares around the labels show the cores studied in this work. The stars correspond to young stellar objects from the survey of Rebull et al. (2010). Solid symbols represent the youngest objects (Class I and Flat), and open symbols represent evolved objects (Class II and III).

tains 39 dense cores revealed in ammonia by Seo et al. (2015) including 19 dense cores previously detected in $N_2H^+$(1–0) emission (Hacar et al. 2013) (see Fig. 1) at different stages of star formation. Tens of starless cores are detected there via continuum emission by *Herschel* (Marsh et al. 2014), and about fifty low-mass protostars in different evolutionary stages were observed with *Spitzer* (see the survey by Rebull et al. 2010). L1495 is a very well studied region, with its physical properties and structure determined by several large observational studies. This includes its gas kinetic temperature (Seo et al. 2015); dust extinction (Schmalzl et al. 2010); low density gas distribution, as traced by $H^{13}CO^+$ (Onishi et al. 2002) and $C^{18}O$ (Hacar et al. 2013); and dense gas distribution and dense cores locations, as traced by $N_2H^+$ (Hacar et al. 2013; Tafalla & Hacar 2015) and $NH_3$ (Seo et al. 2015). Thus, L1495 is an excellent place to test theories of dense core formation within filaments.

Hacar et al. (2013) mapped the whole filament in $C^{18}O$(1–0), SO(3–2), and $N_2H^+$(1–0) using the FCRAO antenna. They found that the filament is not a uniform structure and consists of many fibers. The fibers are elongated structures mostly aligned with the axis of the large-scale filament, with typical lengths of 0.5 pc, coherent velocity fields, and internal velocity dispersions close to the sound speed. The fibers were revealed in the low density gas tracer $C^{18}O$(1–0) kinematically, that is the Gaussian fits of the multiple $C^{18}O$(1–0) components plotted in position-position-velocity space appear as velocity coherent structures. Their distribution resembles the small scale structures revealed with the *getfilaments* algorithm (Men'shchikov 2013) in *Herschel* dust continuum emission when large-scale emission was filtered-out (André et al. 2014). Some of the CO fibers contain dense cores revealed by the high density tracer $N_2H^+$(1–0). Hacar et al. (2013) conclude that fragmentation in the L1495 complex proceeded in a hierarchical manner, from cloud to subregions (bundles) to fibers and then to individual dense cores. In the following study, Tafalla & Hacar (2015) found that the cores tend to cluster in linear groups (chains). Hacar et al. (2013) and Seo et al. (2015) note that some parts of the filament are young (B211 and B216) and others (B213 and B7) are more evolved and actively star-forming. The gas temperature they derived from $NH_3$ is low, 8–15 K with a median value of 9.5 K, with less evolved (B10, B211, and B216) regions only having a median temperature 0.5 K less than more evolved (B7, B213, B218) regions. They found that the gas kinetic temperature decreases towards dense core centres. With $NH_3$, which traces less dense gas than $N_2H^+$, Seo et al. (2015) found 39 ammonia peaks including those 19 found by Hacar et al. (2013). Onishi et al. (2002) presented a large survey of $H^{13}CO^+$(1–0) observed towards $C^{18}O$ emission peaks in L1495.

The previous studies of the gas kinematics in L1495 were focused on the large-scale structure, the filament as a whole and its subregions. Here we focus on the kinematics within the dense gas of the cores traced by $N_2H^+$ and $N_2D^+$, and the kinematics of the surrounding core envelope traced by $H^{13}CO^+$ and $DCO^+$ to study the gas that connects the cores to their host cloud.

The best tracers of the dense gas kinematics are N-bearing species. In dense (> a few $\times 10^4$ cm$^{-3}$) and cold (T≃10 K) regions, CO, CS, and other C-bearing species are heavily frozen onto dust grains (Caselli et al. 1999; Tafalla et al. 2006; Bizzocchi et al. 2014). Nitrogen-bearing species such as $N_2H^+$ and $NH_3$ and their deuterated isotopologues stay in the gas phase up to densities of $10^6$ cm$^{-3}$ (Crapsi et al. 2005, 2007) and become good tracers of dense gas (see also Tafalla et al. 2004). $N_2H^+$ rotational transitions have higher critical densities ($n_{crit} \geq 10^5$ cm$^{-3}$) than the inversion transitions of $NH_3$ ($n_{crit} \sim 10^3$ cm$^{-3}$) and thus $N_2H^+$ traces dense gas better than $NH_3$. Deuterated species also increase their abundance towards the central regions of cores because of the enhanced formation rate of deuterated forms of $H_3^+$, for example $H_2D^+$ (the precursor of deuterated species, such as $DCO^+$, $N_2D^+$, and deuterated ammonia), in zones where CO is mostly frozen onto dust grains (e.g. Dalgarno & Lepp 1984; Caselli et al. 2003). To study the kinematics of the gas in the central parts of the cores, we choose the $N_2D^+$(2–1) transition ($n_{crit} = 2.5 \times 10^6$ cm$^{-3}$, calculated using the data[1] from the LAMDA database, Schöier et al. 2005), and the $N_2H^+$(1–0) line to connect to the work of Hacar et al. (2013), who also mapped $N_2H^+$(1–0), and also to study the deuterium fraction across the cores, which will be presented in a subsequent study (Punanova et al., in prep.). $HCO^+$ follows CO and thus it depletes in the centres of dense cores (e.g. Pon et al. 2009); therefore, $HCO^+$ is a good tracer of core envelopes. To study the kinematics of the core surroundings and provide a connection between the kinematics of the cores and that of the cloud, we choose the $H^{13}CO^+$(1–0) and $DCO^+$(2–1) transitions. This will also enable a future study of the deuterium fraction of the cores and core envelopes (Punanova et al., in prep.).

This paper presents observations of $N_2D^+$(2–1), $N_2H^+$(1–0), $DCO^+$(2–1), and $H^{13}CO^+$(1–0) towards 13 dense cores to study the kinematics of the dense gas along the L1495 filament. In Section 2, the details of the observations are presented. Section 3 describes the data reduction procedure. Section 4 presents the results of the hyperfine structure fitting and velocity gradients and specific angular momenta calculations. In Section 5 we discuss the results and connections between core-scale and cloud-scale kinematics. The conclusions are given in Section 6.

[1] http://home.strw.leidenuniv.nl/ moldata/datafiles/n2h+@xpol.dat

**Table 1.** The observed cores. The names (numbers) of the cores are given following Hacar et al. (2013). The given coordinates are the central positions of the cores from Hacar et al. (2013). The protostellar cores are indicated with asterisks (*).

| Core | $\alpha_{J2000}$ ($^{h\ m\ s}$) | $\delta_{J2000}$ ($^{\circ\ \prime\ \prime\prime}$) |
|---|---|---|
| 2 | 04 17 50 | 27 56 07 |
| 3 | 04 17 56 | 28 12 23 |
| 4 | 04 18 04 | 28 08 14 |
| 6 | 04 18 06 | 28 05 41 |
| 7 | 04 18 10 | 27 35 29 |
| 8 | 04 18 34 | 28 27 37 |
| 10 | 04 19 37 | 27 15 48 |
| 11* | 04 19 44 | 27 13 36 |
| 12 | 04 19 52 | 27 11 42 |
| 13* | 04 19 59 | 27 10 30 |
| 16 | 04 21 21 | 27 00 09 |
| 17 | 04 27 54 | 26 17 50 |
| 19 | 04 28 14 | 26 20 34 |

## 2. Observations

We mapped 13 out of 19 dense cores of the L1495 filamentary structure (see Fig. 1 and Table 1) in $N_2D^+$(2–1) at 154.2 GHz, $N_2H^+$(1–0) at 93.2 GHz, $DCO^+$(2–1) at 144.1 GHz, and $H^{13}CO^+$(1–0) at 86.8 GHz with the IRAM 30 m telescope (IRAM projects 032-14 and 156-14). The $D^{13}CO^+$(2–1) at 141.5 GHz and $HC^{18}O^+$(1–0) at 85.2 GHz lines were observed towards the $DCO^+$(2–1) emission peaks of 9 cores. The observations were performed on 09–14 July, 02–08 December 2014 and 04 June 2015 under acceptable weather conditions, pwv=1–9 mm. The on-the-fly maps and single pointing observations were obtained with the EMIR 090 (3 mm band) and EMIR 150 (2 mm band) heterodyne receivers[2] in position switching mode, and the VESPA backend. The spectral resolution was 20 kHz, the corresponding velocity resolutions were ≃0.07 for the 3 mm band and ≃0.04 km $s^{-1}$ for the 2 mm. The beam sizes were ≃28″ for the 3 mm band and ≃17″ for the 2 mm. The system temperatures were 90–627 K depending on the lines. The exact line frequencies, beam efficiencies, beam sizes, spectral resolutions and sensitivities are given in Table 2. Sky calibrations were obtained every 10–15 minutes. Reference positions were chosen individually for each core to make sure that the positions were free of $N_2H^+$(1–0) emission, using the Hacar et al. (2013) maps. The reference position for core 4 was contaminated with $N_2H^+$(1–0) emission, thus it is not analysed in the paper. Pointing was checked by observing QSO B0316+413, QSO B0415+379, Uranus, and Venus every 2 hours and focus was checked by observing Uranus and Venus every 6 hours.

To connect core-scale and cloud-scale kinematics, we used the fit results of the $C^{18}O$(1–0) observations by Hacar et al. (2013) convolved to 60″, performed with the 14 m FCRAO telescope.

## 3. Data reduction and analysis with Pyspeckit

The data reduction up to the stage of convolved spectral data cubes was performed with the CLASS package[3]. The intensity scale was converted to the main-beam temperature scale according to the beam efficiency values given in Kramer et al. (2013) (see Table 2 for details). The $N_2H^+$(1–0) maps were convolved to a resolution of 27.8″ with 9″ pixel size. The $N_2D^+$(2–1) maps were convolved to the resolution of the $N_2H^+$(1–0) with the same grid spacing to improve the sensitivity for a fair comparison of the kinematics traced by these transitions. The $H^{13}CO^+$(1–0) and $DCO^+$(2–1) maps were convolved to resolutions of 29.9″ and 18″, with pixel sizes of 9″ and 6″, respectively. The rms across the maps in $T_{mb}$ scale is 0.075–0.14 K, 0.04–0.10 K, 0.11–0.21 K, and 0.15–0.50 K for the $N_2H^+$(1–0), $N_2D^+$(2–1), $H^{13}CO^+$(1–0), and $DCO^+$(2–1), respectively. Each map has a different sensitivity (see Table A.3 for details). The undersampled edges of the maps are not used for the analysis. Each data cube contains spectra of one transition towards one core. Some cores lie close to each other so we also produced combined data cubes (cores 4 and 6 and the chain of cores 10–13, see for example Fig. 1 and A.8). Another dataset with all maps convolved to the biggest beam (29.9″) and Nyquist sampling was produced to compare local and total velocity gradients across the cores seen in different species (see Sect. 4.5.2 for details). The rms in $T_{mb}$ scale across the maps smoothed to 29.9″ is in the ranges 0.065–0.12 K, 0.04–0.085 K, 0.11–0.21 K, and 0.05–0.15 K for the $N_2H^+$(1–0), $N_2D^+$(2–1), $H^{13}CO^+$(1–0), and $DCO^+$(2–1), respectively (see Table A.4 for details).

The spectral analysis was performed with the Pyspeckit module of Python (Ginsburg & Mirocha 2011). The $N_2H^+$(1–0), $N_2D^+$(2–1), $H^{13}CO^+$(1–0), and $DCO^+$(2–1) lines have hyperfine splitting with 15, 40, 6, and 6 components, respectively. Thus we performed hyperfine structure (hfs) fitting using the standard routines of Pyspeckit. The routine computes line profiles with the assumptions of Gaussian velocity distributions and equal excitation temperatures for all hyperfine components. It varies four parameters (excitation temperature $T_{ex}$, optical depth $\tau$, central velocity of the main hyperfine component $V_{LSR}$, and velocity dispersion $\sigma$) and finds the best fit with the Levenberg-Marquardt non-linear regression algorithm. The rest frequencies of the main components, the velocity offsets, and the relative intensities of the hyperfine components were taken from Pagani et al. (2009), Schmid-Burgk et al. (2004), Caselli & Dore (2005), and Dore L. (priv. comm.). For the spectra of each transition towards the $N_2H^+$(1–0) emission peak of core 2, the results of Pyspeckit hfs fitting procedure were compared to the results of the CLASS hfs fitting method. The results agree within the errors. We first fitted Gaussians to the $H^{13}CO^+$(1–0) and $DCO^+$(2–1) lines and compared the results to the hfs fit results. We found that the hyperfine structure significantly affects the line profile and should be taken into account to provide accurate line widths (see Appendix A.1 for details).

The general data fitting procedure went as follows. First each spectrum in one data cube (one core, one species) was fitted two times assuming 1) unconstrained $\tau$ (all four parameters $T_{ex}$, $\tau$, $V_{LSR}$ and $\sigma$ were free) and 2) constrained $\tau$ ($\tau$ was fixed at 0.1, which makes the fit close to the assumption that the line is optically thin; the other three parameters were free). Second, the final results data cube was produced by combining the results of the $\tau$-constrained and $\tau$-unconstrained fits. If $\tau/\Delta\tau \geq 3$, where $\Delta\tau$ is the optical depth uncertainty, the results of the $\tau$-

[2] http://www.iram.es/IRAMES/mainWiki/EmirforAstronomers

[3] Continuum and Line Analysis Single-Dish Software http://www.iram.fr/IRAMFR/GILDAS

**Table 2.** Observation parameters.

| Transition | Frequency[a] (GHz) | $F_{eff}$ | $B^h_{eff}$ | HPBW ($''$) | $\Delta v^i_{res}$ (km s$^{-1}$) | rms in $T_{mb}$ (K) | $T_{sys}$ (K) | Mode[j] | Dates |
|---|---|---|---|---|---|---|---|---|---|
| $N_2H^+$(1–0) | 93.1737637[b] | 0.95 | 0.80 | 26.5 | 0.063 | 0.075–0.140 | 155–176 | OTF | 09–14.07.2014 |
| $N_2D^+$(2–1) | 154.2171805[c] | 0.93 | 0.72 | 16.3 | 0.038 | 0.040–0.100 | 203–627 | OTF | 09–14.07.2014 |
| $H^{13}CO^+$(1–0) | 86.7542884[d] | 0.95 | 0.81 | 28.5 | 0.068 | 0.110–0.190 | 102–129 | OTF | 02–08.12.2014 |
| | | | | | | 0.140–0.190 | 122–126 | OTF | 03–04.06.2015 |
| $DCO^+$(2–1) | 144.0772804[e] | 0.93 | 0.73 | 17.2 | 0.041 | 0.150–0.250 | 115–155 | OTF | 02–08.12.2014 |
| | | | | | | 0.310–0.500 | 198–220 | OTF | 03–04.06.2015 |
| $HC^{18}O^+$(1–0) | 85.1622231[f] | 0.95 | 0.81 | 28.5 | 0.069 | 0.026–0.039 | 96–105 | SP | 08.12.2014 |
| | | | | | | 0.020–0.038 | 99–109 | SP | 04.06.2015 |
| $D^{13}CO^+$(2–1) | 141.4651331[g] | 0.93 | 0.74 | 17.5 | 0.041 | 0.028–0.039 | 90–96 | SP | 08.12.2014 |
| | | | | | | 0.029–0.059 | 142–164 | SP | 04.06.2015 |

**Notes.** [a] Frequency of the main hyperfine component; [b] from Pagani et al. (2009); [c] from Pagani et al. (2009) and Dore, L., private communication; [d] from Schmid-Burgk et al. (2004); [e] from Caselli & Dore (2005); [f] from Schmid-Burgk et al. (2004); [g] from Caselli & Dore (2005); [h] $B_{eff}$ and $F_{eff}$ values are available at the 30 m antenna efficiencies web-page http://www.iram.es/IRAMES/mainWiki/Iram30mEfficiencies; [i] $\Delta v_{res}$ is the velocity resolution; [j] Observational mode: OTF is an on-the-fly map, SP is a single-pointing on-off observation towards the $DCO^+$(2–1) emission peak.

unconstrained fit were written to the final result data cube. If $\tau/\Delta\tau < 3$, the $\tau$-constrained fit was chosen and written to the final result data cube. The $\tau/\Delta\tau$ test was done for each pixel. The combination of $\tau$-constrained and $\tau$-unconstrained fits does not cause any sharp change in the velocity dispersion and the centroid velocity maps, although it leads to a small increase ($\simeq$ 0.02 km s$^{-1}$) in velocity dispersion for some cores (see e.g. cores 6 and 8 in Fig. A.9). One can see how constrained opacity affects the fit comparing Fig. A.9 and Fig. A.11, where we show only the opacity maps which have more than three pixels with $\tau$-unconstrained fit. The combined results data cube was written to the final fits file after masking poor data. For the integrated intensity maps, we use all data except the undersampled edges of the maps. The excitation temperature and the optical depth have been measured only for those spectra with a high signal-to-noise ratio (SNR): $I > 5 \cdot rms \cdot \sqrt{N_{ch}} \cdot \Delta v_{res}$, where $I$ is the integrated intensity, $N_{ch}$ is the number of channels in the line, and $\Delta v_{res}$ is the velocity resolution. For $N_{ch}$, we take all channels in the ranges: 2–12, -4–16, 5–10, and 5–10 km s$^{-1}$ for $N_2D^+$(2–1), $N_2H^+$(1–0), and $DCO^+$(2–1) and $H^{13}CO^+$(1–0), respectively. These ranges define the emission above one $rms \cdot \sqrt{N_{ch}} \cdot \Delta v_{res}$ over spectrum averaged over the whole mapped area. For the central velocity and velocity dispersion maps we used the signal-to-noise mask and a mask based on velocity dispersion: the minimum line width must be larger than the thermal line width ($\sigma_T$) for a temperature of 5 K (slightly below the minimum gas temperature in L1495 found by Seo et al. 2015), which is 0.04 km s$^{-1}$ for the given species, and the line width must be defined with an accuracy better than 20% ($\sigma/\Delta\sigma > 5$). If the line is optically thick, the intrinsic line width is found by means of the hfs fit.

In $H^{13}CO^+$(1–0) the hyperfine components are blended and the $\tau$-unconstrained fit is often too ambiguous. For the majority of the $H^{13}CO^+$ spectra, optical depth was defined with an uncertainty $\Delta\tau > \tau/3$, thus we used the $\tau$-constrained fit ($\tau$=0.1) to gauge the central velocities and the velocity dispersions. The $N_2H^+$(1–0) and $H^{13}CO^+$(1–0) spectra show the presence of a second component towards cores 3, 8, 13, and 16 (see Fig. 2 and A.5). When the second component is more than one line width away from the main line, it is assumed to represent an independent velocity component and is not considered in our analysis. When the second component is closer than one line width and blended with the main line so the fitting routine can not resolve them, we consider the two components as one line.

The $DCO^+$(2–1) spectra towards all of the observed cores show double or multiple peaked lines caused by either additional velocity components or self-absorption. We estimate the possible flux loss due to self-absorption of the line by observing the transition of the rare isotopologue $D^{13}CO^+$(2–1) towards the $DCO^+$(2–1) emission peaks. We compare the estimates of column densities ($N_{col}$) of $DCO^+$ obtained with both isotopologues and find that the column densities are the same within the error bars. We thus conclude that the double-peaked lines are most likely two blended velocity components (see Sect. A.2 for details). The velocity dispersions produced by $\tau$-constrained fits of self-absorbed or blended lines are overestimated, because the fit considers only one velocity component and the optical depth is assumed to be 0.1. We compared the velocity dispersion values obtained with the $\tau$-constrained and $\tau$-unconstrained fits where the optical depth is defined with $\tau > 3\Delta\tau$. We found that the $\tau$-constrained fit produces a velocity dispersion 1–3.3 times larger than the $\tau$-unconstrained fit, with an average dispersion 1.56 times larger. As such, the velocity dispersions derived from these fits should be considered as upper limits for our $DCO^+$ data. Thus, for all of the $DCO^+$ spectra we used the $\tau$-constrained fit ($\tau$=0.1) to define the velocity dispersion upper limits and the central velocities. Significant asymmetry of the $DCO^+$(2–1) line towards core 7 (see Fig. 2) produced a systematic difference of $\simeq$0.1 km s$^{-1}$ in its centroid velocity compared to the other species.

## 4. Results

### 4.1. Distribution of gas emission

Figure 2 presents the spectra of all four species towards the $N_2H^+$(1–0) emission peak of each core. For core 4, where the reference position was contaminated with $N_2H^+$(1–0) emission, we present the spectra towards the $N_2D^+$(2–1) emission peak.

There are starless (2, 3, 4, 6, 7, 8, 10, 12, 16, 17 and 19) and protostellar (11 and 13) cores among the observed targets. The protostellar cores host class 0–I young stellar objects (YSOs).

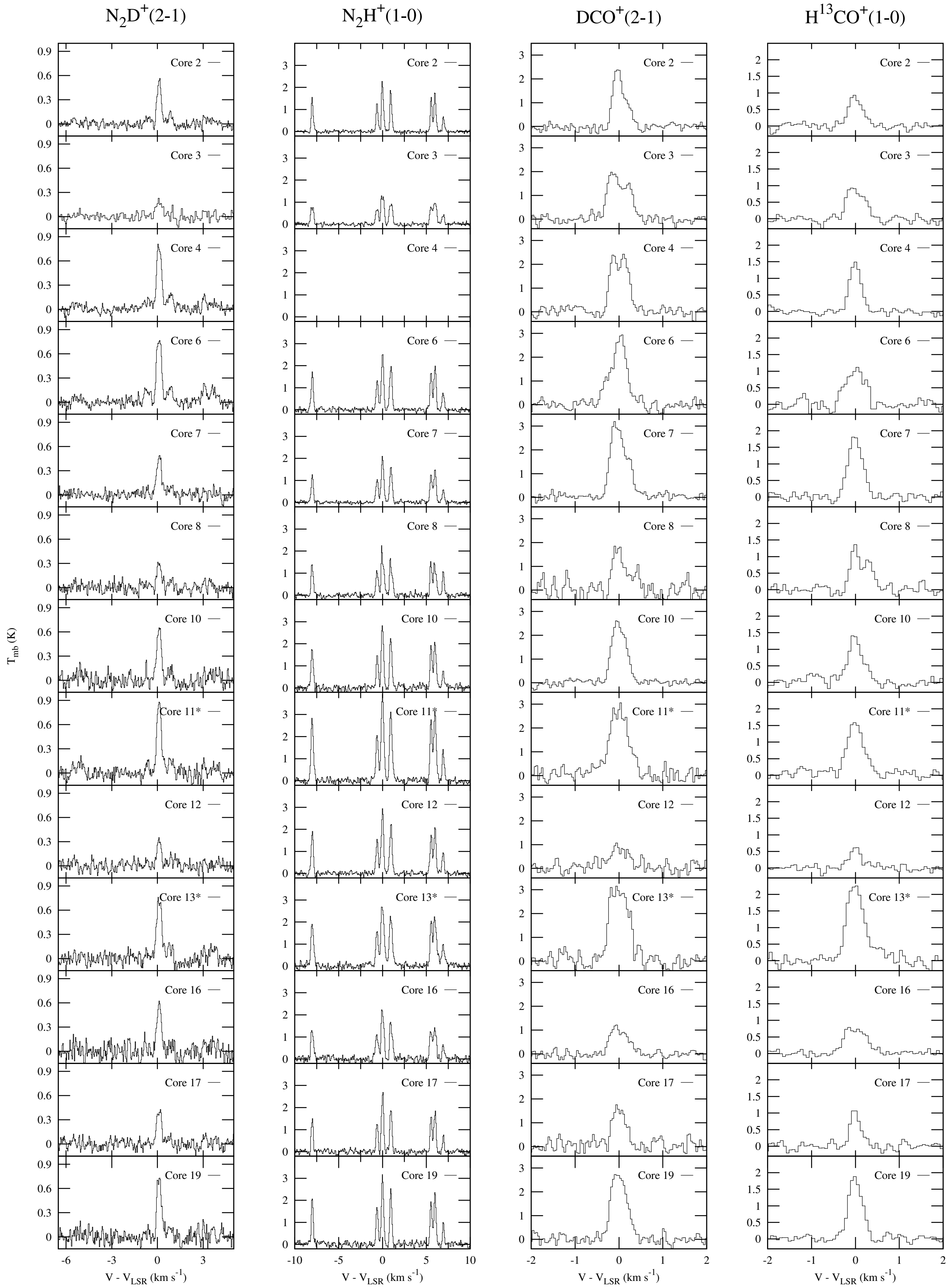

**Fig. 2.** Spectra of all observed transitions towards the $N_2H^+(1–0)$ intensity peak for each core. The $V_{LSR}$ come from individual fits for each line. For core 4, where the reference position was contaminated with $N_2H^+(1–0)$ emission, we present the spectra towards the $N_2D^+(2–1)$ emission peak. The cores with an asterisk (*) near the title contain protostars.

**Table 3.** The velocity dispersions ($\sigma$) of the observed lines.

| Transition | $\sigma$ (km s$^{-1}$) | $\sigma_{median}$ (km s$^{-1}$) | $\Delta\sigma$ (km s$^{-1}$) |
|---|---|---|---|
| $N_2D^+$(2–1) | 0.04–0.26 | 0.10 | 0.01 |
| $N_2H^+$(1–0) | 0.07–0.29 | 0.12 | 0.003 |
| $DCO^+$(2–1) | 0.05–0.30 | 0.16 | 0.01 |
| $H^{13}CO^+$(1–0) | 0.06–0.46 | 0.16 | 0.02 |

Two YSOs (class II and III) are close in projection to but not associated with core 8. Two cores (2 and 7) are isolated from other cores and YSOs, the other cores are clustered in chains with other starless and protostellar cores and YSOs (see Fig. 1). The integrated intensity maps are shown in Fig. A.8. All four species are detected towards all observed cores (with SNR> 5) except for $N_2D^+$(2–1) towards core 3 (detected with SNR≃ 3).

The $N_2H^+$(1–0) and $N_2D^+$(2–1) emission peaks usually match within one beam size (cores 2, 6, 7, 8, 10, 11, 13, 17, 19), however in cores 12 and 16 the peaks are offset by one or two beam widths (~30–60″, see Fig. 3). Often the $N_2H^+$(1–0) and $N_2D^+$(2–1) emission areas have different shapes, with the $N_2D^+$(2–1) being more compact. Emission peaks of $N_2H^+$(1–0) and $N_2D^+$(2–1) are associated with a position of a YSO towards the protostellar cores (11 and 13). One core (12) has two $N_2D^+$(2–1) emission peaks with the $N_2H^+$(1–0) emission peak in between (see cores 10–13 in Fig. A.8).

$H^{13}CO^+$(1–0) usually has several emission peaks within one core (3, 4, 6, 8, 10, 12, 17) and avoids the $N_2D^+$(2–1) emission peaks (2, 3, 4, 6, 10, 12, 16, 17), which is a sign of possible depletion. Towards the protostellar cores 11 and 13, $H^{13}CO^+$(1–0) and $DCO^+$(2–1) are centrally concentrated and their emission peaks are within a beam size to the YSOs. The $DCO^+$(2–1) emission follows the shape of the $N_2D^+$(2–1) emission, but it is more extended (4, 6, 7, 10, 11, 13, 16, 17) and in a few cases avoids $N_2D^+$(2–1) (2, 3, 8, 12) (see Fig. A.8).

Towards the cores 7, 11 and 13 all four species have very similar emission distribution and close peak positions.

### 4.2. Velocity dispersion

The velocity dispersions ($\sigma$) of all transitions are determined from the hfs fits. The maps of the velocity dispersions of the $N_2D^+$(2–1), $N_2H^+$(1–0), $DCO^+$(2–1), and $H^{13}CO^+$(1–0) lines are shown in Fig. A.9. The ranges of the velocity dispersions of the different lines are given in Table 3 with median values ($\sigma_{median}$) and typical uncertainties ($\Delta\sigma$) for easier comparison. The velocity dispersion increases going from higher density gas tracers ($N_2D^+$ and $N_2H^+$) to lower density gas tracers ($DCO^+$ and $H^{13}CO^+$), as expected (Fuller & Myers 1992). The fact that the $N_2D^+$(2–1) velocity dispersion is on average lower than the rest of the lines may be the cause of the observed small velocity centroid shifts (see Sect. 4.4 and Fig. 5). Tracers of the lower density parts of the core may show more material along the line of sight with slightly different velocities than those traced by $N_2D^+$ and produce the small shifts (this possibility was shown with simulations in Bailey et al. 2015).

The velocity dispersion increases towards YSOs (cores 3, 8, 11, 12, 13, 17, 19, there is a protostellar core 18 between cores 17 and 19 see e.g. Fig. 1). Nevertheless, the line widths stay very narrow across the cores and are dominated by thermal motions (see Sect. 4.3 for details). The $N_2D^+$(2–1) velocity dispersions become large (compared to the thermal linewidth) only towards the protostar in core 13 and on the edges of cores 6, 7, and 19. The $N_2H^+$(1–0) velocity dispersions become large (compared to the thermal linewidth) only towards the protostars in cores 11 and 13, and on the edges of cores 6, 7, 8, and 16. The $N_2H^+$(1–0) and $H^{13}CO^+$(1–0) lines towards cores 3, 8, 13 and 16 also have hints of a second velocity component, not clearly resolved with our observations, which may increase the velocity dispersions towards those positions (see Sect. A.2 for details). A small fraction of the $DCO^+$(2–1) and $H^{13}CO^+$(1–0) spectra show a transonic velocity dispersions towards each core. The $DCO^+$(2–1) lines show asymmetric, double, or multiple peaked lines which are probably unresolved multiple velocity components towards all cores (see Sect. A.2 for details).

### 4.3. Non-thermal motions

Figure 4 shows the ratio of the non-thermal components $\sigma_{NT}$ of the $N_2D^+$(2–1), $N_2H^+$(1–0), $DCO^+$(2–1), $H^{13}CO^+$(1–0), and $C^{18}O$(1–0) lines in each pixel of the maps and the thermal velocity dispersion of a mean particle, $\sigma_T$, as a function of core radius measured as a distance from the pixel to the $N_2H^+$(1–0) emission peak. The $C^{18}O$(1–0) velocity dispersions come from the fit results in Hacar et al. (2013). For the $C^{18}O$(1–0) plot, we take only the CO components whose $V_{LSR}$ coincide with the $V_{LSR}$ of the dense gas. The non-thermal components are derived from the observed velocity dispersion $\sigma_{obs}$ via

$$\sigma_{NT}^2 = \sigma_{obs}^2 - \frac{kT_k}{m_{obs}}, \tag{1}$$

where $k$ is Boltzmann's constant, $T_k$ is the kinetic temperature, and $m_{obs}$ is the mass of the observed molecule. The formula is adopted from Myers et al. (1991), taking into account that $\sigma^2 = \Delta v^2/(8\ln(2))$, where $\Delta v$ is the full width at half maximum of the line, FWHM. To measure the non-thermal component, we use the kinetic temperature determined by Seo et al. (2015) from ammonia observations. We use the same temperature for all five lines, the temperature towards the $N_2H^+$(1–0) peak, for each core. The variations in the kinetic temperature across the mapped core areas are within 1–2 K, which produce an uncertainty of 6–12% in the $\sigma_{NT}/\sigma_T$ ratio. Since the kinetic temperature of the gas in the studied cores usually increases towards their edges, we can expect the right hand sides of the distributions in Fig. 4 to shift to lower values by 6–12%.

The thermal velocity dispersions $\sigma_T$ for a mean particle with mass 2.33 amu are 0.17–0.21 km s$^{-1}$ for typical temperatures of 8–12 K across the cores of L1495. The majority of all four high density tracers' lines are subsonic. However going from tracers of more dense gas to tracers of less dense gas, the fraction of transonic ($1 < \sigma_{NT}/\sigma_T < 2$) lines increases: 0.8% of the $N_2D^+$(2–1) lines, 2.6% of the $N_2H^+$(1–0) lines, 19% of the $DCO^+$(2–1) lines, and 24% of the $H^{13}CO^+$(1–0) lines are transonic, while 67% of the $C^{18}O$(1–0) lines show transonic or supersonic line widths. One should remember that all considered points are on the line of sight through the cores and the ambient cloud (filament), and we consider that all high density tracers belong to dense cores (with $H^{13}CO^+$ providing also a connection to the cloud) and $C^{18}O$ represents mostly the cloud. The median $\sigma_{NT}/\sigma_T$ ratio also increases from high to low density, being 0.49, 0.58, 0.84, 0.81, and 1.16 for $N_2D^+$(2–1), $N_2H^+$(1–0), $DCO^+$(2–1), $H^{13}CO^+$(1–0), and $C^{18}O$(1–0), respectively. The maximum non-thermal to thermal velocity dispersion ratio is 1.4 for $N_2D^+$(2–1), 1.5 for $N_2H^+$(1–0), 1.6 for $DCO^+$(2–1), 2.4 for $H^{13}CO^+$(1–0), and 3.9 for $C^{18}O$(1–0). The ratio slightly decreases with radius for $N_2D^+$(2–1) and $DCO^+$(2–

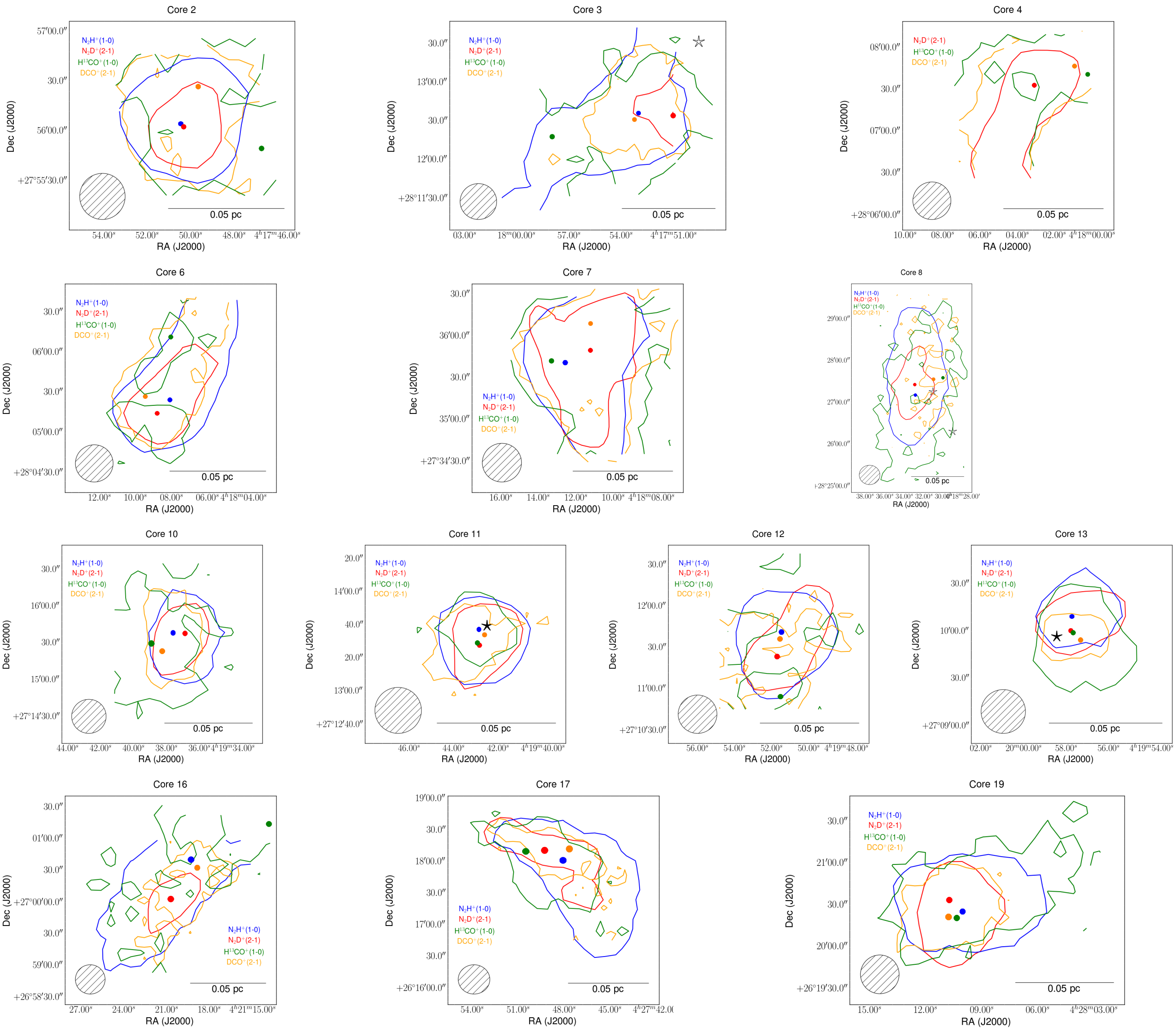

**Fig. 3.** Integrated intensities of the $N_2D^+$(2–1), $N_2H^+$(1–0), $DCO^+$(2–1), and $H^{13}CO^+$(1–0) lines across the cores. The contours show 60% of the intensity peak, filled circles show the emission peaks. Stars show the positions of young stellar objects (YSOs) from Rebull et al. (2010): black stars are young, flat and class I objects, white stars are more evolved, class II and III objects. The 27.8″ beam size of $N_2H^+$(1–0) is shown in the bottom left of each panel.

1), but stays relatively constant for $N_2H^+$(1–0) and $H^{13}CO^+$(1–0). The highest density in the points distribution occupies the same $\sigma_{NT}/\sigma_T$ range in the $N_2H^+$ and $N_2D^+$ plots (0.4–0.6), and another zone for $H^{13}CO^+$ and $DCO^+$ (0.65–0.85). We also estimate the $\sigma_{NT}/\sigma_T$ ratio for the $DCO^+$ lines with well-constrained $\tau$ (11% of all data points). The median ratio is 0.55 (which is even smaller than the ratio for $N_2H^+$). However, eliminating the lines with the unconstrained $\tau$ (89% of all data points), we exclude both multiple component lines and optically thin lines which are present mainly in the core outskirts. Nevertheless, with possibly overestimated velocity dispersions of $N_2H^+$(1–0) and $H^{13}CO^+$(1–0) and to a greater extent $DCO^+$(2–1), the lines stay subsonic and split between more narrow $N_2D^+$/$N_2H^+$ with 0.5 $\sigma_T$ and less narrow $DCO^+$/$H^{13}CO^+$ with 0.8 $\sigma_T$.

### 4.4. Velocity field: $V_{LSR}$ and velocity gradients

The centroid velocities $V_{LSR}$ of all the transitions are determined from the hfs fits. The maps of the centroid velocities of the $N_2D^+$(2–1), $N_2H^+$(1–0), $DCO^+$(2–1), and $H^{13}CO^+$(1–0) lines across the cores are presented in Fig. A.10. The range of the $V_{LSR}$ seen in all lines (6.3–7.6 km $s^{-1}$) is narrower than the velocity range seen across the entire filament in $NH_3$ (5.0–7.2 km $s^{-1}$, Seo et al. 2015) and $C^{18}O$ (4.5–7.5 km $s^{-1}$, Hacar et al. 2013), which trace both dense and diffuse gas. The velocity fields traced by the four lines are usually similar within one core. The dense gas in core 13, which hosts the class 0 protostar IRAS 04166+2706 (Santiago-García et al. 2009), has a peculiar velocity field in that the different species have significantly different velocity field morphologies. It is probably affected by the outflow of the protostar.

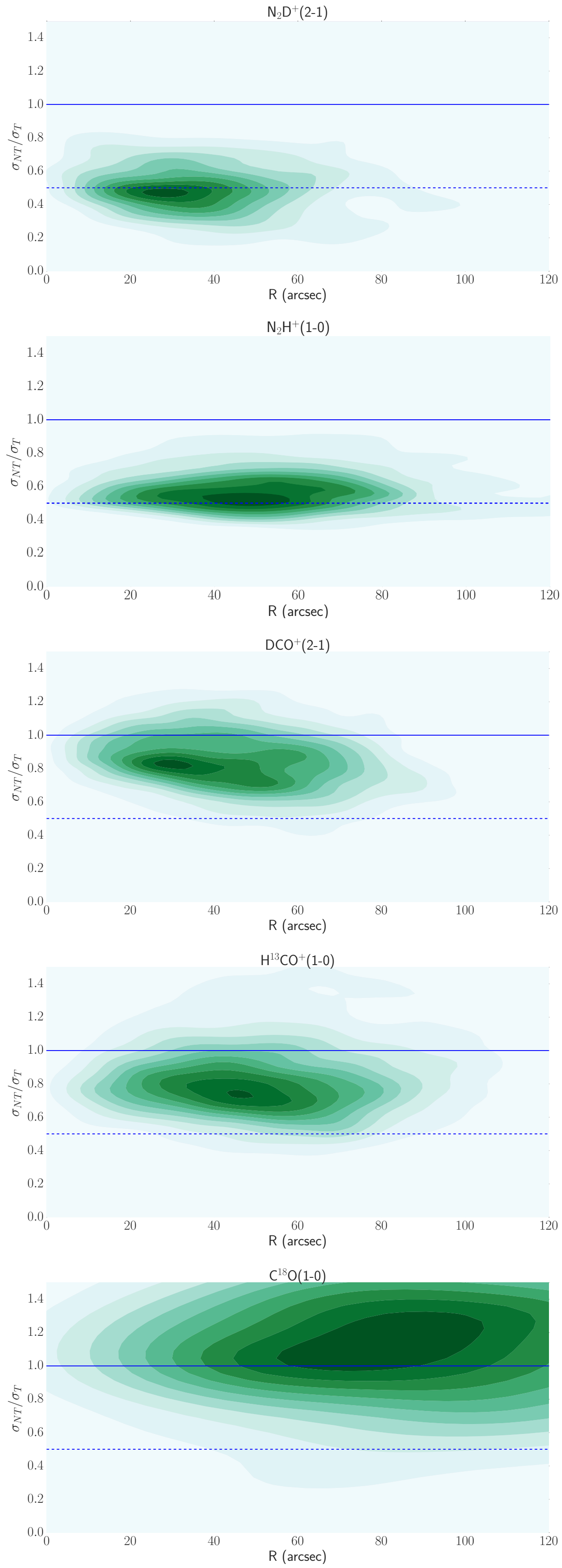

**Fig. 4.** Ratio of non-thermal components of $N_2D^+$(2–1), $N_2H^+$(1–0), $DCO^+$(2–1), $H^{13}CO^+$(1–0), and $C^{18}O$(1–0) to the thermal line width of a mean particle as a function of radius (distance from the $N_2H^+$(1–0) intensity peak) for all cores. The solid and dashed blue lines show the ratios equal to 1 and 0.5. The colourscale represents the density of the points.

Figure 5 presents velocity profiles along the fiber directions defined in Hacar et al. (2013), from core 19 in the south-east to core 8 in the north-west. Some cores (3, 7, 8, 10, 17, and 19) are located next to the places where the fibers change their directions. Core 2 is not associated with any fiber, such that we use a -45° position angle measured from the north to east for the filament direction there. The velocity changes both along and across the fibers. To better show the difference in velocity between $N_2H^+$ and $N_2D^+$ we plot data only for these molecules in Fig. A.12. Figure A.12 shows that the $N_2D^+$(2–1) velocity is less spread than that of $N_2H^+$(1–0) and the other lines towards cores 6, 8, 10, 11, 12, 13, and 16 (all observed cores in more evolved subregions B213 and B7, and core 6). Comparison of the velocity maps in Fig. A.10 shows that the smaller area of the $N_2D^+$(2–1) emission can not explain the difference. Here the densest gas is not participating in the oscillative distribution of material along the fibers as seen in $N_2H^+$ and other tracers. The highest dispersions of the centroid velocities (≃0.6 km $s^{-1}$) appear at the protostellar cores (11 and 13) and cores close to a protostar (10, 12, 17, and 19).

The $V_{LSR}$ of $DCO^+$(2–1) towards core 7 is systematically lower than that of other tracers. The $DCO^+$(2–1) line here is blended with a second velocity component and the $V_{LSR}$ has a systematic shift because of the line asymmetry (see Fig. 2). The $N_2D^+$(2–1) velocity along the filament is lower than that of $N_2H^+$ and the other species in the majority of data points. The lower $V_{LSR}$ values are associated with the smaller $N_2D^+$ velocity dispersion values (compared to those of $N_2H^+$) which confirms our suggestion that $N_2H^+$ traces more material along the line of sight with possibly different velocities.

### 4.5. Total gradients and specific angular momentum

Assuming that the cores are in solid body rotation, we estimate total and local velocity gradients across the cores following the method described in Goodman et al. (1993) for total gradients and applied for local gradients by Caselli et al. (2002a) (see Sect. 4.6 for local gradients). The results of the total velocity gradient calculation provide the average velocity across the core $< V_{LSR} >$, the magnitude of the velocity gradient $G$, and the position angle $\theta_G$. The total gradients are calculated using all available points weighted by $1/\Delta^2_{V_{LSR}}$, where $\Delta_{V_{LSR}}$ is the uncertainty of the central velocity. We also calculate the specific angular momentum as $J/M \equiv p\Omega R^2$, where $p = 0.4$ is a geometry factor appropriate for spheres, $\Omega$ is the angular velocity, derived from the velocity gradient analysis, $R$ is the radius, with the assumption that $R = \sqrt{S/\pi}$, where $S$ is the emitting area (see e.g. Phillips 1999) which is here the area of the velocity map. Not all the cores are spherical, however we assume spherical geometry to be consistent in our approach to treat all of them. The assumption of sperical geometry introduces a systematic difference by 20% to the resulting specific angular momentum for the elongated cores if we consider the rotation axis parallel to the core axis. The total gradient direction, which in our assumption is a direction of core rotation, is not always parallel to either of the core axes, in fact the angles are random, so we can not use the same approach (cylinder or disk shape) to all cores. In this situation the best approach is to assume spherical geometry.

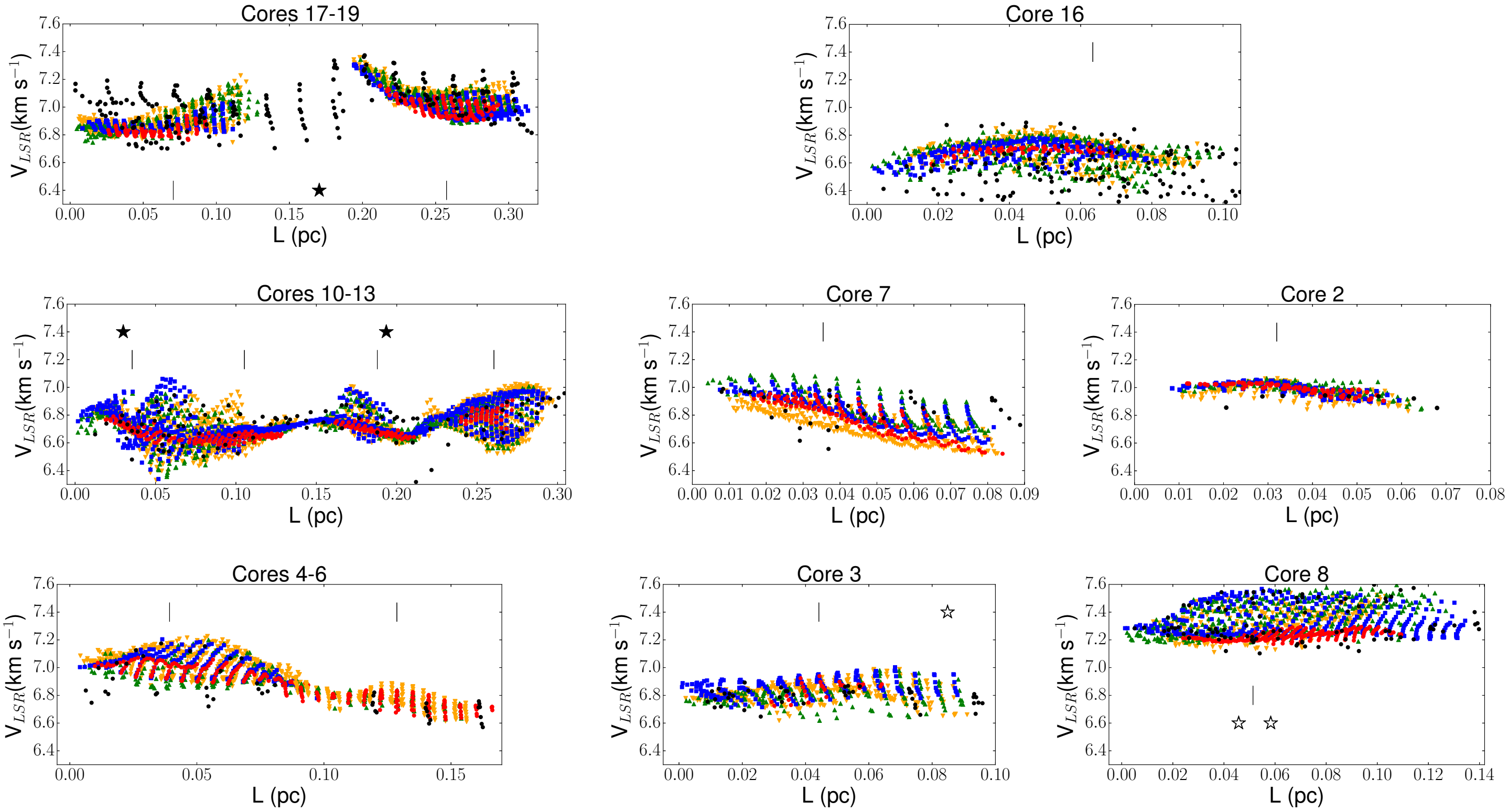

**Fig. 5.** $V_{LSR}$ along the filament direction (from core 19 in the south-east to core 8 in the north-west). The transitions are shown with colours: $N_2D^+$(2–1) – red circles, $N_2H^+$(1–0) – blue squares, $DCO^+$(2–1) – orange flipped triangles, $H^{13}CO^+$(1–0) – green triangles, and $C^{18}O$(1–0) – black circles. The vertical bars show the $N_2H^+$(1–0) emission peaks. Stars show the positions of YSOs from Rebull et al. (2010): black stars are flat and class I objects, white stars are class II and III objects.

### 4.5.1. Total gradients and specific angular momentum measured over all detected emission

At first, $< V_{LSR} >$, $G$, $\theta_G$, $R$ and $J/M$ are calculated for all emitting areas for each species (the numbers are given in Table B.1). The total gradients of the four species with their position angles are shown in Fig. B.1. One can expect that higher density tracers have smaller gradient values than lower density tracers, as the decrease in velocity gradient values should trace the loss of the corresponding specific angular momentum towards the small scales (Crapsi et al. 2007; Belloche 2013). That means that velocity gradients should increase in a sequence $N_2D^+ \rightarrow N_2H^+ \rightarrow DCO^+ \rightarrow H^{13}CO^+$. Only core 19 obeys this sequence. Four cores (4, 6, 8, and 16) increase their gradients in a sequence $N_2D^+ \rightarrow N_2H^+ \rightarrow H^{13}CO^+ \rightarrow DCO^+$ (although for core 4 there is no $N_2H^+$ data and gradients of $DCO^+$ and $H^{13}CO^+$ differ within the uncertainties, by $\simeq$1%). Three cores (11, 12, and 13), belonging to one core chain, increase their gradients in a sequence $N_2H^+ \rightarrow N_2D^+ \rightarrow H^{13}CO^+ \rightarrow DCO^+$. Core 17 has the sequence $N_2H^+ \rightarrow N_2D^+ \rightarrow DCO^+ \rightarrow H^{13}CO^+$; core 10 has the sequence $N_2D^+ \rightarrow H^{13}CO^+ \rightarrow N_2H^+ \rightarrow DCO^+$. Cores 2 and 3 have the sequence $H^{13}CO^+ \rightarrow DCO^+ \rightarrow N_2H^+ \rightarrow N_2D^+$ (core 2 has equal $DCO^+$ and $H^{13}CO^+$ gradients, and core 3 has an unreliable detection of $N_2D^+$ with SNR=3). Core 7 has the sequence $DCO^+ \rightarrow H^{13}CO^+ \rightarrow N_2H^+ \rightarrow N_2D^+$. The differences between the gradient values are significant taking the errors into account. If we assume that $N_2H^+$ and $N_2D^+$ equally well trace the dense central part of a core, and $DCO^+$ and $H^{13}CO^+$ equally well trace the more diffuse envelope of a core, we have 10 out of 13 cores which follow the expectations about total gradient increase. There are three cores (2, 3, and 7) which show a total velocity gradient increase towards the denser gas (similar to L1544, Caselli et al. 2002b). Cores 2 and 7 are isolated starless cores, whereas core 3 has an evolved YSO nearby. Among them only core 7 shows a coherent velocity field and very likely is in solid body rotation.

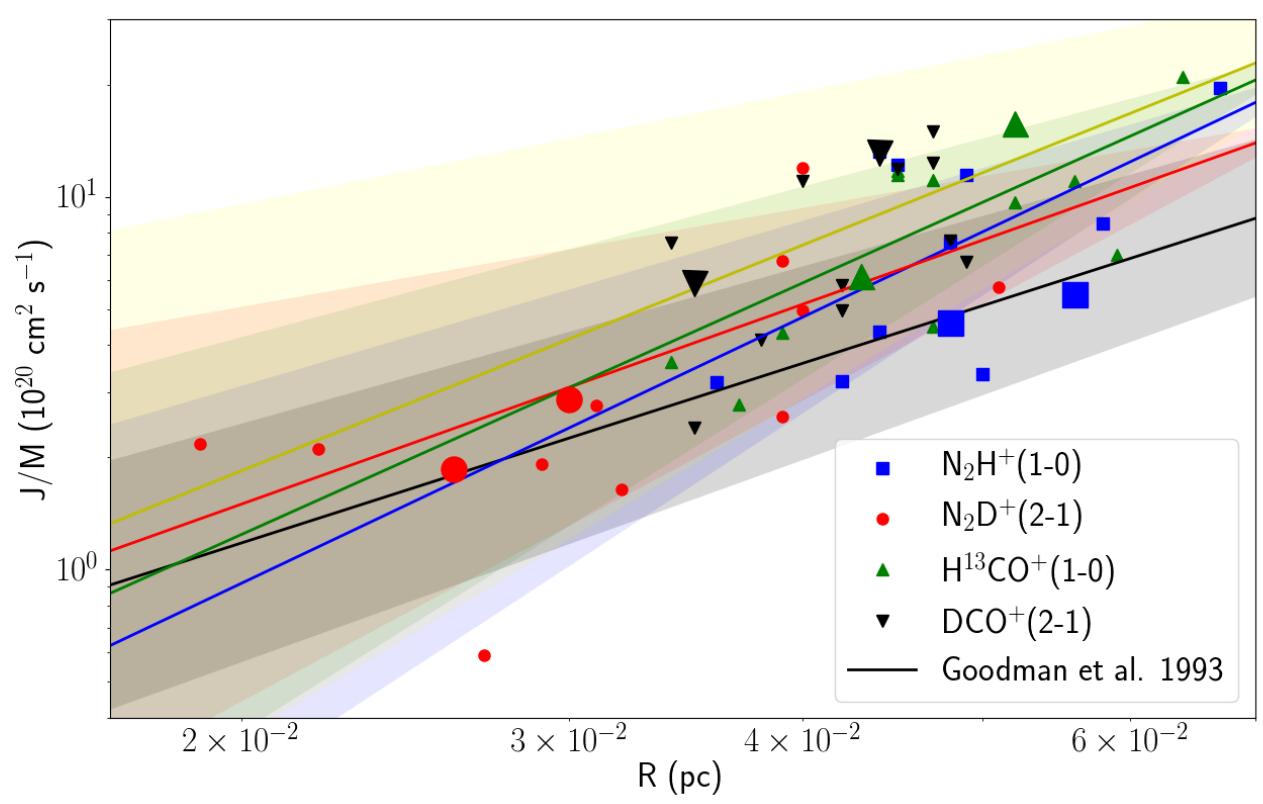

**Fig. 6.** Specific angular momentum as a function of core radius, measured with different lines. Large symbols show protostellar cores, small symbols show starless cores. The lines are the best fits of a power-law function $aR^b$ calculated for each species ($N_2H^+$(1–0) – blue, $N_2D^+$(2–1) – red, $H^{13}CO^+$(1–0) – green, $DCO^+$(2–1) – yellow), $a$ and $b$ values are given in the main text. The black line shows the relation found by Goodman et al. (1993). Thick light-colour strips represent the accuracies of the fits ($N_2H^+$(1–0) – light blue, $DCO^+$(2–1) – light orange, Goodman et al. (1993) – grey).

Specific angular momentum varies in a range (0.6–21.0)$\times10^{20}$ cm$^2$ s$^{-1}$ with a typical error of (0.02–0.21)$\times10^{20}$ cm$^2$ s$^{-1}$ at core radii between 0.019–0.067 pc. It is similar to the results for other dense cores ($J/M$ = (6.2–30.9) $\times$ $10^{20}$ cm$^2$ s$^{-1}$ at radii 0.06–0.60 pc and $J/M$ = (0.04–25.7) $\times$ $10^{20}$ cm$^2$ s$^{-1}$ at radii 0.018–0.095 pc, Goodman et al. 1993; Caselli et al. 2002a, respectively). Figure 6 shows the correlation between specific angular momentum and radius of the core. For each species we find a best-fit power law correlation $aR^b$: for $N_2D^+$(2–1) $J/M = 10^{23\pm1}R^{1.8\pm0.8}$ cm$^2$ s$^{-1}$; for $N_2H^+$(1–0) $J/M = 10^{24\pm1}R^{2.4\pm0.9}$ cm$^2$ s$^{-1}$; for $DCO^+$(2–1) $J/M = 10^{24\pm1}R^{2.0\pm1.1}$ cm$^2$ s$^{-1}$; and for $H^{13}CO^+$(1–0) $J/M = 10^{24\pm1}R^{2.2\pm0.7}$ cm$^2$ s$^{-1}$. These numbers agree with those found by Goodman et al. (1993) with ammonia observations: $J/M = 10^{22.8\pm0.2}R^{1.6\pm0.2}$ cm$^2$ s$^{-1}$. We note that the radii used to calculate $J/M$ here are larger than the radii at FWHM of emission. The significant difference between FWHM and equivalent radius of emission (see Fig. B.4) shows that the common approach to use the FWHM as size of a core may not be correct. However if we use FWHM as a size of the core we also see the decrease in the specific angular momentum at higher densities in the majority of the cores (see Table B.4). All data used to calculate $J/M$ have signal-to-noise ratio > 5 except for the $N_2D^+$ data for core 3 with SNR$\simeq$ 3.

#### 4.5.2. Total gradients and specific angular momentum measured over the maps with uniform area and resolution

To compare the total velocity gradients for the four species on the same size scale, we convolve all maps to the spatial resolution of 29.9″, corresponding to the $H^{13}CO^+$(1–0) beam size, with Nyquist grid spacing, and use only the area where the emission in all four species is detected, which matches where $N_2D^+$(2–1) is detected, because it has the most compact emission. The results of the gradient calculations are given in Table B.2.

We compare the results for $N_2D^+$(2–1) with the results obtained with the other species in Fig. B.2. The average centroid velocity $< V_{LSR} >$ of $N_2D^+$(2–1) agrees with the $< V_{LSR} >$ of the other species within 1.5%, however for most of the cores the $N_2D^+$(2–1) velocity is systematically lower (see Fig. B.2, a). The total centroid velocity gradient $G$ of $N_2D^+$(2–1) for the majority of cores is also lower than $G$ measured with other species. The range of the total gradients seen in all species is 0.66–7.35 km s$^{-1}$ pc$^{-1}$ with the errors in a range 0.02–0.14 km s$^{-1}$ pc$^{-1}$. The $N_2D^+$(2–1) line is most similar to the $N_2H^+$(1–0) line, with the maximum difference being 1.9 km s$^{-1}$pc$^{-1}$ at core 8, and the median difference being 0.5 km s$^{-1}$pc$^{-1}$. The difference with $H^{13}CO^+$(1–0) and $DCO^+$(2–1) is more significant. The maximum difference between $N_2D^+$(2–1) and $H^{13}CO^+$(1–0) is 2.8 km s$^{-1}$ pc$^{-1}$ at core 16. The median difference is 1.1 km s$^{-1}$ pc$^{-1}$. The maximum difference between $N_2D^+$(2–1) and $DCO^+$(2–1) is 3.8 km s$^{-1}$ pc$^{-1}$ at core 16. The median difference is 1.2 km s$^{-1}$ pc$^{-1}$.

The directions of the gradients seen in different species are generally in better agreement. The typical errors of the determination of the position angle are 0.4–3.0°. The best agreement is between $N_2D^+$(2–1) and $DCO^+$(2–1); the biggest difference is 94° at core 16, and the median difference is 10°. The biggest difference between $N_2D^+$(2–1) and $H^{13}CO^+$(1–0) is 118° at core 16, and the median difference is 11°. The biggest difference between $N_2D^+$(2–1) and $N_2H^+$(1–0) is 68° at core 16, and the median difference is 12°. Thus, the median difference in the total gradient directions between the species is the same within the uncertainties.

The specific angular momentum varies in a range of (0.7–12.2)$\times10^{20}$ cm$^2$ s$^{-1}$ with a typical error of (0.02–0.31)$\times10^{20}$ cm$^2$ s$^{-1}$. $J/M$ measured in $N_2D^+$(2–1) agrees better with $J/M$ measured in $N_2H^+$(1–0) (median difference is 14%, maximum difference is 114% at core 8). $J/M$ measured in $N_2D^+$(2–1) is systematically lower than $J/M$ measured in $H^{13}CO^+$(1–0) and $DCO^+$(2–1) (median differences with $N_2D^+$(2–1) are 40 and 39%, maximum differences 425 and 578% at core 16) with the exception of core 7.

We also compare the results for $DCO^+$(2–1) and $H^{13}CO^+$(1–0) in Fig. B.3. The average centroid velocities of the two lines agree within 0.7%. The total centroid velocity gradient of the two species stays along the one-to-one correlation with no systematic difference. A median scatter of $G$ is 0.9 km s$^{-1}$ pc$^{-1}$ and the maximum difference is 3.6 km s$^{-1}$ pc$^{-1}$ at core 3. The directions of the gradients of $DCO^+$(2–1) and $H^{13}CO^+$(1–0) are in good agreement, with the difference in position angle varying from 0.4–24° and having a median difference of 6°. The specific angular momentum of the two lines agrees within (0.03–2.2)$\times10^{20}$ cm$^2$ s$^{-1}$ with a maximum difference at core 8.

### 4.6. Local gradients

The local velocity gradients are presented in Fig. B.5 with black arrows, plotted over integrated intensity colour maps of the corresponding molecular transition along with red arrows which present local velocity gradients of $C^{18}O$(1–0) calculated from the Hacar et al. (2013) fit results (a comparison with the $C^{18}O$ local velocity gradients is given in Sect. 5.2). The maps are convolved to a 29.9″ beam with Nyquist sampling. To calculate a gradient in a local position, we use all pixels closer than two pixel size distance, weighted according to their distance to the given position and their uncertainty of the central velocity:

$$W = \frac{1}{\Delta^2_{\mathrm{V_{LSR}}}} \cdot \exp\left\{-d^2/\left[2\left(\frac{\theta_{Gauss}}{2.354}\right)^2\right]\right\}, \qquad (2)$$

where $W$ is the weight, $\Delta_{\mathrm{V_{LSR}}}$ is the uncertainty of the central velocity, $d$ is the distance from the weighted pixel to the given position, and $\theta_{Gauss} = 2$ is the FWHM of the weighting function.

Cores 2, 4, and 16 increase their velocities towards the core centres, while core 17 shows velocities decreasing towards the core centre (the arrows shown in Fig. B.5 point towards higher $V_{LSR}$). For a given core, the velocity fields of the different species are typically fairly similar.

There are, however, some variations from one species to the other, in particular towards the protostellar cores. Core 13 has the most complex velocity field with the $N_2H^+$(1–0) data showing four different velocity gradient directions in various areas across the core. Only two of these gradients can be seen in the $N_2D^+$(2–1) data due to the smaller detected extent of the $N_2D^+$(2–1) emission. The significant change of the gradient direction on a small spatial scale, which is also present in core 11 and less prominent in cores 3, 8, and 12 is characteristic for a protostellar core, first revealed by Crapsi et al. (2004) towards another Taurus core, L1521F. This would imply that the last three cores known as starless may be relatively highly evolved. The difference between $N_2D^+/N_2H^+$ and $DCO^+/H^{13}CO^+$ can represent some differential gas motions between the core envelope and the central regions.

## 5. Discussion

### 5.1. Velocity dispersions

Pineda et al. (2010) define the coherent dense core as a region with nearly constant subsonic non-thermal motions. The linewidth distribution we see in the dense gas tracers, $N_2D^+$(2–1) and $N_2H^+$(1–0), shows constant subsonic ($\simeq \sigma_T/2$) non-thermal motions (see Fig. 4), with $N_2D^+$ having $\simeq$20% lower non-thermal components than $N_2H^+$, consistent with these observations tracing the coherent centres of dense cores in L1495.

Pineda et al. (2010) observed a sharp transition to coherence, where the line widths of $NH_3$(1,1) increased by a factor of at least two over a distance of less then 0.04 pc, which corresponds to their beam size. We do not see any reliable sign of the transition to coherence in any of the tracers that we have observed. One can explain this by the fact that the $N_2D^+$(2–1) line, which traces the high density gas, has a significantly higher critical density than the inversion transitions of $NH_3$. Thus, the $N_2D^+$ line intensity drops faster towards the edge of the core because the excitation temperature decreases more rapidly than the one of the $NH_3$ inversion line. Similarly, for most of the cores, the next highest density gas tracers, the $N_2H^+$(1–0) and $DCO^+$(2–1) lines (both having critical densities higher than that of $NH_3$(1,1)), are detected over larger areas and still do not show any sudden increase in line width, as seen in $NH_3$. Towards those cores where the emission strongly decreases away from the core centres (cores 6 and 17 for $N_2H^+$ and cores 8, 11, 13, 16, and 17 for $DCO^+$), the line widths at the edges do not increase significantly from the line widths towards the centre of the cores, or even decrease. However, even the $NH_3$ maps by Seo et al. (2015), which cover larger areas than our maps, still do not show a significant increase in the velocity dispersions towards the core edges.

We see however supersonic line widths in $H^{13}CO^+$, with the line widths reaching 2.4 $\sigma_T$ in some places. Its transonic non-thermal motions lie mostly between 1 and 1.5 $\sigma_T$, consistent with the typical observed $C^{18}O$ non-thermal component (see Fig. 4). $H^{13}CO^+$ tends to be slightly (cores 2, 6, 8, 10) or significantly (cores 4, 12, 16) depleted towards the core centres and better traces the core envelopes compared to $N_2D^+$ and $N_2H^+$. The fact that the median velocity dispersions of $N_2D^+$(2–1) and the $N_2H^+$(1–0) are 1.3–1.6 times lower than the median velocity dispersions of $DCO^+$(2–1) and $H^{13}CO^+$(1–0) indicates that the velocity dispersion is indeed increasing outwards.

### 5.2. Connection to the filament scale

To search for a connection from the cores and their envelopes to the filament scale, we estimate local and total $V_{LSR}$ gradients of the CO and compare the velocity field patterns and total gradients to the patterns seen in each species we observed. For the CO data we use the fit results of the $C^{18}O$(1–0) mapping done by Hacar et al. (2013), which created maps convolved to a 60″ beam size. In their work, Hacar et al. (2013) found multiple velocity components of the $C^{18}O$(1–0) line and revealed the fiber structure of the L1495 filament. They classify the fibers as "fertile", if they host dense cores and coincide with $N_2H^+$ emission, and "sterile" if they do not. For our study, we take only the $C^{18}O$(1–0) components that have a $V_{LSR}$ close to the $V_{LSR}$ of our lines (6.3–7.6 km s$^{-1}$).

We show the maps of local velocity gradients in Fig. B.5. Here, black arrows represent the velocity gradients of $N_2D^+$(2–1), $N_2H^+$(1–0), $DCO^+$(2–1), and $H^{13}CO^+$(1–0) and red arrows represent the velocity gradients of $C^{18}O$(1–0). For the majority of the cores, the $C^{18}O$ velocity gradient value $G$ is lower than the $G$ of the other lines. The velocity field of the dense gas is affected by protostars both in protostellar cores (11 and 13 as seen in all species) and cores sitting next to protostellar cores and YSOs (3, 8, 10, 12, 16, 17, and 19 as seen mostly in $H^{13}CO^+$, see Sect. 4.6). While the $C^{18}O$ velocity fields do not show the same level of complexity as that of the dense gas tracers, the velocity field of the $C^{18}O$ is by no means uniform and exhibits changes in direction in cores 3, 8, 10, 16, and 19.

In cores 3, 6, 12, and 19 the velocity pattern of $C^{18}O$ matches the patterns of the higher density tracers, with the agreement being better between $C^{18}O$, $N_2H^+$, and $N_2D^+$. In cores 4, 11, 13, and 17, $C^{18}O$ has uniform velocity patterns which match some velocity patterns seen in $N_2H^+$, $DCO^+$, and $H^{13}CO^+$. In the northern half of core 8 the velocity pattern of $C^{18}O$ matches the velocity patterns of $N_2D^+$, $DCO^+$, and $H^{13}CO^+$ (the velocity increases to the north-west) while in the southern half of the core the $C^{18}O$ velocity gradient is perpendicular to the gradients of the other species. Here, the dense gas tracers all increase in velocity towards the south-west, towards a YSO, whereas the CO velocity increases to the north-west. In core 16 the velocity of the high density tracers increases towards the core centre and the velocity of the $C^{18}O$ increases towards a position $\sim 1'$ to the north of the core, so the pattern looks similar but shifted, although this could be insignificant given the 60″ beam size of the $C^{18}O$ data. In core 2 the velocity of the high density tracers also increases towards the core centre and the velocity of the $C^{18}O$ uniformly increases towards the south-west of the core. Core 10 has curved velocity fields both in the $C^{18}O$ and high density tracers, although the $C^{18}O$ velocity field is roughly perpendicular to the velocity fields of the other tracers towards most locations. Core 7 has coherent velocity fields both in $C^{18}O$ and high density tracers (local gradients arrows almost parallel in $C^{18}O$ and divergent in the other species), oriented with an angle to each other which changes from 90° to 0°. In general, $C^{18}O$ show similar velocity patterns in the starless cores as high density tracers, with less complexity. $C^{18}O$ does not show any affect from embedded protostars in the protostellar cores, which makes the $C^{18}O$ velocity gradients almost perpendicular to those of the high density tracers.

The total gradients are shown in Fig. B.1. The difference between $C^{18}O$ and the higher density tracers is more significant in the overall gradient directions due to the additional complexity of the higher density tracers' velocity fields. Cores 4 and 6 show a good correlation between $C^{18}O$ and the other species, with the largest difference being the 8° separation between the $C^{18}O$ and $H^{13}CO^+$ gradients in core 6. Cores 3, 6, 7, 8, 10, 11, 12, 17, and 19 show directions for the gradient of $C^{18}O$ which differ less than 90° (20–83°) from that of the dense gas tracers. The $C^{18}O$ gradient differs by more than 90° from that of at least one dense gas tracer in cores 2, 13, and 16, although the total velocity gradient of $C^{18}O$ towards core 2 has a large error of 40° owing to the small number of data points. Core 13 has very complex velocity fields of the higher density tracers due to the strong affect of the embedded protostar. Core 16 has complex velocity patterns both in $C^{18}O$ and the higher density tracers which affect the total gradient. For the majority of the cores. The $C^{18}O$(1–0) velocity field coincides with the high density tracers' but does not show the variations near YSOs (cores 8, 11, 12, 13, 17, and 19) seen in high density tracers.

The value of the $C^{18}O$ total velocity gradient (0.3–3.3 km s pc$^{-1}$ measured with a typical error of 0.03–0.17 km s pc$^{-1}$) is usually lower than that of the other species (0.6–6.1 km s pc$^{-1}$ in cores 2, 3, 6, 7, 8, 10, 13, and 16) or lower than that of only $DCO^+$ and $H^{13}CO^+$ (cores 4, 11, 12, 17, and 19). This may mean that the denser material is spinning up in the process of core formation.

### 5.3. Velocity gradient and large-scale polarization

Caselli et al. (2002b) found that in another Taurus core, L1544, the directions of the total gradients observed in $DCO^+$(2–1) and $N_2H^+$(1–0) are similar to the direction of the magnetic field. In Fig. 7 we compare the total velocity gradients of our $N_2D^+$(2–1), $N_2H^+$(1–0), $DCO^+$(2–1), and $H^{13}CO^+$(1–0) maps and $C^{18}O$(1–0) from Hacar et al. (2013) with the polarization directions measured with optical (Heyer et al. 1987) and infrared (Goodman et al. 1992; Chapman et al. 2011) observations. The magnetic field is parallel to the polarization direction. Chapman et al. (2011) point out that where the filament turns sharply to the north (above core 7 or starting from the B10 subregion and farther up in Fig. 1), the magnetic field changes sharply from being perpendicular to the filament to being parallel to the filament.

To quantify the alignment of the total velocity gradients with the polarization directions, we plot the minimum angle between the corresponding position angles (see left panel of Fig. 8). In Fig. 8 the grey and cyan strips show the directions within 10° of being parallel or perpendicular to the magnetic field, respectively, because the typical uncertainty of the position angle difference is of the order of 10°. There is no big difference between the species in the alignment with the polarization directions. We compare the alignment of the polarization directions with a random angle distribution (see right panel of Fig. 8). The alignment between the total velocity gradient and polarization directions is comparable to a random distribution: 27% of the gradient directions are parallel to the polarization directions and 14% of the random directions are parallel to the polarization directions; 13% of the gradient directions are perpendicular to the polarization directions and 14% of the random directions are perpendicular to the polarization directions.

We also compare the gradient directions with the directions of the fibers which host the cores and test if the alignment is significant compared to a random distribution of the angles (see Fig. 9). The alignment between the total gradient directions and the fibers' directions is comparable to a random distribution: 10% of the gradient directions are parallel to the fibers' directions and 10% of the random directions are parallel to the fibers' directions; 7% of the gradient directions are perpendicular to the fibers' directions and 8% of the random directions are perpendicular to the fibers' directions. Figure 10 compares cumulative distributions of the angles between the total gradients, polarization, fiber directions, and random directions.

### 5.4. Dynamical state of the cores

Six out of 13 cores (6, 7, 8, 10, 12, and 19, all starless) show relatively coherent velocity fields consistent with solid body rotation. Five of the listed above cores show significantly lower total gradients in higher density tracers ($N_2D^+$ and $N_2H^+$) than in lower density tracers ($DCO^+$ and $H^{13}CO^+$) on a core scale, which implies that denser material is spinning down. This does not contradict to the fact that all four core tracers show higher total velocity gradients than $C^{18}O$. At the level of cloud–core transition, the core material, namely it's external envelope traced by $DCO^+$ and $H^{13}CO^+$ is spinning up, consistent with conservation of specific angular momentum. However within the core at higher densities the central material traced by $N_2D^+$ and $N_2H^+$ is spinning down. The decrease in specific angular momentum with density may be caused by such mechanisms as magnetic braking, gravitational torques, and the transfer of angular momentum into the orbital motions of fragments if further material fragmentation takes place in the cores as explained by Belloche (2013). However our current data do not allow us to investigate the possible fragmentation and the dynamics of the material within the cores at smaller scales because of the limited angular resolution. These five cores may be more evolved than the other cores not listed here. Three starless cores (3, 8, and 12) show complex local velocity gradient patterns resembling those of protostellar cores, thus they may contain a very low luminosity object, as in the case of L1521F (Crapsi et al. 2004; Bourke et al. 2006).

Emission peaks of $N_2H^+$(1–0) and $N_2D^+$(2–1) are associated with the position of a YSO towards both protostellar cores, 11 and 13. This means that these YSOs are very young or very low luminosity, so that they have not yet affected the chemistry of their surrounding. One core (12) has two $N_2D^+$(2–1) emission peaks with the $N_2H^+$(1–0) emission peak in between (see Fig. A.8), which could be a sign of further core fragmentation or binary formation.

Four cores (6, 7, 12, and 17) show that the $N_2D^+$(2–1) emission is more elongated and compact than the emission from the other lines. This may mean that the $N_2D^+$ is tracing the higher density gas, which may be contracting along the magnetic field lines, thus producing the elongated structure.

## 6. Conclusions

This paper presents maps of four high density tracers, $N_2D^+$(2–1), $N_2H^+$(1–0), $DCO^+$(2–1), and $H^{13}CO^+$(1–0), towards 13 dense cores (starless and protostellar) along the L1495 filament. We use $N_2D^+$(2–1) and $N_2H^+$(1–0) as tracers of the core central regions, and $DCO^+$(2–1) and $H^{13}CO^+$(1–0) as tracers of the core envelope. We measure velocity dispersions, local and total velocity gradients, and specific angular momenta. We connect the core-scale kinematics traced by these high-density tracers to the filament-scale kinematics traced by the $C^{18}O$(1–0) observations presented in Hacar et al. (2013). We study the variations in the dense gas kinematics along the filament. Our main findings are:

1. All studied cores show similar kinematic properties along the 10 pc-long filament. They have similar central velocities (6.3–7.6 km s$^{-1}$), similar velocity dispersions (mostly subsonic), same order of total velocity gradient magnitudes (0.6–6 km s pc$^{-1}$), same order of specific angular momentum magnitudes (~$10^{20}$ cm$^2$ s$^{-1}$). This is consistent with the overall coherent structure of the whole filament (Hacar et al. 2013).
2. $N_2D^+$(2–1) shows the most centrally concentrated structure, followed by $N_2H^+$(1–0), $DCO^+$(2–1), and $H^{13}CO^+$(1–0) which is consistent with chemical models of centrally concentrated dense cores with large amount of CO freeze-out onto dust grains.
3. The cores show very uniform velocity dispersions. The $N_2D^+$(2–1) and $N_2H^+$(1–0) lines are subsonic, close to 0.5 $\sigma_T$ with thermal linewidth $\sigma_T$=0.17–0.21 km s$^{-1}$. The $DCO^+$(2–1) and $H^{13}CO^+$(1–0) lines are mostly subsonic, close to 0.8 $\sigma_T$, and partly transonic. The non-thermal contribution to the velocity dispersion increases from higher to lower density tracers. The median $\sigma_{NT}/\sigma_T$ are 0.49, 0.58,

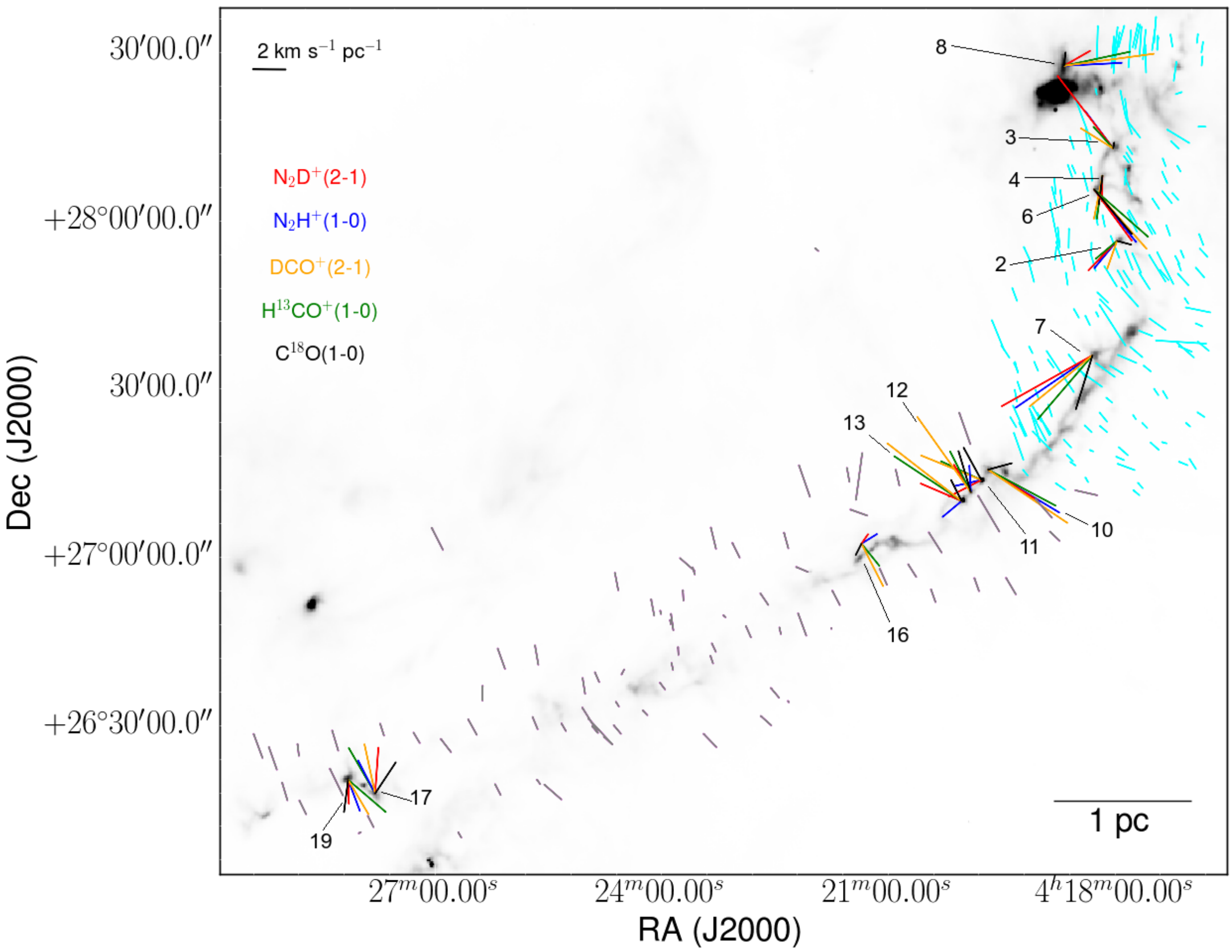

**Fig. 7.** Polarization in L1495 measured with optical (Heyer et al. 1987) (grey segments) and infrared (Goodman et al. 1992; Chapman et al. 2011) observations (cyan segments) and total velocity gradients across the cores ($N_2D^+$ – red, $N_2H^+$ – blue, $DCO^+$ – orange, $H^{13}CO^+$ – green, and $C^{18}O$ – black). The colourscale shows the dust continuum emission at 500 $\mu m$ obtained with *Herschel*/SPIRE (Palmeirim et al. 2013).

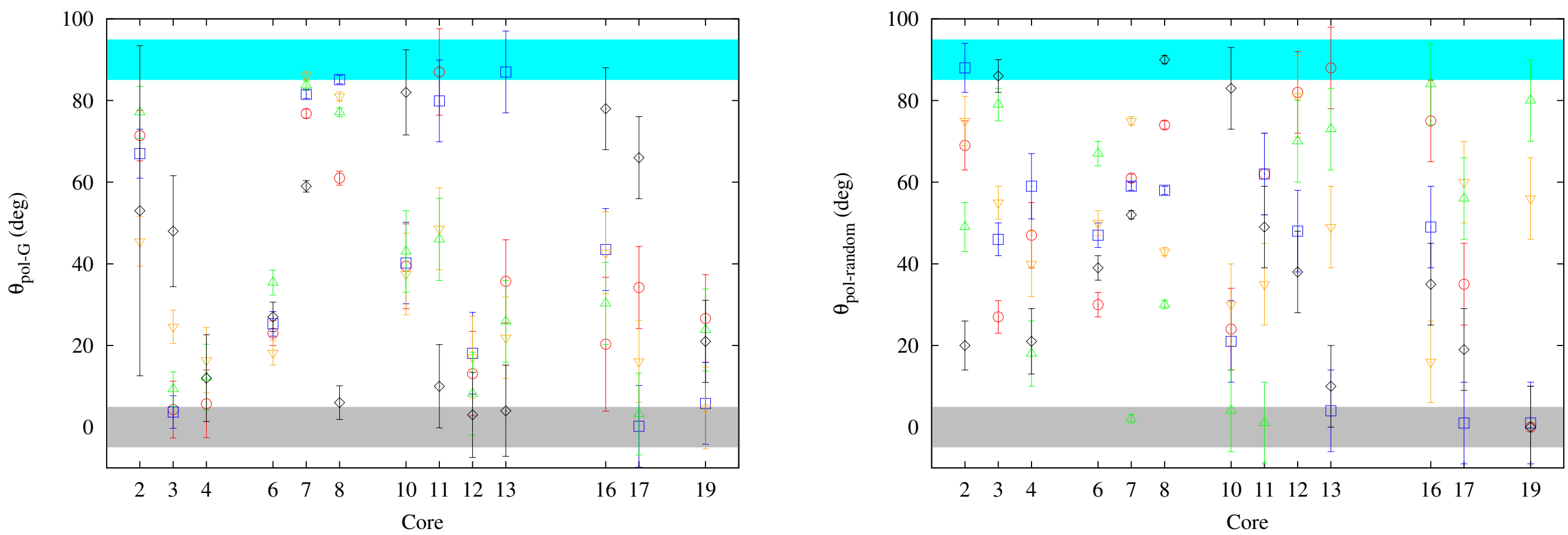

**Fig. 8.** Left: difference between polarization angles $\theta_{pol}$ and position angles of the total gradients $\theta_G$ for each core ($N_2D^+$ – red circles, $N_2H^+$ – blue squares, $DCO^+$ – orange flipped triangles, $H^{13}CO^+$ – green triangles, and $C^{18}O$ – black diamonds). Right: difference between $\theta_{pol}$ and a set of random angles $\theta_{random}$ from 0 to 90°. Gray strip corresponds to parallel directions ±10°, cyan strip corresponds to perpendicular directions ±10°.

0.84, 0.81, and 1.16 for $N_2D^+$(2–1), $N_2H^+$(1–0), $DCO^+$(2–1), $H^{13}CO^+$(1–0), and $C^{18}O$(1–0), respectively. The non-thermal contribution increases by ≃20% from $N_2D^+$ to $N_2H^+$, and by ≃40% from $N_2H^+$ to $DCO^+$ and $H^{13}CO^+$ and further to $C^{18}O$.

4. The total velocity gradients of the observed core tracers show a variety of directions and values. For 10 out of 13 cores the higher density tracers $N_2D^+$(2–1) and $N_2H^+$(1–0) show lower gradients than the lower density tracers $DCO^+$(2–1) and $H^{13}CO^+$(1–0), implying a loss of specific angular mo-mentum at small scales due to magnetic breaking, gravitational torques, and/ or the transfer of angular momentum into the orbital motions of fragments, as theoretically expected (e.g. Belloche 2013). The other 3 cores present different behaviour: core 7, the most isolated and located at the position where the magnetic field changes direction, presents a higher gradient in the higher density tracers; cores 2 and 3 present a complex kinematic field both in the low and high density tracers which prevents a meaningful determination of the total velocity gradient. The change of magnitude and direction of the total velocity gradients depending on the tracer indicates that internal motions change at different depths within the cloud.

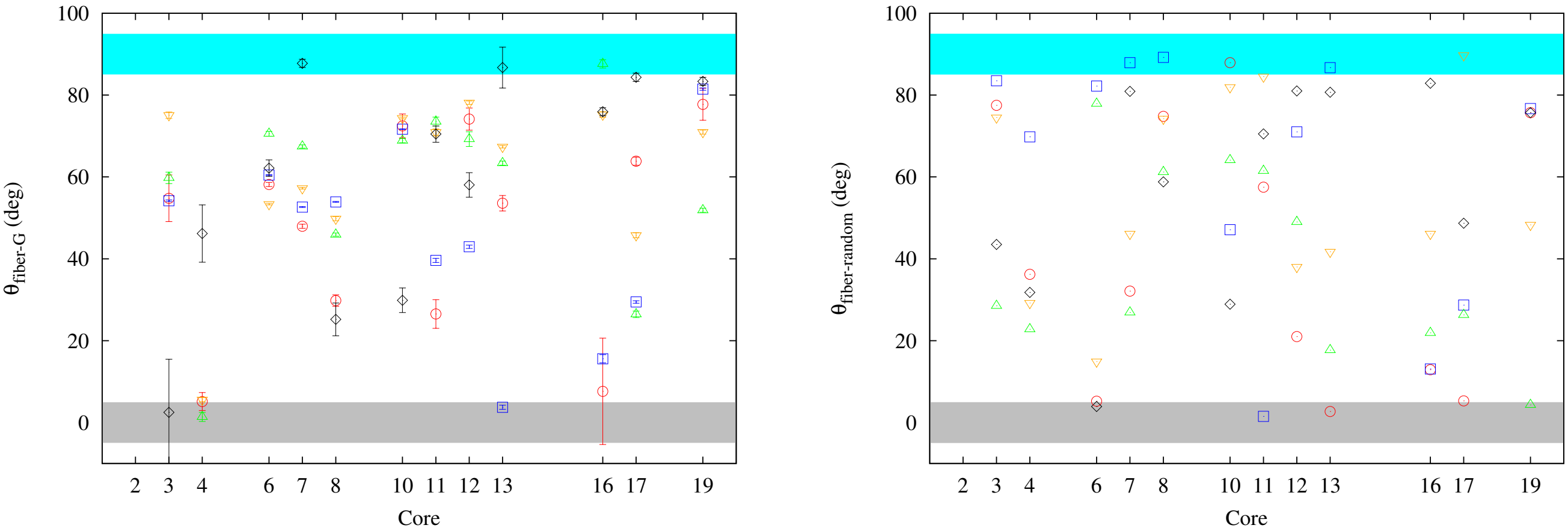

**Fig. 9.** Left: difference between fiber position angles $\theta_{fiber}$ and position angles of the total gradients $\theta_G$ for each core ($N_2D^+$ – red circles, $N_2H^+$ – blue squares, $DCO^+$ – orange flipped triangles, $H^{13}CO^+$ – green triangles, and $C^{18}O$ – black diamonds). Right: difference between $\theta_{fiber}$ and a set of random angles $\theta_{random}$ from 0 to 90°. Gray strip corresponds to parallel directions ±10°, cyan strip corresponds to perpendicular directions ±10°.

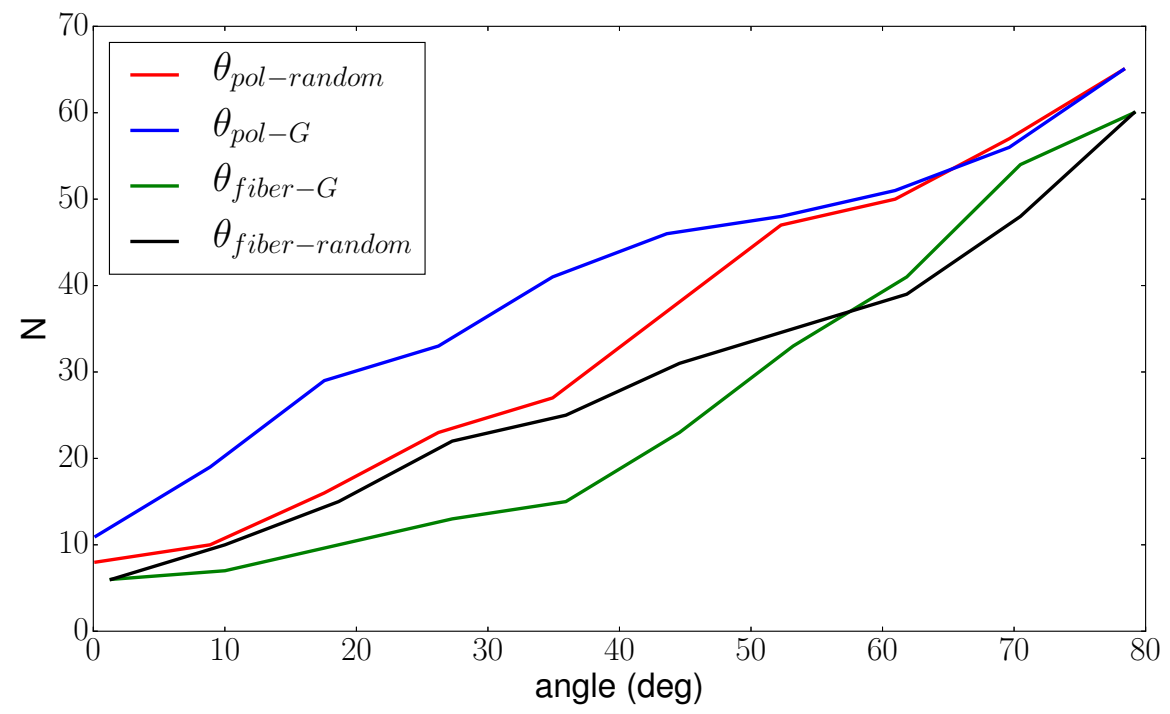

**Fig. 10.** Cumulative distribution of angles between the random angles and polarization (red), the total gradients and polarization (blue), the total gradients and the fiber directions (green), and the random angles and the fiber directions (black).

5. The above point is strengthened by looking at local velocity gradients, which show complex patterns, especially in comparison with the velocity field traced by $C^{18}O$ (although with lower spatial resolution). Half of the cores (6 out of 13) show a velocity pattern consistent with solid body rotation. The velocity fields are mainly similar in different high density tracers, however $DCO^+$(2–1) and $H^{13}CO^+$(1–0) show more local variations because they also trace some of the additional motions within the extended core envelope. The presence of a YSO at a distance less than 0.1 pc from a core locally affects the core's velocity field.
6. $C^{18}O$ traces the more extended cloud material whose kinematics is not affected by the presence of dense cores: it's total velocity gradient is always lower than that of $DCO^+$ and $H^{13}CO^+$, the core envelope tracers. For the majority of the cores, the $C^{18}O$ velocity field coincides with that of the high density tracers, but does not show the variations near YSOs (cores 8, 11, 12, 13, 17, and 19) seen in the high density tracers, thus suggesting that the gas traced by $C^{18}O$ is not affected by protostellar feedback. At the level of cloud-core transition, the core's external envelope traced by $DCO^+$ and $H^{13}CO^+$ is spinning up, consistent with conservation of angular momentum during core contraction.
7. The specific angular momenta of the cores vary in the range (0.6–21.0)$\times10^{20}$ cm$^2$ s$^{-1}$, consistent with previous observations of dense cores (Goodman et al. 1993; Caselli et al. 2002a). The specific angular momentum decreases at higher densities, which shows the importance of local magnetic fields to the small scale (central $\simeq$ 0.02 – 0.06 pc) dynamics of the cores.
8. The distributions of angles between the total velocity gradient and the large scale magnetic field are consistent with being random as well as the distributions of angles between the velocity gradient and filament direction. This suggests that the large scale magnetic fields traced by optical and near-infrared polarization measurements is not important in shaping the high density regions of the filament.

*Acknowledgements.* The authors thank the anonymous referee for valuable comments which helped to improve the manuscript. The authors acknowledge the financial support of the European Research Council (ERC; project PALs 320620); Andy Pon acknowledges that partial salary support was provided by a CITA National Fellowship. This work is part of the research programme VENI with project number 639.041.644, which is financed by the Netherlands Organisation for Scientific Research (NWO). MT and AH thank the Spanish MINECO for support under grant AYA2016-79006-P.

## Appendix A: Line fitting

### *Appendix A.1: Comparison of hyperfine splitting and Gaussian fit results for the $H^{13}CO^+$(1–0) and $DCO^+$(2–1) transitions*

Here we compare the line widths obtained by fitting the $H^{13}CO^+$(1–0) and $DCO^+$(2–1) line with single Gaussians and taking into account the hyperfine structure (hfs). We find that the Gaussian method gives systematically larger line widths than the hfs method. The hyperfine structure of the $H^{13}CO^+$ and $DCO^+$ rotational transitions were previously studied by Schmid-Burgk et al. (2004) and Caselli & Dore (2005). However the impact of hyperfine splitting in the $DCO^+$ transitions higher than the (1–0) transition and in $H^{13}CO^+$ was assumed negligible (e.g. Volgenau et al. 2006; Hirota & Yamamoto 2006). We implemented hfs fits of the $H^{13}CO^+$(1–0) and $DCO^+$(2–1) transitions in Pyspeckit. The line widths calculated with the hfs fits are systematically smaller than with the Gaussian fit. For this comparison, we use all of the spectra towards core 11 which have a SNR>5, produce fits that give uncertainties to the derived line width under 10%, and for which there is no sign of any asymmetry or multiple peaks. On average, the Gaussian line widths are 25% wider than the hfs derived line widths for $H^{13}CO^+$(1–0) and 22% wider for $DCO^+$(2–1) (see Fig. A.1).

### *Appendix A.2: Multiple velocity components and self-absorption*

All spectra were fitted assuming one velocity component. To check the fit quality, we inspected by eye any spectrum where the residual after fitting divided by rms was less than 0.8 or more than 1.2. Ratio less than 0.8 implies an "overfit" when routine fit the noise features instead of the spectral line. A ratio higher than 1.2 implies the fit does not reproduce the line correctly. A significant difference between the data and the fit could be due to the presence of a second line component, self-absorption of the line, asymmetric wings caused by gas flows or non-LTE effects. The $N_2D^+$(2–1) fits do not show significant differences between the fits and the data. The $N_2H^+$(1–0) and $H^{13}CO^+$(1–0) spectra show the presence of a second component towards cores 3, 8, 13, and 16 (see Fig. 2 and A.5). The $DCO^+$(2–1) spectra show double-peaked lines caused by either multiple velocity components or self-absorption towards all cores. Many $N_2H^+$(1–0) spectra also show non-LTE effects in the intensities of the hfs components, similar to those described in Caselli et al. (1995). Since the non-LTE effects do not affect the velocity dispersion and central velocity estimates, they are not taken into account in this work.

$N_2H^+$, $DCO^+$, and $H^{13}CO^+$ show the same line asymmetries, blue and red wings, towards cores 2, 6, 7, 10, 11, 13, and 19 (see Fig. A.5). The asymmetries also appear in the lines of the rare isotopologues $D^{13}CO^+$ and $HC^{18}O^+$ towards cores 2, 4, 7, 10, and 13 (see Fig. A.6). The wings cause an increase of the velocity dispersion. For example, on the north edge of core 10 the isolated component of the $N_2H^+$(1–0) hyperfine structure traces a gas flow which appears to connect the core and the envelope (see Fig. A.3). The other observed tracers also show an increase of the line width at the same place towards core 10 (see Fig. A.9).

$DCO^+$(2–1) and $H^{13}CO^+$(1–0) have no isolated components in their hyperfine structure. Because of the hyperfine components are close, it is not easy to find an unambiguous $\tau$-unconstrained fit with the hfs fitting method, and the Gaussian fits do not properly fit the wings of the lines, thereby producing larger line widths. If the hfs structure makes the line asymmetric, the $V_{LSR}$ of the Gaussian fit will also be altered from the correct value (see Fig. A.1 and A.4). Thus in the case of asymmetric or double-peaked lines in $DCO^+$ and $H^{13}CO^+$ we used a $\tau$-constrained fit ($\tau$=0.1) to gauge the central velocities. The velocity dispersions produced by the $\tau$-constrained fit of the self-absorbed or blended lines are overestimated, as the fit considers one velocity component and the line is assumed to be optically thin ($\tau$=1). We note that the assumption of the $\tau$-constrained fit overestimates the majority of the velocity dispersion measurements for our $DCO^+$ data. We compared the velocity dispersion values obtained with $\tau$-constrained and $\tau$-unconstrained fits where the optical depth is defined with $\tau > 3\Delta\tau$. The $\tau$-constrained fit produces a velocity dispersion 1–3.3 times larger

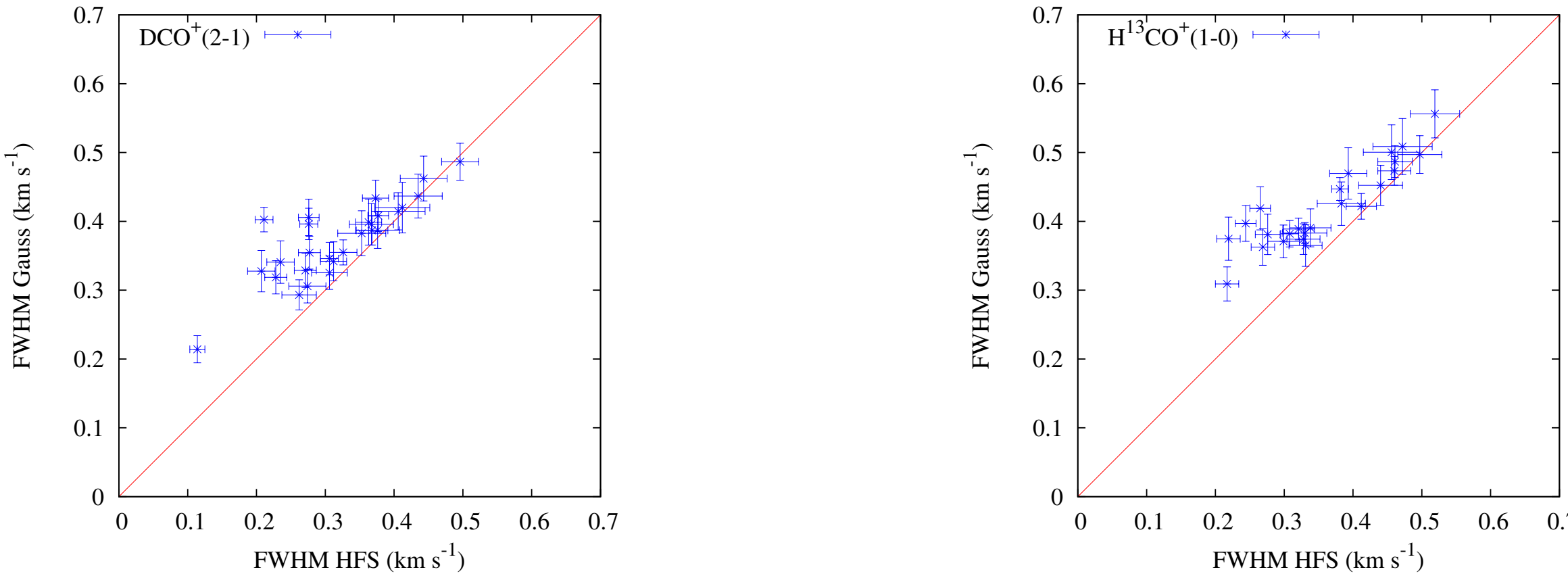

**Fig. A.1.** Comparison of line widths (FWHM) obtained with Gaussian and hfs fits towards core 11 for $DCO^+$(2–1) (left) and for $H^{13}CO^+$(1–0) (right). The red lines show one-to-one correlation.

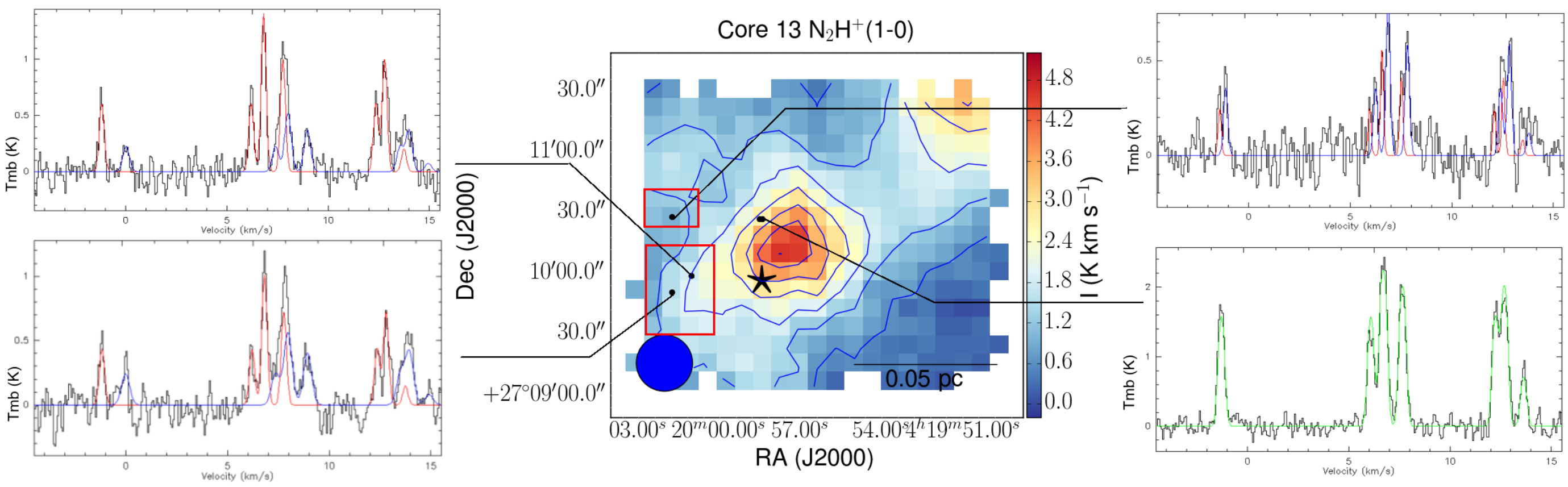

**Fig. A.2.** The integrated intensity map of $N_2H^+$(1–0) towards core 13. The red boxes in the map show the areas where the additional velocity components are present. Left: the second velocity component (blue fit) is well separated (by ≃1 km $s^{-1}$) from the main velocity component (red fit). Right top: the second velocity component (red fit) is partially blended with the main velocity component (blue fit, the separation is ≃0.3 km $s^{-1}$ which corresponds to a line width $\sigma$ = 0.13 km $s^{-1}$). Right bottom: an example spectrum where only one velocity component is detected with the best fit shown as the green line.

than the $\tau$-unconstrained fit, with an average dispersion 1.56 times larger.

Since $DCO^+$(2–1) and $H^{13}CO^+$(1–0) have no isolated components, it is not possible to distinguish between self-absorbed lines and blended second velocity components. However we control if the flux was lost due to self-absorption of the lines by observing transitions of the rare isotopologues $D^{13}CO^+$(2–1) and $HC^{18}O^+$(1–0) towards the $DCO^+$(2–1) emission peaks (which would suffer self-absorption more than $H^{13}CO^+$, because of its higher critical densities and because it is more abundant in dense, CO depleted gas) in nine cores. We compare the estimates of the column densities ($N_{col}$) of $DCO^+$ and $HCO^+$ obtained with the rearer isotopologues in Fig. A.7. To estimate the column densities, we assume that both lines are optically thin and the fractional abundances of O and C in the species are the same as the fractional abundances of the elements, $^{16}O/^{18}O = 560 \pm 25$, $^{12}C/^{13}C = 77 \pm 7$ (Wilson & Rood 1994). The comparison shows that both $DCO^+$(2–1) and $H^{13}CO^+$(1–0) do not suffer significant self-absorption. Thus the double-peaked lines are most likely two blended velocity components which trace different layers of an envelope and need better sensitivity and spectral resolution to be resolved.

We consider that a line has two velocity components if the components are separated by more than one line width (such as in core 13, see Fig. A.2). Multiple velocity components are only detected towards an area roughly the size of one to two beams towards core 13. The second components which are well separated from the main line (see Fig. A.2, left) were missed by the one-component fit procedure and were not taken into account in this work. We tried to avoid the blended components, however, in case they are close (see Fig. A.2, right top), they were fit as one velocity component which increased velocity dispersion in those points.

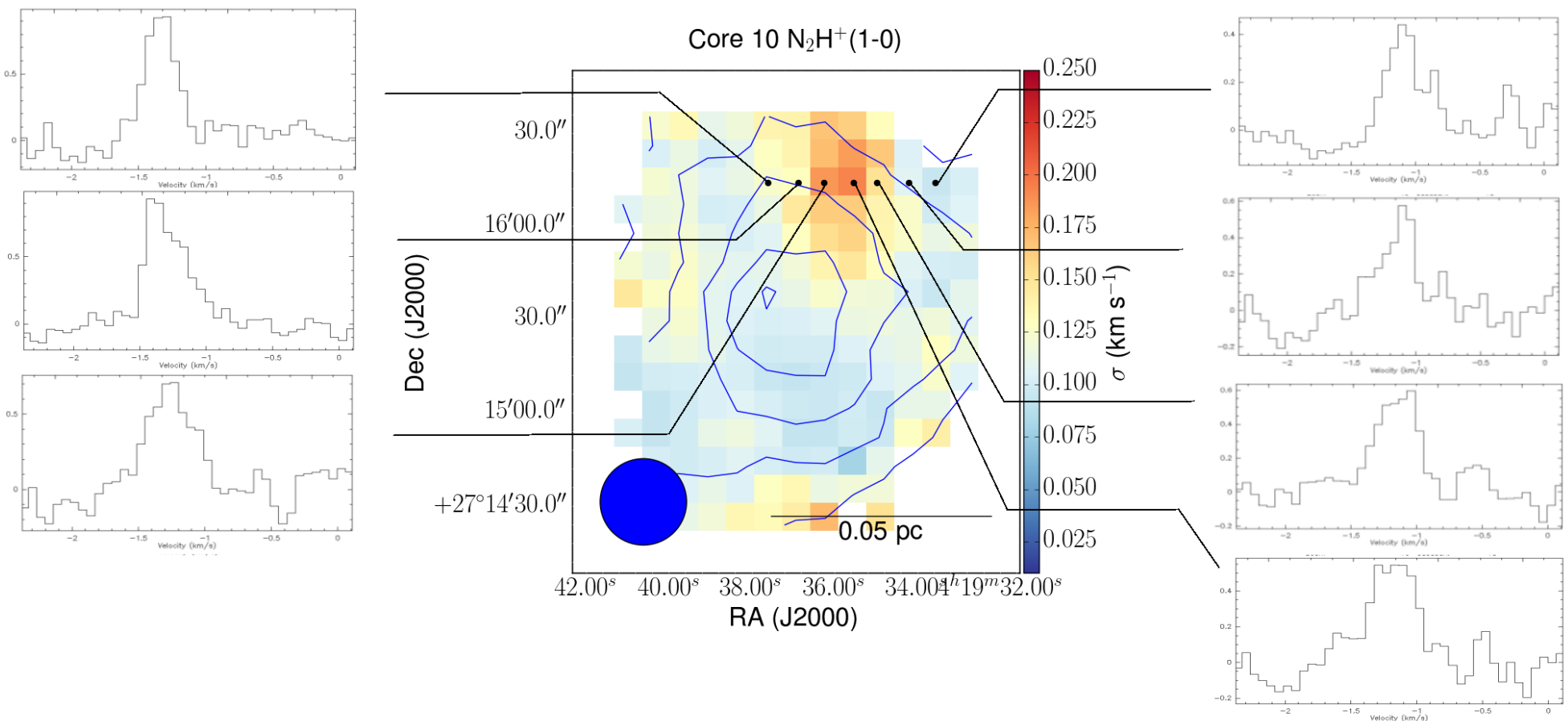

**Fig. A.3.** The velocity dispersion map of $N_2H^+$(1–0) towards core 10 with a sample of the gas flow on the edge of the core connecting it to the envelope and the isolated components spectra.

**Table A.1.** Results of a Gaussian fit of $D^{13}CO^+$(2–1) towards the $DCO^+$(2–1) emission peaks of the cores.

| Core | $I^a$ (K km $s^{-1}$) | $V_{LSR}$ (km $s^{-1}$) | $\Delta v$ (km $s^{-1}$) | $\sigma$ (km $s^{-1}$) | $I^b_{peak}$ (K) | Base rms (K) | Line rms (K) |
|---|---|---|---|---|---|---|---|
| *2 | – | – | – | – | – | 0.028 | – |
| *3 | – | – | – | – | – | 0.039 | – |
| 4 | 0.061 ± 0.008 | 6.663 ± 0.021 | 0.329 ± 0.050 | 0.140 ± 0.021 | 0.173 | 0.045 | 0.028 |
| 6 | 0.060 ± 0.006 | 6.855 ± 0.017 | 0.385 ± 0.047 | 0.163 ± 0.020 | 0.146 | 0.029 | 0.021 |
| 7 | 0.062 ± 0.006 | 6.717 ± 0.019 | 0.353 ± 0.039 | 0.150 ± 0.017 | 0.164 | 0.037 | 0.035 |
| 10 | 0.035 ± 0.006 | 6.769 ± 0.030 | 0.324 ± 0.064 | 0.138 ± 0.027 | 0.102 | 0.039 | 0.020 |
| 11 | 0.031 ± 0.006 | 6.753 ± 0.064 | 0.592 ± 0.115 | 0.251 ± 0.049 | 0.050 | 0.029 | 0.025 |
| 13 | 0.244 ± 0.013 | 6.746 ± 0.014 | 0.531 ± 0.032 | 0.225 ± 0.014 | 0.431 | 0.059 | 0.078 |
| 19 | 0.034 ± 0.011 | 6.850 ± 0.050 | 0.376 ± 0.202 | 0.160 ± 0.086 | 0.086 | 0.037 | 0.027 |

**Notes.** * are non-detections; [(a)] Integrated intensity; [(b)] Peak intensity of the line.

**Table A.2.** Results of an hfs fit of $HC^{18}O^+$(1–0) towards the $DCO^+$(2–1) emission peaks of the cores.

| Core | $I^a$ (K km $s^{-1}$) | $V_{LSR}$ (km $s^{-1}$) | $\Delta v$ (km $s^{-1}$) | $\sigma$ (km $s^{-1}$) | $\tau$ | Base rms (K) | Line rms (K) |
|---|---|---|---|---|---|---|---|
| 2 | 0.118 ± 0.025 | 6.972 ± 0.020 | 0.196 ± 0.051 | 0.083 ± 0.022 | 0.1 | 0.027 | 0.002 |
| 3 | 0.119 ± 0.017 | 6.798 ± 0.024 | 0.375 ± 0.070 | 0.159 ± 0.030 | 0.1 | 0.024 | 0.020 |
| 4N | 0.151 ± 0.019 | 6.654 ± 0.017 | 0.248 ± 0.032 | 0.105 ± 0.014 | 0.1 | 0.026 | 0.023 |
| 6S | 0.160 ± 0.017 | 6.841 ± 0.021 | 0.364 ± 0.041 | 0.155 ± 0.017 | 0.1 | 0.028 | 0.015 |
| 7 | 0.284 ± 0.028 | 6.715 ± 0.014 | 0.281 ± 0.031 | 0.119 ± 0.013 | 0.1 | 0.039 | 0.027 |
| 10 | 0.207 ± 0.021 | 6.777 ± 0.013 | 0.275 ± 0.035 | 0.117 ± 0.015 | 0.1 | 0.026 | 0.015 |
| 11 | 0.229 ± 0.020 | 6.839 ± 0.031 | 0.468 ± 0.043 | 0.199 ± 0.018 | 2.5 ± 0.3 | 0.020 | 0.015 |
| 13 | 0.306 ± 0.025 | 6.687 ± 0.015 | 0.388 ± 0.041 | 0.165 ± 0.017 | 0.1 | 0.038 | 0.035 |
| 19 | 0.161 ± 0.020 | 6.835 ± 0.015 | 0.264 ± 0.039 | 0.112 ± 0.017 | 0.1 | 0.025 | 0.010 |

**Notes.** [(a)] Integrated intensity.

*Appendix A.3: The results of hyperfine splitting fits*

## Appendix B: Results of velocity gradients and specific angular momentum calculations

**Table A.3.** Characteristics of the $N_2H^+$(1–0), $N_2D^+$(2–1), $H^{13}CO^+$(1–0), and $DCO^+$(2–1) maps convolved to 27.8″, 27.8″, 29.9″, and 18″ beams, respectively.

| Core | $N_2H^+$(1–0) rms in $T_{mb}$ (K) | $N_2D^+$(2–1) rms in $T_{mb}$ (K) | $H^{13}CO^+$(1–0) rms in $T_{mb}$ (K) | $DCO^+$(2–1) rms in $T_{mb}$ (K) |
|---|---|---|---|---|
| 2 | 0.080 | 0.040 | 0.110 | 0.150 |
| 3 | 0.140 | 0.050 | 0.140 | 0.210 |
| 4 | – | 0.060 | 0.110 | 0.150 |
| 6 | 0.080 | 0.080 | 0.190 | 0.230 |
| 7 | 0.075 | 0.060 | 0.110 | 0.200 |
| 8 | 0.110 | 0.060 | 0.190 | 0.500 |
| 10 | 0.110 | 0.080 | 0.150 | 0.150 |
| 11 | 0.110 | 0.080 | 0.140 | 0.330 |
| 12 | 0.080 | 0.060 | 0.120 | 0.220 |
| 13 | 0.115 | 0.080 | 0.140 | 0.310 |
| 16 | 0.125 | 0.100 | 0.140 | 0.250 |
| 17 | 0.125 | 0.090 | 0.130 | 0.250 |
| 19 | 0.115 | 0.090 | 0.140 | 0.210 |

**Table A.4.** Characteristics of the $N_2H^+$(1–0), $N_2D^+$(2–1), $H^{13}CO^+$(1–0), and $DCO^+$(2–1) maps convolved to a 29.9″ beam.

| Core | $N_2H^+$(1–0) rms in $T_{mb}$ (K) | $N_2D^+$(2–1) rms in $T_{mb}$ (K) | $H^{13}CO^+$(1–0) rms in $T_{mb}$ (K) | $DCO^+$(2–1) rms in $T_{mb}$ (K) |
|---|---|---|---|---|
| 2 | 0.070 | 0.040 | 0.120 | 0.070 |
| 3 | 0.120 | 0.040 | 0.150 | 0.100 |
| 4 | – | 0.050 | 0.110 | 0.070 |
| 6 | 0.080 | 0.050 | 0.205 | 0.085 |
| 7 | 0.065 | 0.050 | 0.110 | 0.080 |
| 8 | 0.090 | 0.055 | 0.210 | 0.150 |
| 10 | 0.090 | 0.070 | 0.155 | 0.050 |
| 11 | 0.090 | 0.070 | 0.140 | 0.120 |
| 12 | 0.075 | 0.050 | 0.120 | 0.080 |
| 13 | 0.090 | 0.070 | 0.150 | 0.115 |
| 16 | 0.100 | 0.085 | 0.150 | 0.090 |
| 17 | 0.100 | 0.080 | 0.140 | 0.100 |
| 19 | 0.090 | 0.080 | 0.130 | 0.080 |

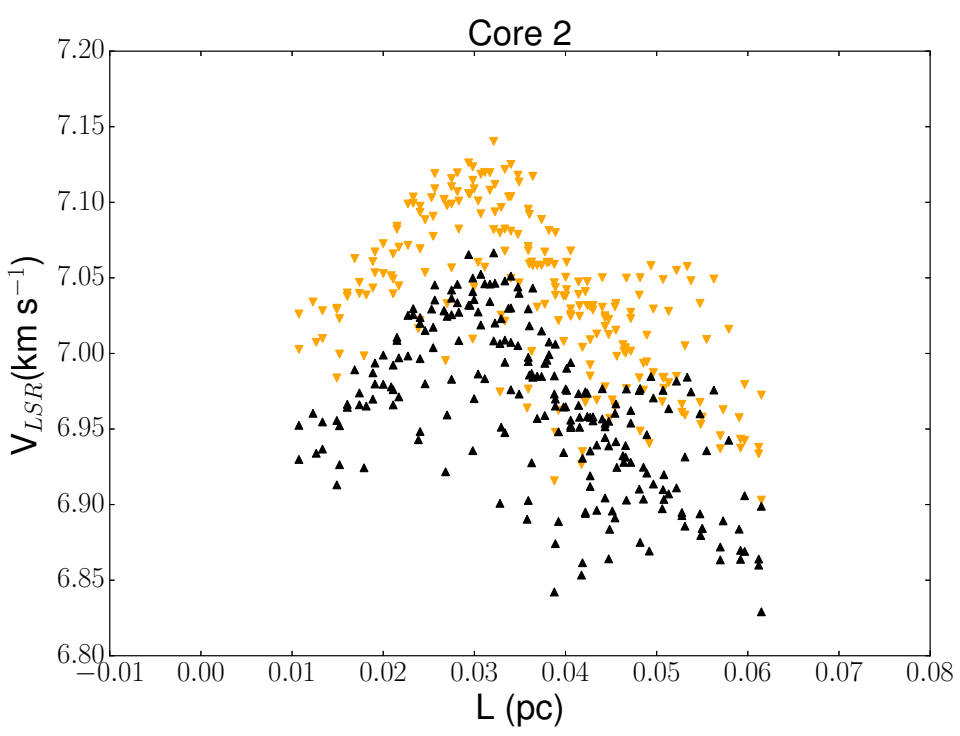

**Fig. A.4.** Comparison of the central velocity ($V_{LSR}$) of the $DCO^+$(2–1) line measured with Gaussian (yellow) and hfs (black) fits for core 2 plotted along the filament direction.

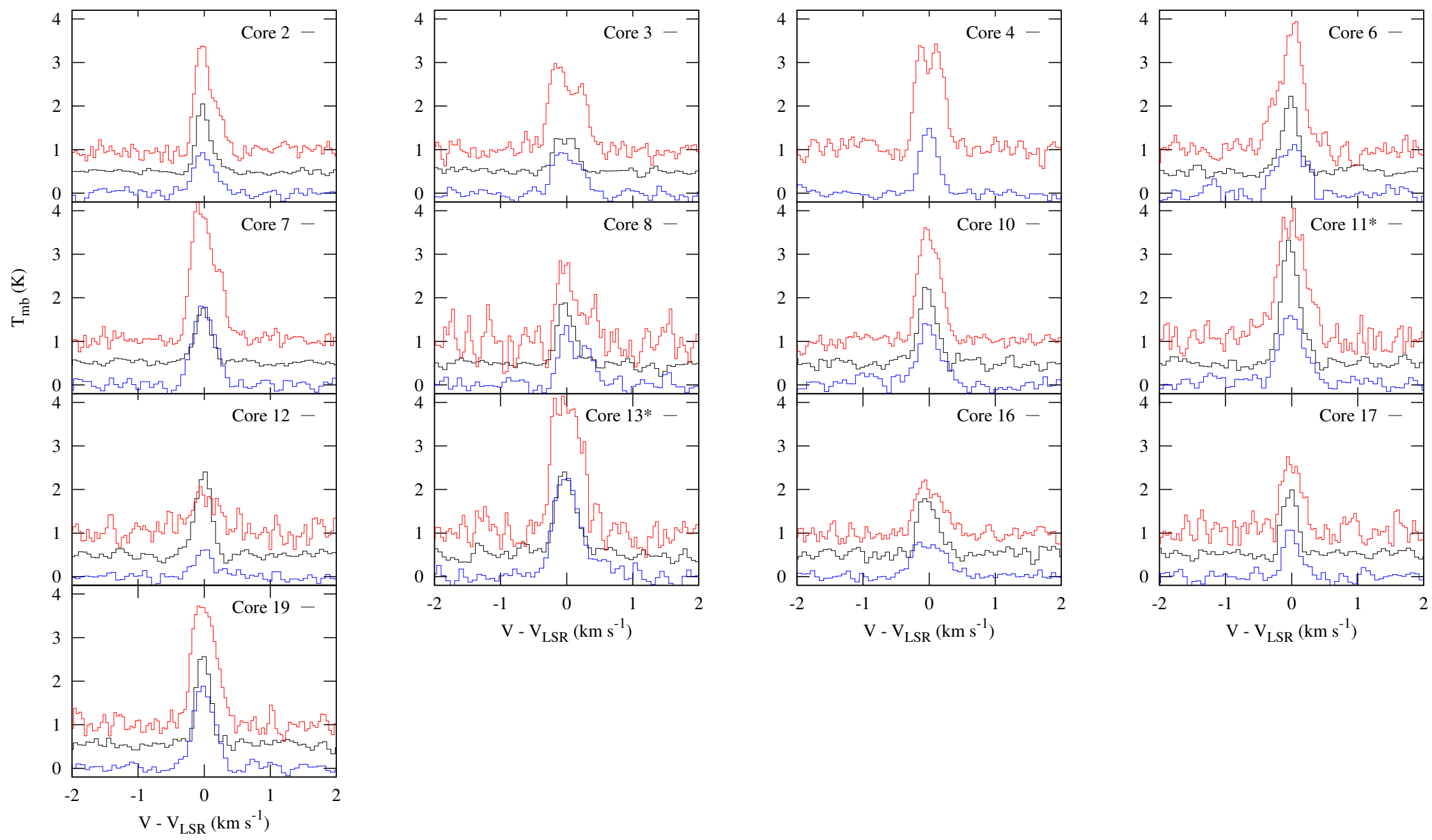

**Fig. A.5.** Spectra of the isolated components of $N_2H^+$(1–0) (black), $DCO^+$(2–1) (red) and $H^{13}CO^+$(1–0) (blue) towards the $N_2H^+$(1–0) emission peaks. For convenient comparison, the $DCO^+$(2–1) spectra are shifted up by 1 K and the $N_2H^+$(1–0) spectra are shifted up by 0.5 K. The cores with an asterisk (*) near the title contain embedded protostrars.

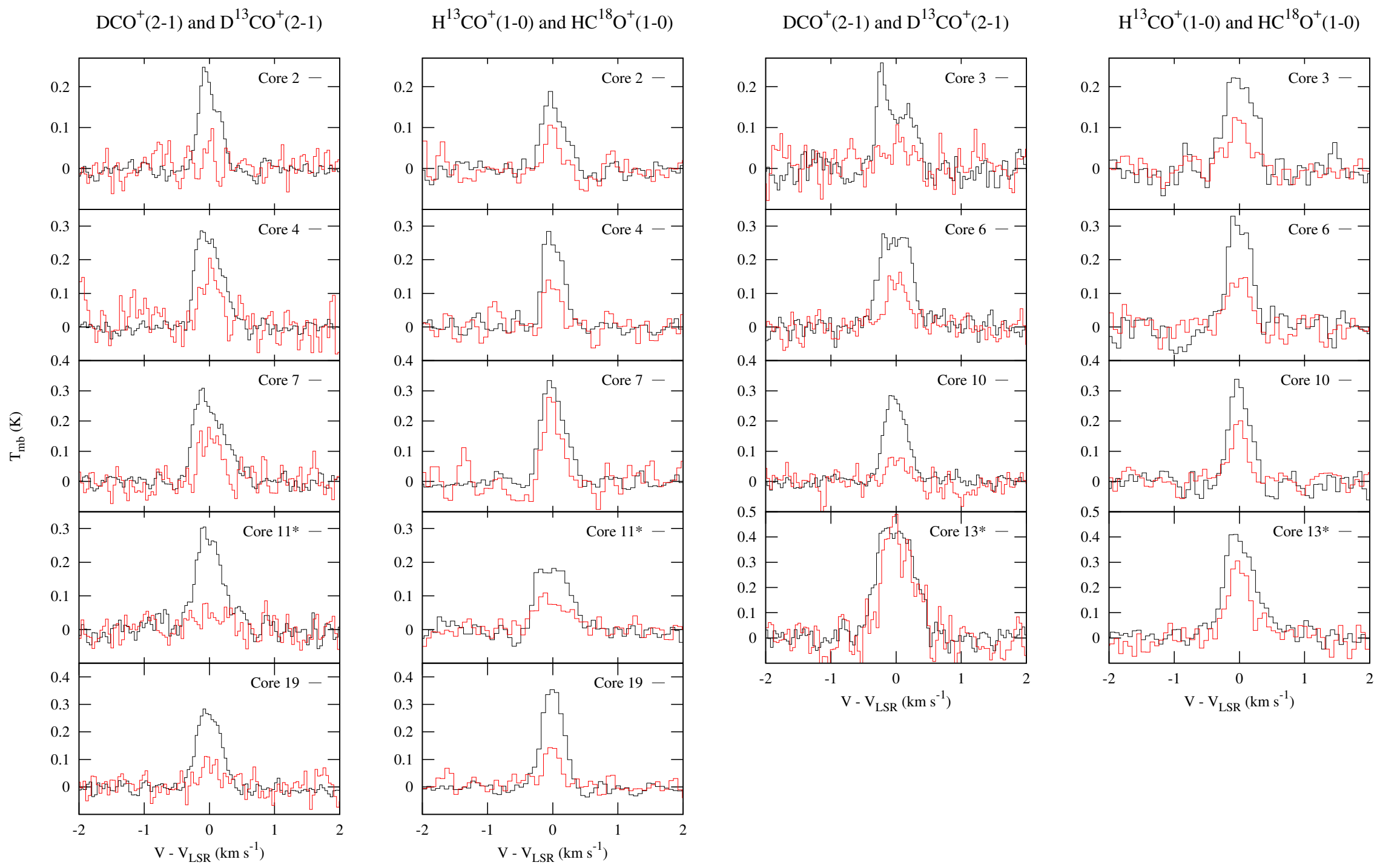

**Fig. A.6.** Spectra of $DCO^+$(2–1) and $H^{13}CO^+$(1–0) (black) compared to $D^{13}CO^+$(2–1) and $HC^{18}O^+$(1–0) respectively, towards the $DCO^+$(2–1) emission peaks. For convenient comparison the $DCO^+$(2–1) spectra are divided by 10 and the $H^{13}CO^+$(1–0) spectra are divided by 5. The cores with an asterisk (*) near the title contain embedded protostrars.

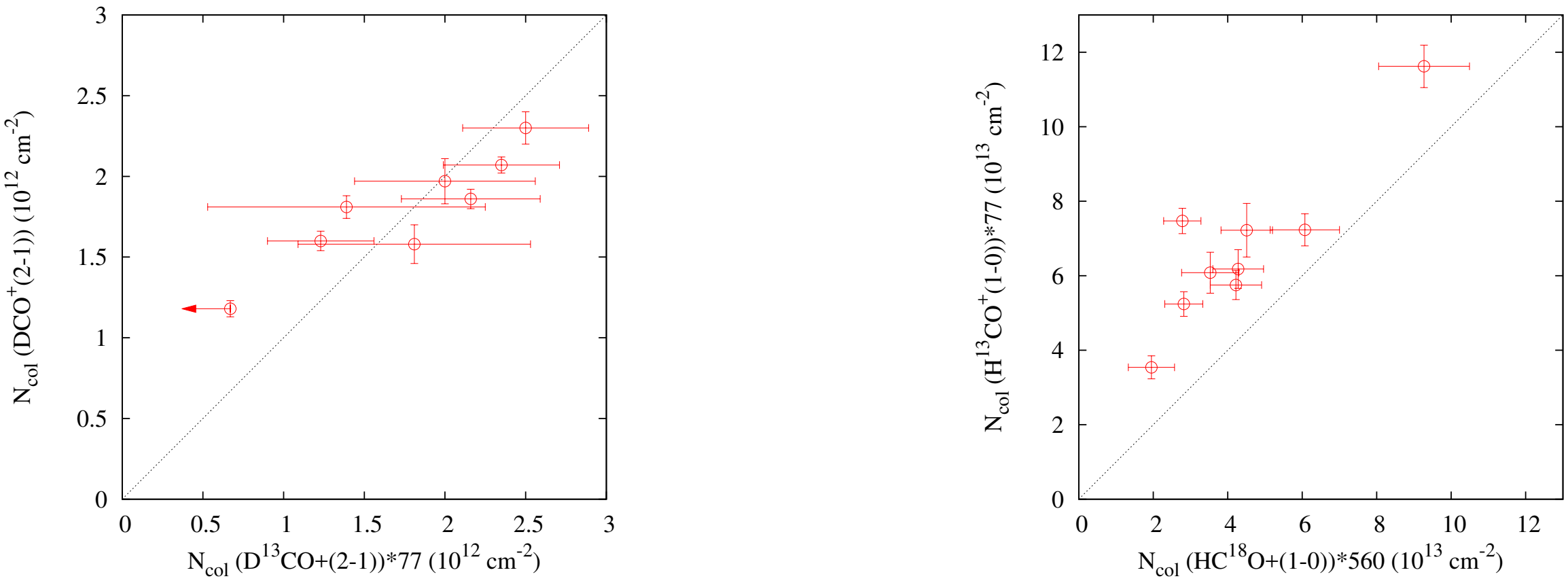

**Fig. A.7.** Comparison of the column densities of $DCO^+$ (left) and $HCO^+$ (right) towards the $DCO^+$(2–1) emission peaks obtained with an assumption of optically thin lines. The dotted lines show one-to-one correlations.

**Table B.1.** Total velocity gradients $G$, position angles $\theta_G$ (measured west to north), equivalent radii[a] measured across the cores using all emitting areas of each species and corresponding specific angular momenta $J/M$.

| Core | $G$ (km s$^{-1}$ pc$^{-1}$) | $\theta_G$ (degrees) | R (pc) | $J/M$ ($10^{20}$ cm$^2$ s$^{-1}$) | $G$ (km s$^{-1}$ pc$^{-1}$) | $\theta_G$ (degrees) | R (pc) | $J/M$ ($10^{20}$ cm$^2$ s$^{-1}$) |
|---|---|---|---|---|---|---|---|---|
| | | $N_2H^+$(1–0) | | | | $N_2D^+$(2–1) | | |
| 2 | 2.012± 0.010 | -134.0 ± 0.3 | 0.036 | 3.19 ± 0.02 | 2.348 ± 0.055 | -138.4 ± 1.5 | 0.031 | 2.75 ± 0.06 |
| 3 | 2.623 ± 0.017 | 129.7 ± 0.2 | 0.048 | 7.54 ± 0.05 | 5.135 ± 0.452 | 130.3 ± 5.7 | 0.019 | 2.18 ± 0.19 |
| 4 | – | – | – | – | 1.377 ± 0.032 | -92.7 ± 2.2 | 0.039 | 2.58 ± 0.06 |
| 6 | 3.804 ± 0.010 | -48.7 ± 0.1 | 0.049 | 11.43 ± 0.03 | 3.571 ± 0.047 | -51.0 ± 0.5 | 0.039 | 6.75 ± 0.09 |
| 7 | 5.423 ± 0.001 | -149.5 ± 0.1 | 0.044 | 13.20 ± 0.02 | 6.125 ± 0.054 | -154.2 ± 0.5 | 0.040 | 11.94 ± 0.11 |
| 8 | 3.523 ± 0.008 | 1.9 ± 0.1 | 0.067 | 19.63 ± 0.05 | 1.789 ± 0.061 | 26.0 ± 1.4 | 0.051 | 5.72 ± 0.20 |
| 10 | 4.857 ± 0.015 | -28.8 ± 0.1 | 0.045 | 12.18 ± 0.04 | 3.419 ± 0.180 | -29.6 ± 2.9 | 0.022 | 2.11 ± 0.11 |
| 11 | 1.632 ± 0.012 | -169.1 ± 0.5 | 0.048 | 4.59 ± 0.03 | 2.151 ± 0.100 | -156.0 ± 3.5 | 0.026 | 1.86 ± 0.09 |
| 12 | 1.435 ± 0.009 | 92.9 ± 0.4 | 0.042 | 3.20 ± 0.02 | 1.795 ± 0.130 | 124.1 ± 2.7 | 0.029 | 1.92 ± 0.14 |
| 13 | 1.432 ± 0.007 | -144.0 ± 0.5 | 0.056 | 5.48 ± 0.03 | 2.639 ± 0.123 | 158.7 ± 1.9 | 0.030 | 2.86 ± 0.13 |
| 16 | 1.069 ± 0.011 | 28.5 ± 1.0 | 0.050 | 3.34 ± 0.03 | 0.648 ± 0.075 | 51.7 ± 13.0 | 0.027 | 0.59 ± 0.07 |
| 17 | 2.048 ± 0.006 | 120.2 ± 0.3 | 0.058 | 8.47 ± 0.02 | 2.470 ± 0.045 | 85.8 ± 1.0 | 0.040 | 4.96 ± 0.09 |
| 19 | 1.809 ± 0.011 | -66.8 ± 0.3 | 0.044 | 4.35 ± 0.03 | 1.284 ± 0.099 | -87.6 ± 3.9 | 0.032 | 1.64 ± 0.13 |
| | | $H^{13}CO^+$(1–0) | | | | $DCO^+$(2–1) | | |
| 2 | 1.616 ± 0.062 | -144.1 ± 2.1 | 0.037 | 2.77 ± 0.11 | 1.616 ± 0.029 | -112.5 ± 1.1 | 0.035 | 2.40 ± 0.04 |
| 3 | 1.670 ± 0.053 | 135.3 ± 1.4 | 0.047 | 4.48 ± 0.14 | 2.233 ± 0.036 | 150.6 ± 0.8 | 0.042 | 4.95 ± 0.08 |
| 4 | 2.336 ± 0.037 | -99.2 ± 1.1 | 0.039 | 4.31 ± 0.07 | 2.370 ± 0.016 | -103.4 ± 0.5 | 0.038 | 4.15 ± 0.03 |
| 6 | 4.144 ± 0.057 | -38.6 ± 0.6 | 0.047 | 11.12 ± 0.15 | 4.495 ± 0.026 | -45.8 ± 0.2 | 0.047 | 12.33 ± 0.07 |
| 7 | 4.764 ± 0.030 | -134.7 ± 0.4 | 0.045 | 11.73 ± 0.07 | 4.672 ± 0.014 | -144.9 ± 0.2 | 0.045 | 11.87 ± 0.04 |
| 8 | 4.091 ± 0.040 | 9.9 ± 0.4 | 0.064 | 20.99 ± 0.21 | 5.503 ± 0.089 | 6.0 ± 0.5 | 0.040 | 11.01 ± 0.18 |
| 10 | 4.497 ± 0.047 | -26.0 ± 0.5 | 0.045 | 11.47 ± 0.12 | 5.548 ± 0.019 | -31.5 ± 0.2 | 0.047 | 15.00 ± 0.05 |
| 11 | 2.633 ± 0.073 | 157.0 ± 1.1 | 0.043 | 6.14 ± 0.17 | 3.859 ± 0.066 | 159.6 ± 0.8 | 0.035 | 5.86 ± 0.10 |
| 12 | 2.515 ± 0.128 | 119.2 ± 1.8 | 0.034 | 3.61 ± 0.18 | 5.159 ± 0.071 | 128.1 ± 0.4 | 0.034 | 7.56 ± 0.10 |
| 13 | 4.714 ± 0.050 | 148.9 ± 0.5 | 0.052 | 15.76 ± 0.17 | 5.435 ± 0.045 | 144.9 ± 0.3 | 0.044 | 13.14 ± 0.11 |
| 16 | 1.602 ± 0.040 | -48.3 ± 1.0 | 0.059 | 6.98 ± 0.17 | 2.624 ± 0.051 | -60.7 ± 0.7 | 0.042 | 5.81 ± 0.11 |
| 17 | 2.906 ± 0.029 | 123.2 ± 0.8 | 0.052 | 9.67 ± 0.10 | 2.719 ± 0.027 | 103.9 ± 0.6 | 0.048 | 7.62 ± 0.08 |
| 19 | 2.838 ± 0.029 | -37.2 ± 0.5 | 0.056 | 11.03 ± 0.11 | 2.214 ± 0.026 | -56.3 ± 0.5 | 0.049 | 6.69 ± 0.08 |

**Notes.** [a] Equivalent radius here is the radius of a circle which has the same area as the emitting area (see Subsect. 4.5).

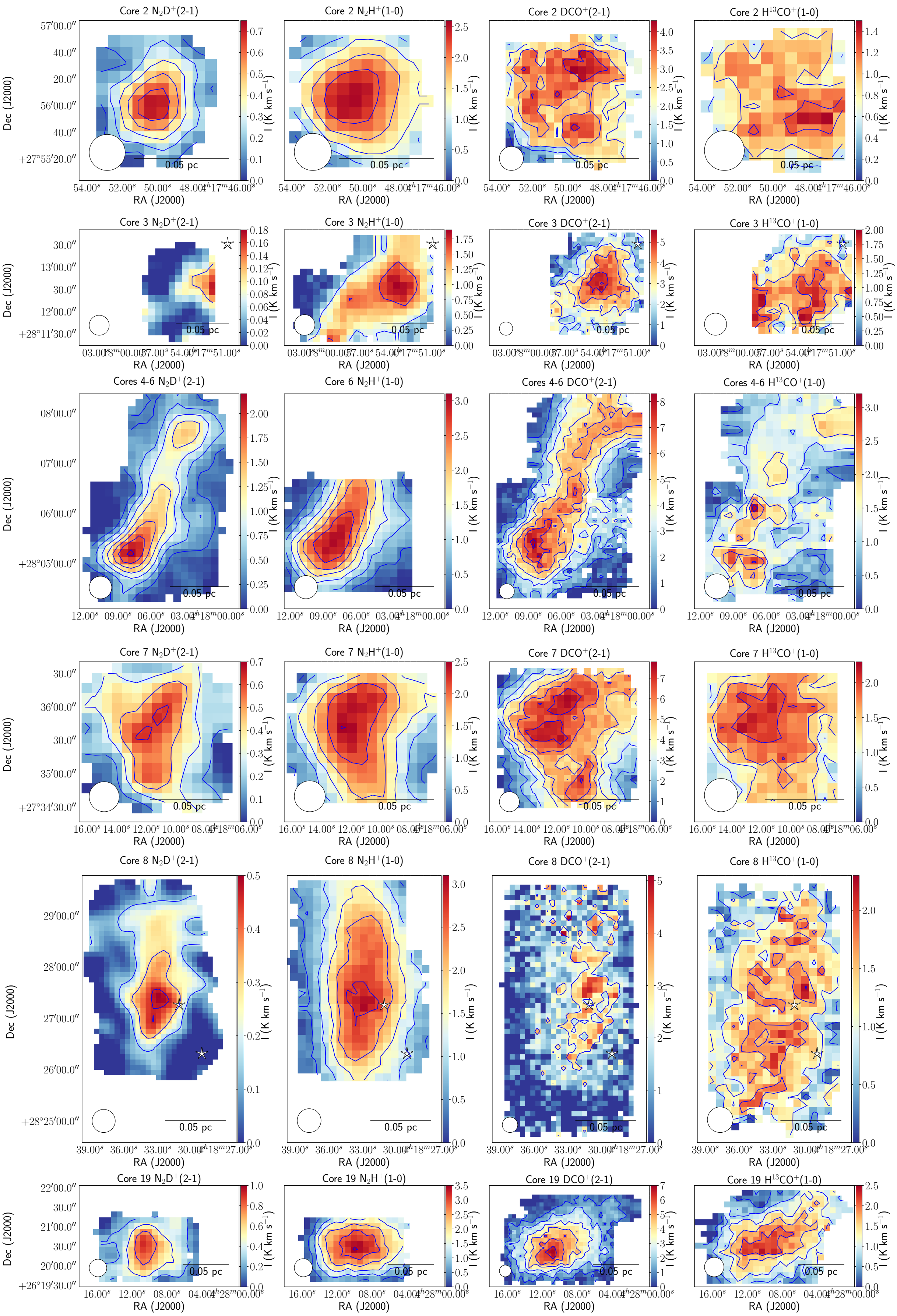

**Fig. A.8.** Integrated intensities of the $N_2D^+$(2–1), $N_2H^+$(1–0), $DCO^+$(2–1), and $H^{13}CO^+$(1–0) lines across the cores. First contour is at the $5\sigma$ level and the contour step is $5\sigma$. Stars show the positions of young stellar objects (YSOs) from Rebull et al. (2010): black stars are young, flat and class I objects, white stars are more evolved, class II and III objects.

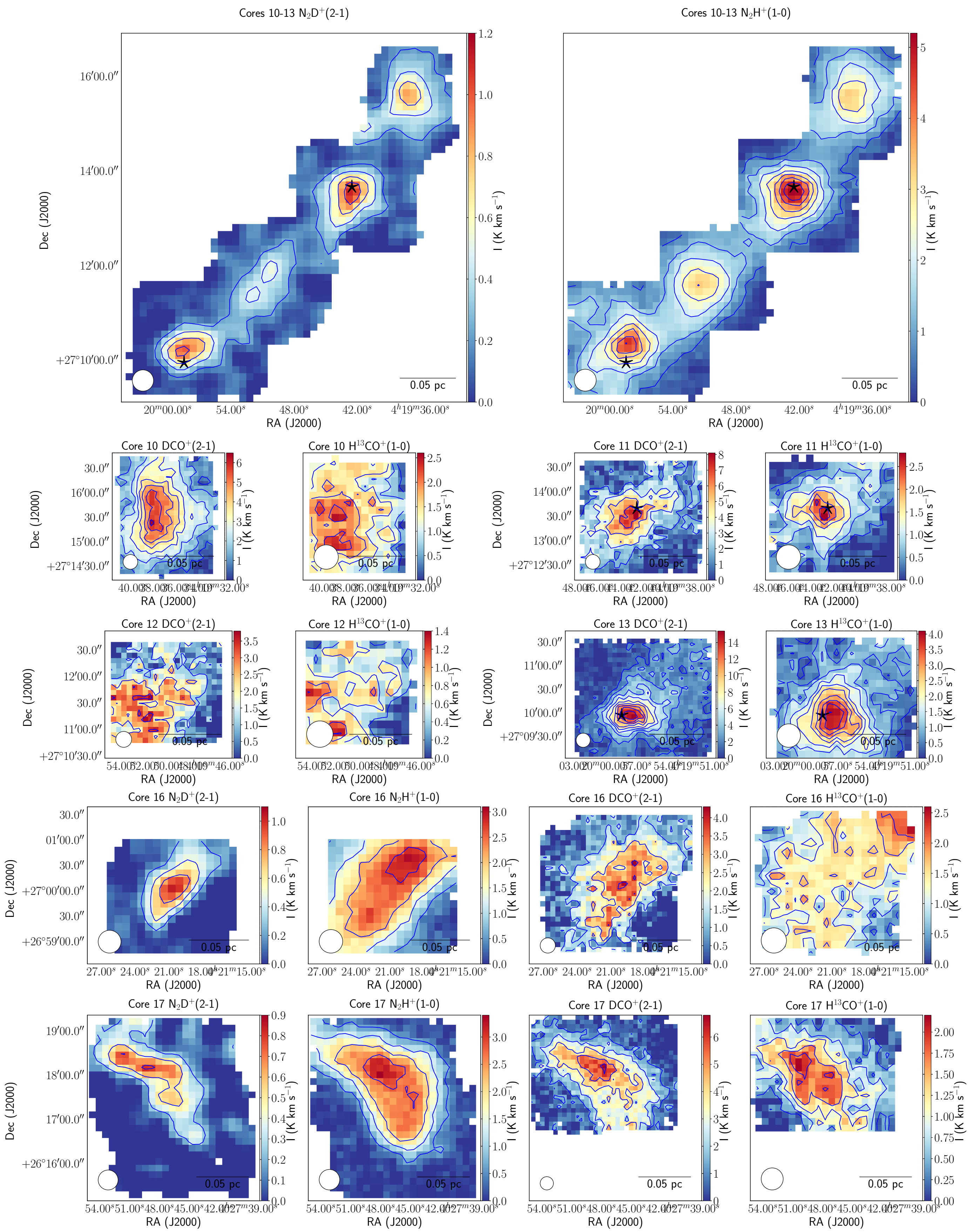

**Fig. A.8.** continued.

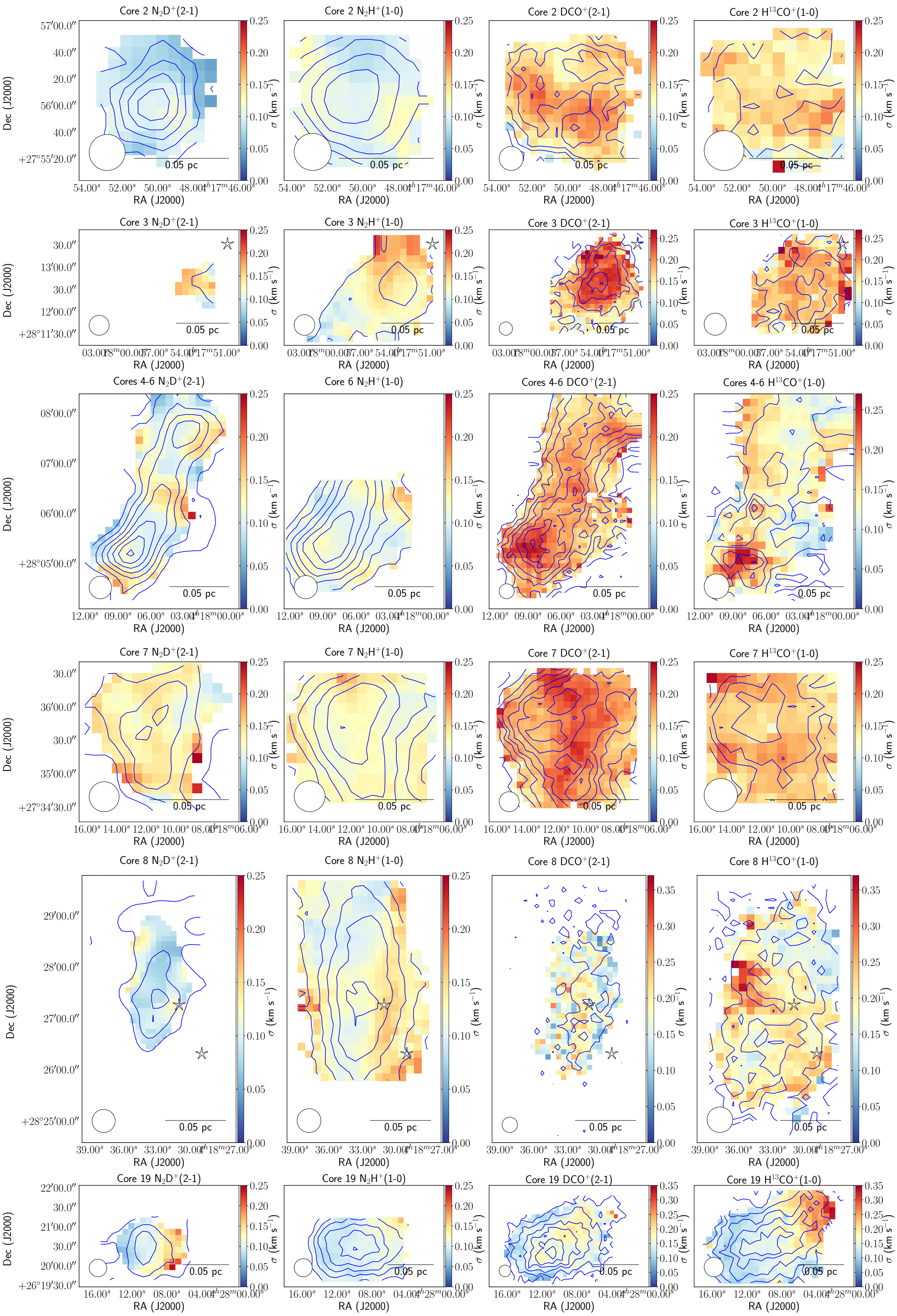

**Fig. A.9.** Velocity dispersions ($\sigma$) of the $N_2D^+$(2–1), $N_2H^+$(1–0), $DCO^+$(2–1), and $H^{13}CO^+$(1–0) lines across the cores. Blue contours show the integrated intensities of the corresponding species. First contour is at the 5$\sigma$ level and the contour step is 5$\sigma$. Stars show the positions of young stellar objects (YSOs) from Rebull et al. (2010): black stars are young, flat and class I objects, white stars are more evolved, class II and III objects.

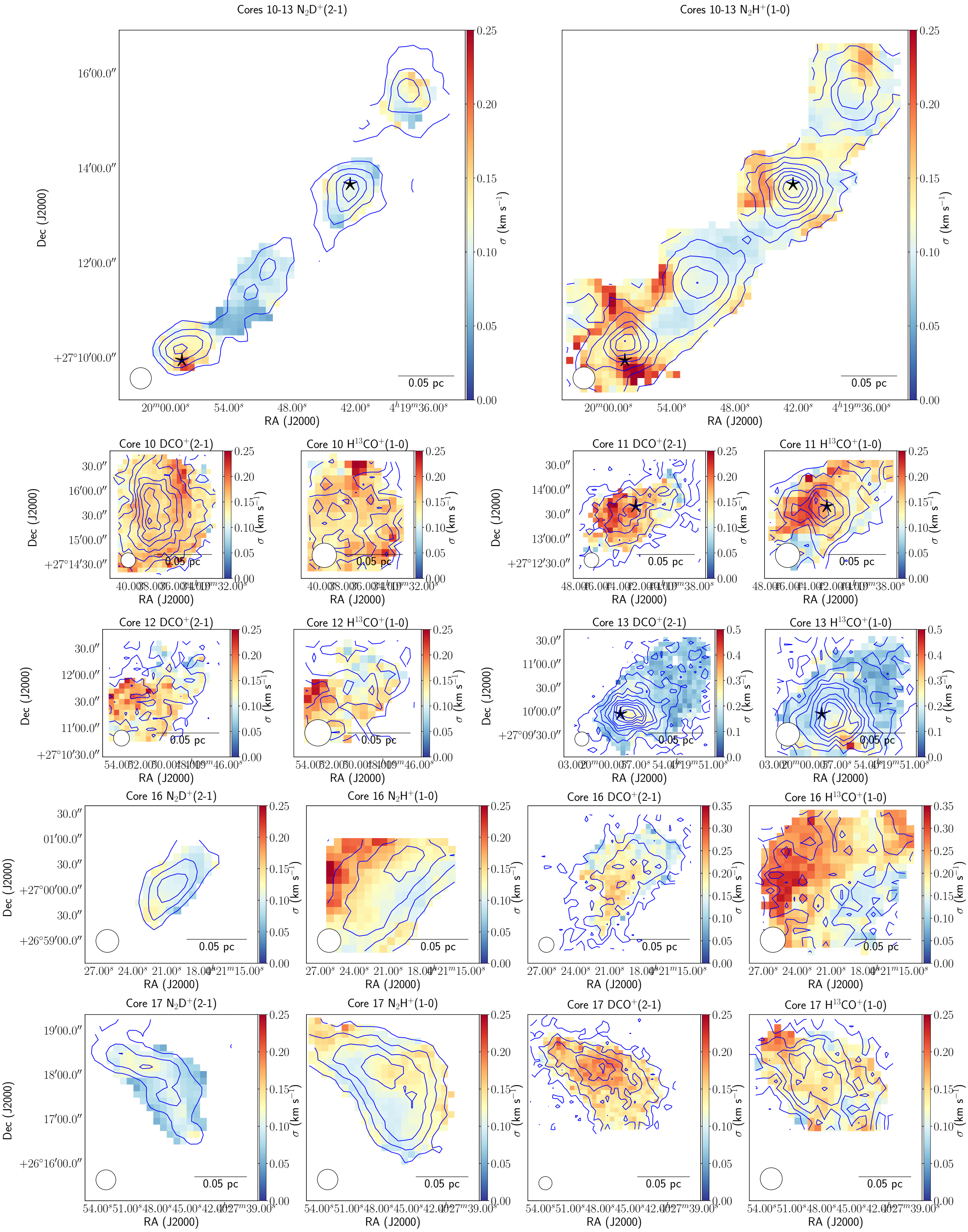

**Fig. A.9.** continued.

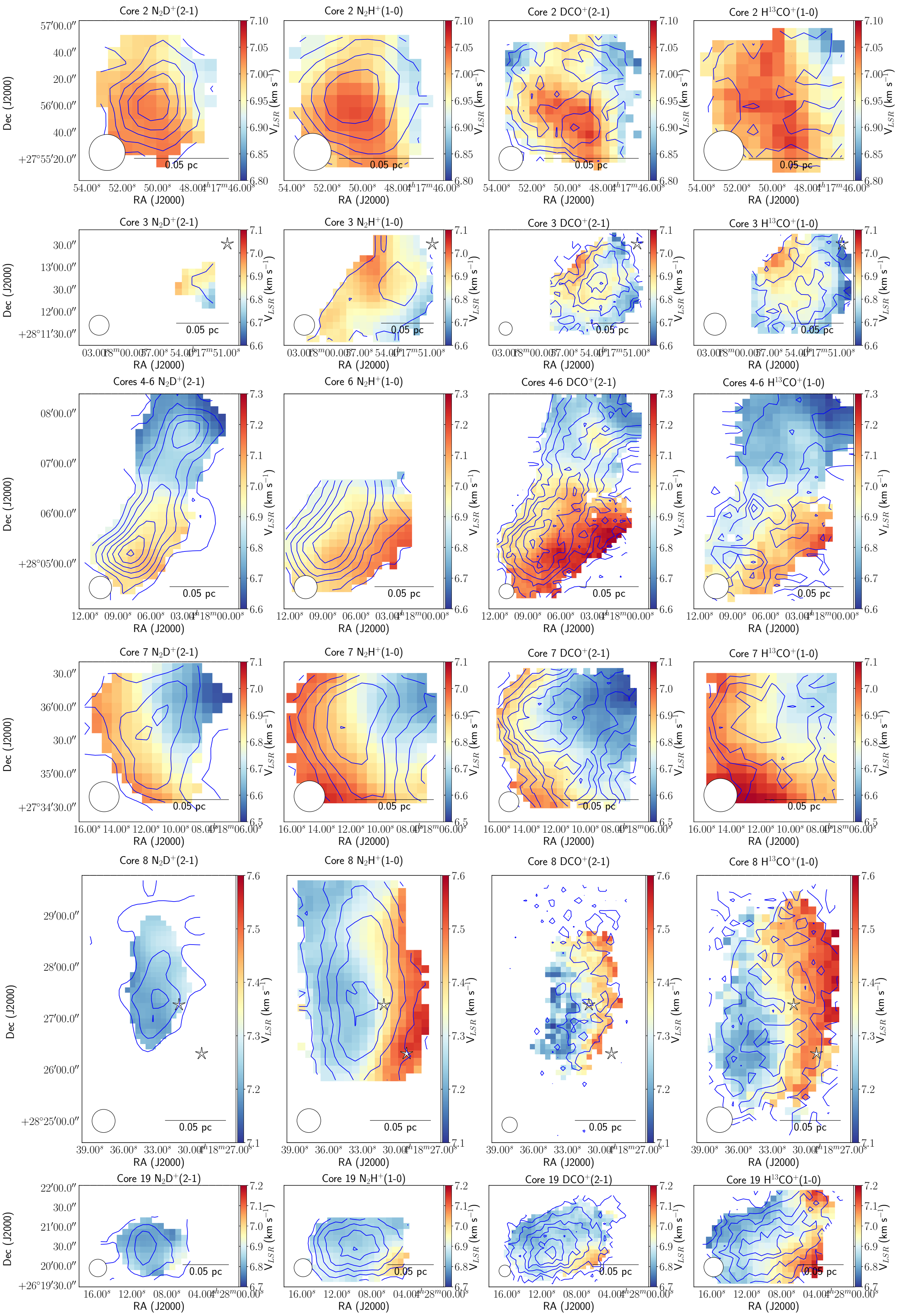

**Fig. A.10.** Central velocities ($V_{LSR}$) of the $N_2D^+$(2–1), $N_2H^+$(1–0), $DCO^+$(2–1), and $H^{13}CO^+$(1–0) lines across the cores. Blue contours show the integrated intensities of the corresponding species. First contour is at the $5\sigma$ level and the contour step is $5\sigma$. Stars show the positions of young stellar objects (YSOs) from Rebull et al. (2010): black stars are young, flat and class I objects, white stars are more evolved, class II and III objects.

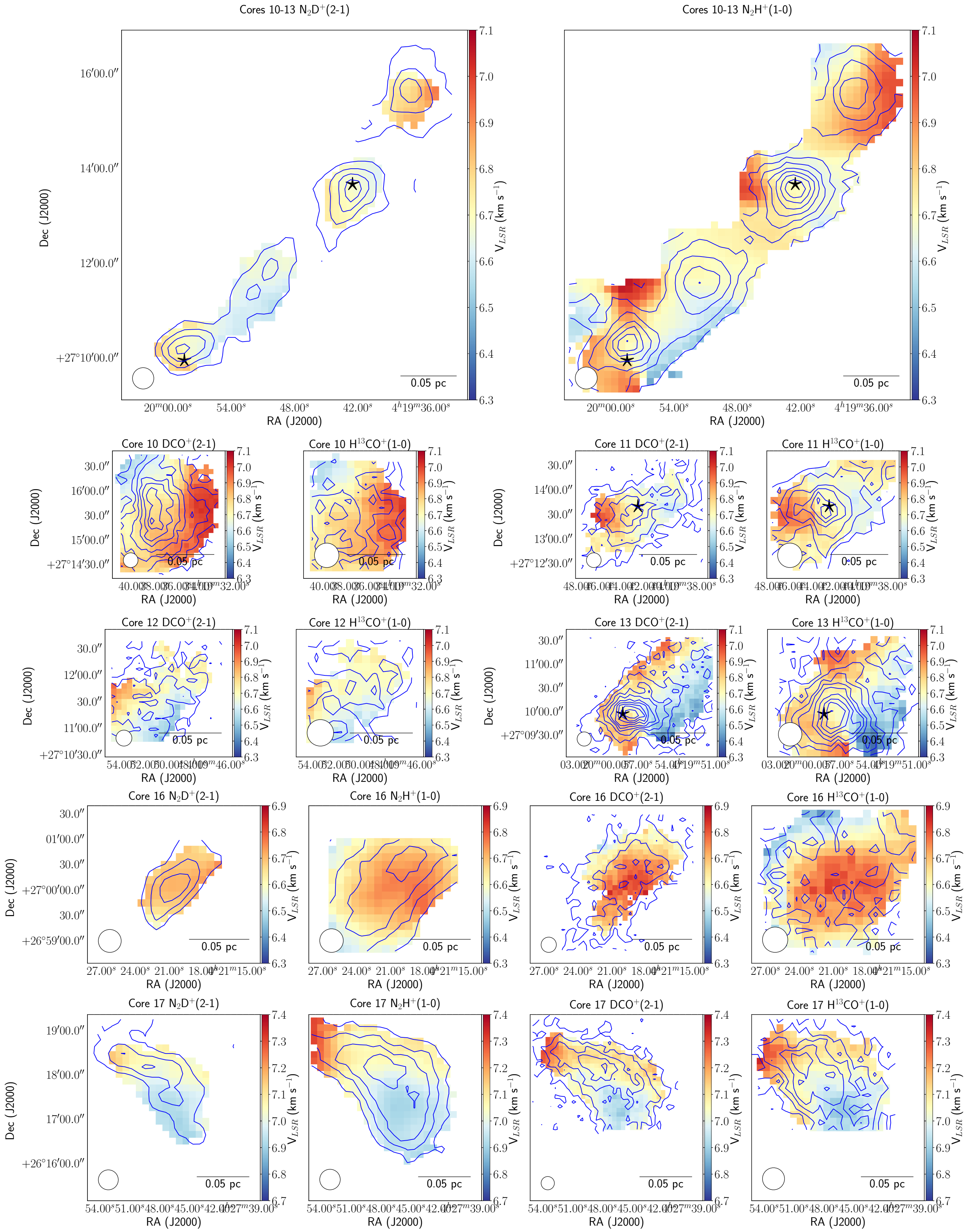

**Fig. A.10.** continued.

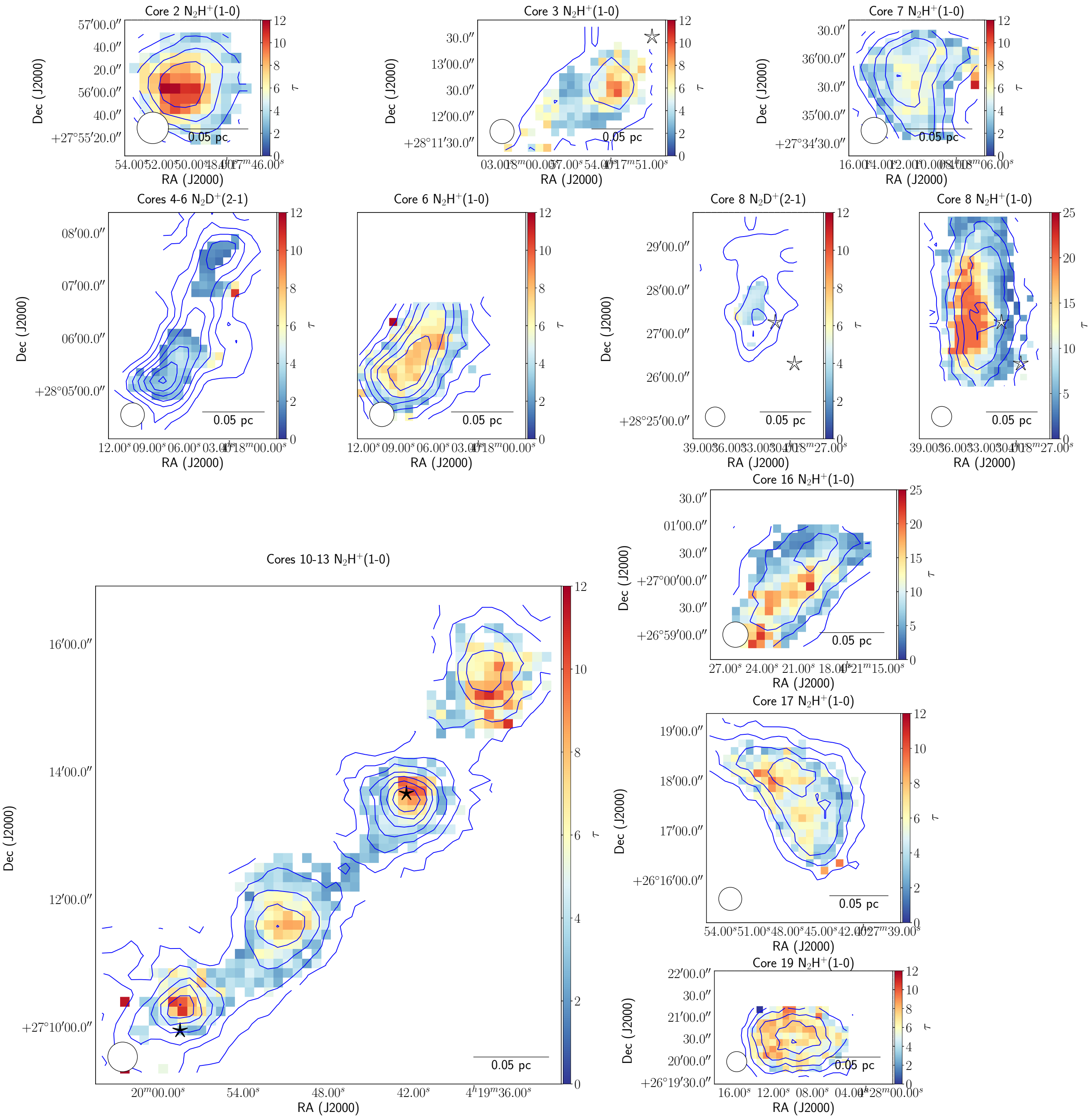

**Fig. A.11.** Opacities of the $N_2D^+$(2–1) and $N_2H^+$(1–0) lines across the cores. We show only the maps which have more than three pixels with $\tau$-unconstrained fit. Blue contours show the integrated intensities of the corresponding species. First contour is at the $5\sigma$ level and the contour step is $5\sigma$. Stars show the positions of young stellar objects (YSOs) from Rebull et al. (2010): black stars are young, flat and class I objects, white stars are more evolved, class II and III objects.

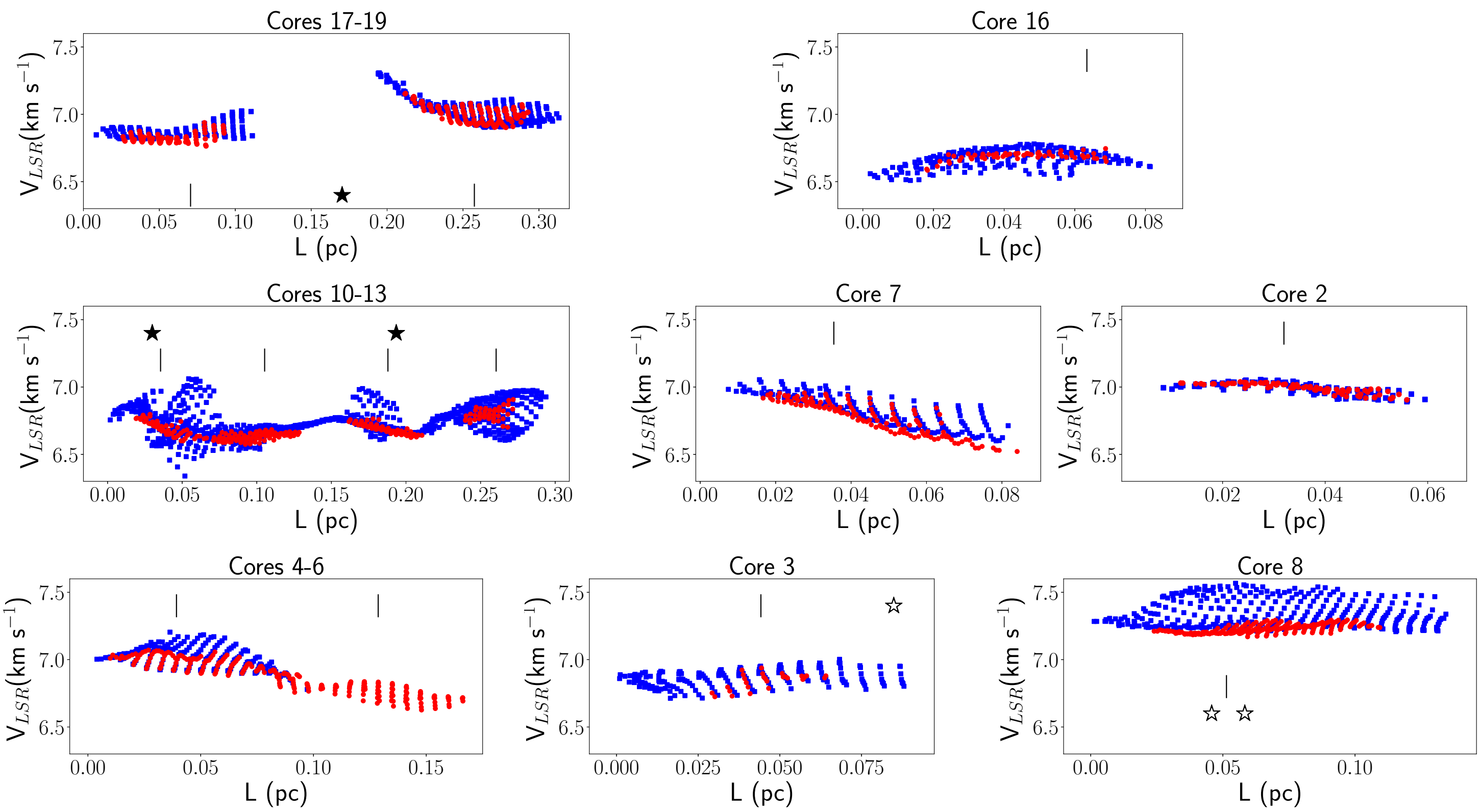

**Fig. A.12.** $V_{LSR}$ along the filament direction (from core 19 in the south-east to core 8 in the north-west). The transitions are shown with colours: $N_2D^+$(2–1) – red circles and $N_2H^+$(1–0) – blue squares. The vertical bars show the $N_2H^+$(1–0) emission peaks. Stars show the positions of YSOs from Rebull et al. (2010): black stars are flat and class I objects, white stars are class II and III objects.

**Table B.2.** Total velocity gradients $G$, position angles $\theta_G$ (measured west to north), equivalent radii[a], measured across the cores maps convolved to the largest beam of 29.9″ using the emitting area of $N_2D^+$(2–1), and corresponding specific angular momenta $J/M$.

| Core | $G$ (km s$^{-1}$ pc$^{-1}$) | $\theta_G$ (degrees) | R (pc) | $J/M$ ($10^{20}$ cm$^2$ s$^{-1}$) | $G$ (km s$^{-1}$ pc$^{-1}$) | $\theta_G$ (degrees) | R (pc) | $J/M$ ($10^{20}$ cm$^2$ s$^{-1}$) |
|---|---|---|---|---|---|---|---|---|
| | | $N_2H^+$(1–0) | | | | $N_2D^+$(2–1) | | |
| 2 | 2.251 ± 0.017 | -134.7 ± 0.5 | 0.032 | 2.91 ± 0.02 | 2.586 ± 0.073 | -142.7 ± 1.9 | 0.032 | 3.35 ± 0.09 |
| 3 | 2.787 ± 0.100 | 161.3 ± 1.6 | 0.018 | 1.06 ± 0.04 | 2.689 ± 0.867 | 119.0 ± 18.8 | 0.018 | 1.02 ± 0.33 |
| 4 | – | – | | – | 1.337 ± 0.055 | -87.9 ± 3.4 | 0.038 | 2.44 ± 0.10 |
| 6 | 3.832 ± 0.016 | -48.9 ± 0.2 | 0.040 | 7.44 ± 0.03 | 3.443 ± 0.075 | -51.5 ± 0.8 | 0.040 | 6.68 ± 0.14 |
| 7 | 5.445 ± 0.012 | -151.0 ± 0.1 | 0.040 | 10.77 ± 0.02 | 5.929 ± 0.070 | -153.7 ± 0.7 | 0.041 | 12.18 ± 0.14 |
| 8 | 3.608 ± 0.017 | 7.3 ± 0.1 | 0.044 | 8.79 ± 0.04 | 1.687 ± 0.121 | 17.4 ± 2.1 | 0.044 | 4.11 ± 0.30 |
| 10 | 4.634 ± 0.035 | -33.9 ± 0.3 | 0.027 | 4.06 ± 0.03 | 3.666 ± 0.248 | -37.0 ± 3.6 | 0.027 | 3.21 ± 0.22 |
| 11 | 2.053 ± 0.018 | -143.6 ± 0.8 | 0.031 | 2.42 ± 0.02 | 2.197 ± 0.134 | -156.8 ± 4.5 | 0.031 | 2.59 ± 0.16 |
| 12 | 1.875 ± 0.018 | 109.3 ± 0.5 | 0.030 | 2.14 ± 0.02 | 1.629 ± 0.191 | 124.0 ± 4.7 | 0.030 | 1.86 ± 0.22 |
| 13 | 3.861 ± 0.042 | 150.6 ± 0.4 | 0.031 | 4.70 ± 0.05 | 2.741 ± 0.185 | 159.3 ± 3.0 | 0.031 | 3.34 ± 0.23 |
| 16 | 0.602 ± 0.044 | -33.3 ± 2.2 | 0.029 | 0.64 ± 0.05 | 0.661 ± 0.115 | 34.8 ± 17.9 | 0.029 | 0.70 ± 0.12 |
| 17 | 2.049 ± 0.010 | 114.8 ± 0.5 | 0.038 | 3.59 ± 0.02 | 2.538 ± 0.080 | 84.1 ± 1.8 | 0.038 | 4.44 ± 0.14 |
| 19 | 1.523 ± 0.017 | -72.8 ± 0.5 | 0.034 | 2.20 ± 0.03 | 1.224 ± 0.144 | -89.1 ± 6.0 | 0.034 | 1.77 ± 0.21 |
| | | $H^{13}CO^+$(1–0) | | | | $DCO^+$(2–1) | | |
| 2 | 2.577 ± 0.128 | -149.7 ± 2.7 | 0.031 | 3.14 ± 0.16 | 2.392 ± 0.035 | -132.0 ± 0.8 | 0.032 | 3.09 ± 0.05 |
| 3 | 7.354 ± 0.784 | -164.0 ± 6.9 | 0.016 | 2.24 ± 0.24 | 3.732 ± 0.245 | -156.4 ± 4.6 | 0.018 | 1.42 ± 0.09 |
| 4 | 2.410 ± 0.061 | -97.9 ± 1.7 | 0.037 | 4.04 ± 0.10 | 2.150 ± 0.015 | -103.9 ± 0.6 | 0.038 | 3.93 ± 0.03 |
| 6 | 4.274 ± 0.135 | -40.1 ± 1.0 | 0.040 | 8.30 ± 0.26 | 4.638 ± 0.030 | -43.5 ± 0.2 | 0.039 | 8.82 ± 0.06 |
| 7 | 4.964 ± 0.056 | -136.3 ± 0.7 | 0.038 | 9.07 ± 0.10 | 4.623 ± 0.015 | -146.5 ± 0.2 | 0.041 | 9.50 ± 0.03 |
| 8 | 4.779 ± 0.126 | 13.8 ± 0.8 | 0.044 | 11.64 ± 0.31 | 3.877 ± 0.093 | 7.1 ± 0.7 | 0.044 | 9.44 ± 0.23 |
| 10 | 3.998 ± 0.196 | -25.5 ± 2.1 | 0.027 | 3.50 ± 0.17 | 4.904 ± 0.040 | -37.1 ± 0.3 | 0.027 | 4.29 ± 0.04 |
| 11 | 3.300 ± 0.166 | -178.4 ± 2.9 | 0.030 | 3.77 ± 0.19 | 4.325 ± 0.079 | 172.6 ± 0.9 | 0.031 | 5.10 ± 0.09 |
| 12 | 2.747 ± 0.300 | 120.8 ± 3.6 | 0.029 | 2.82 ± 0.31 | 3.924 ± 0.086 | 129.0 ± 0.7 | 0.030 | 4.48 ± 0.10 |
| 13 | 3.704 ± 0.152 | 157.6 ± 1.9 | 0.031 | 4.51 ± 0.19 | 3.681 ± 0.066 | 157.0 ± 0.8 | 0.031 | 4.48 ± 0.08 |
| 16 | 3.475 ± 0.185 | -84.1 ± 2.8 | 0.029 | 3.70 ± 0.20 | 4.483 ± 0.098 | -59.7 ± 0.8 | 0.029 | 4.78 ± 0.10 |
| 17 | 3.681 ± 0.112 | 96.3 ± 2.1 | 0.032 | 4.76 ± 0.15 | 3.160 ± 0.049 | 86.7 ± 0.8 | 0.037 | 5.41 ± 0.08 |
| 19 | 3.125 ± 0.080 | -50.7 ± 1.3 | 0.034 | 4.52 ± 0.12 | 2.250 ± 0.036 | -58.7 ± 0.8 | 0.034 | 3.25 ± 0.05 |

**Notes.** [a] Equivalent radius here is the radius of a circle which has the same area as the emitting area (see Subsect. 4.5).

**Table B.3.** The parameters of 2D Gaussians fitted to the $N_2D^+$(2–1) integrated intensity maps: centres of the ellipses, major and minor axes, position angles and aspect ratios.

| Core | $\alpha_{J2000}$ ($^{h\ m\ s}$) | $\delta_{J2000}$ ($^{\circ}\ '\ ''$) | $FWHM^{a}_{mj}$ ($'$) | $FWHM_{mn}$ ($'$) | $FWHM_{mj}$ (pc) | $FWHM_{mn}$ (pc) | $FWHM_{av}$ (pc) | $\theta$ ($^{\circ}$) | $a$ |
|---|---|---|---|---|---|---|---|---|---|
| 2 | 04:17:50.12 | 27:56:03.67 | 0.918 | 0.728 | 0.037 | 0.030 | 0.034 | 162.553 | 1.26 |
| 3[b] | 04:17:54.15 | 28:12:30.07 | 3.239 | 1.423 | 0.132 | 0.058 | 0.095 | 134.217 | 2.28 |
| 4 | 04:18:03.77 | 28:07:15.40 | 2.328 | 1.041 | 0.095 | 0.042 | 0.069 | 150.323 | 2.24 |
| 6 | 04:18:07.63 | 28:05:34.74 | 1.919 | 0.816 | 0.078 | 0.033 | 0.056 | 145.081 | 2.35 |
| 7 | 04:18:11.44 | 27:35:41.40 | 1.932 | 1.316 | 0.079 | 0.054 | 0.066 | 5.461 | 1.47 |
| 8 | 04:18:32.99 | 28:27:46.29 | 2.462 | 0.865 | 0.100 | 0.035 | 0.068 | 175.419 | 2.85 |
| 10 | 04:19:37.19 | 27:15:31.84 | 1.497 | 0.925 | 0.061 | 0.038 | 0.049 | 140.137 | 1.62 |
| 11 | 04:19:42.77 | 27:13:27.51 | 1.160 | 0.739 | 0.047 | 0.030 | 0.039 | 137.747 | 1.57 |
| 12 | 04:19:51.15 | 27:11:33.36 | 1.905 | 0.798 | 0.078 | 0.032 | 0.055 | 142.323 | 2.39 |
| 13 | 04:19:57.02 | 27:10:08.35 | 1.052 | 0.536 | 0.043 | 0.022 | 0.032 | 110.592 | 1.96 |
| 16 | 04:21:20.46 | 27:00:06.37 | 1.327 | 0.451 | 0.054 | 0.018 | 0.036 | 139.372 | 2.94 |
| 17 | 04:27:48.93 | 26:18:06.43 | 2.446 | 0.759 | 0.100 | 0.031 | 0.065 | 49.841 | 3.22 |
| 19 | 04:28:10.12 | 26:20:23.98 | 1.605 | 1.274 | 0.065 | 0.052 | 0.059 | 127.836 | 1.26 |

**Notes.** [a] FWHM of the cores: $FWHM_{mj}$ – major axis, $FWHM_{mn}$ – minor axis, $FWHM_{av}$ – average of the major and minor axes.
[b] The $N_2D^+$(2–1) data for core 3 is not sufficient to fit a Gaussian, the $N_2H^+$(1–0) data used here.

**Table B.4.** The total velocity gradients $G$ and FWHM radii $R_{FWHM}$, measured across the cores maps convolved to the largest beam of 29.9″. The $R_{FWHM}$ are calculated using the $N_2D^+$(2–1) maps which trace the densest gas[a].

| Core | $G$ (km s$^{-1}$ pc$^{-1}$) | | | | $R_{FWHM}$ (pc) |
|---|---|---|---|---|---|
| | $N_2D^+$(2–1) | $N_2H^+$(1–0) | $DCO^+$(2–1) | $H^{13}CO^+$(1–0) | |
| 2 | 2.07 ± 0.14 | 1.89 ± 0.03 | 2.99 ± 0.07 | 2.20 ± 0.28 | 0.019 |
| 4 | 0.51 ± 0.07 | – | 1.57 ± 0.036 | 1.79 ± 0.13 | 0.026 |
| 6 | 3.78 ± 0.13 | 4.47 ± 0.03 | 6.36 ± 0.07 | 5.76 ± 0.37 | 0.024 |
| 7 | 6.53 ± 0.10 | 6.08 ± 0.02 | 5.12 ± 0.02 | 5.60 ± 0.09 | 0.032 |
| 8 | 1.68 ± 0.16 | 3.97 ± 0.03 | 3.84 ± 0.19 | 5.15 ± 0.27 | 0.030 |
| 10 | 4.19 ± 0.40 | 4.79 ± 0.06 | 4.83 ± 0.06 | 4.74 ± 0.37 | 0.021 |
| 11 | 1.88 ± 0.26 | 1.95 ± 0.04 | 3.96 ± 0.21 | 2.89 ± 0.32 | 0.017 |
| 12 | 2.01 ± 0.28 | 1.99 ± 0.03 | 4.79 ± 0.13 | 2.93 ± 0.40 | 0.025 |
| 13 | 3.15 ± 0.26 | 3.75 ± 0.09 | 4.52 ± 0.09 | 3.38 ± 0.25 | 0.018 |
| 16 | 0.80 ± 0.22 | 0.72 ± 0.06 | 4.15 ± 0.22 | 1.00 ± 0.46 | 0.018 |
| 17 | 3.39 ± 0.15 | 3.12 ± 0.02 | 2.14 ± 0.07 | 2.42 ± 0.23 | 0.026 |
| 19 | 1.54 ± 0.17 | 1.63 ± 0.02 | 2.40 ± 0.05 | 3.31 ± 0.11 | 0.027 |

**Notes.** [(a)] We do not show the data for core 3 since the poor $N_2D^+$(2–1) detection does not allow to find FWHM.

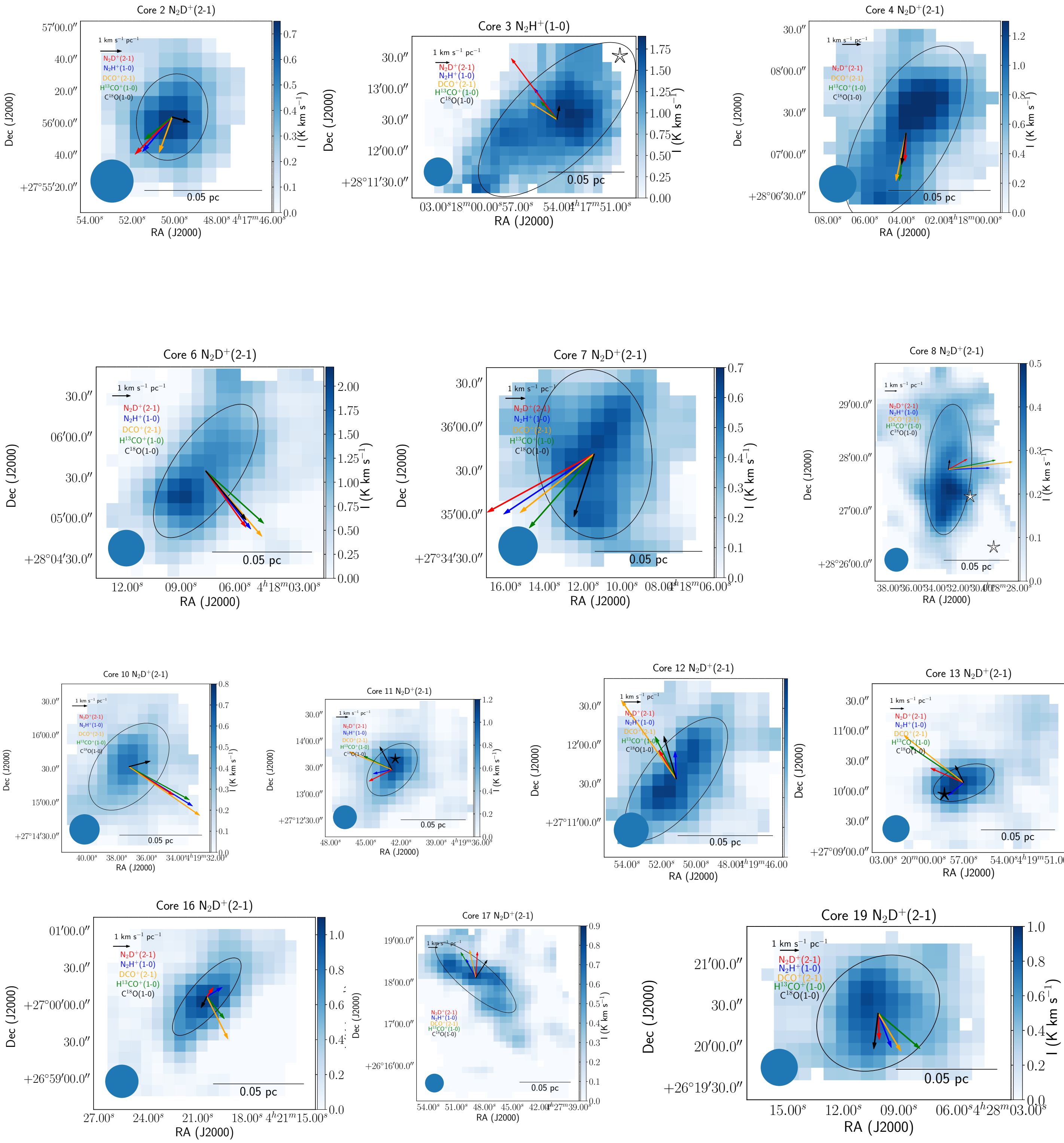

**Fig. B.1.** Total velocity gradients of the $N_2D^+$(2–1) (red), $N_2H^+$(1–0) (blue), $DCO^+$(2–1) (orange), $H^{13}CO^+$(1–0) (green), and $C^{18}O$(1–0) (black) lines across the cores. The colorscale shows the integrated intensities of $N_2D^+$(2–1) across the cores except core 3, where the integrated intensity of $N_2H^+$(1–0) is shown. The ellipses show the 2D Gaussian FWHM fitted to the $N_2D^+$(2–1) integrated intensity maps ($N_2H^+$(1–0) for core 3). Stars show the positions of young stellar objects (YSOs) from Rebull et al. (2010): black stars are young, flat and class I objects, white stars are more evolved, class II and III objects.

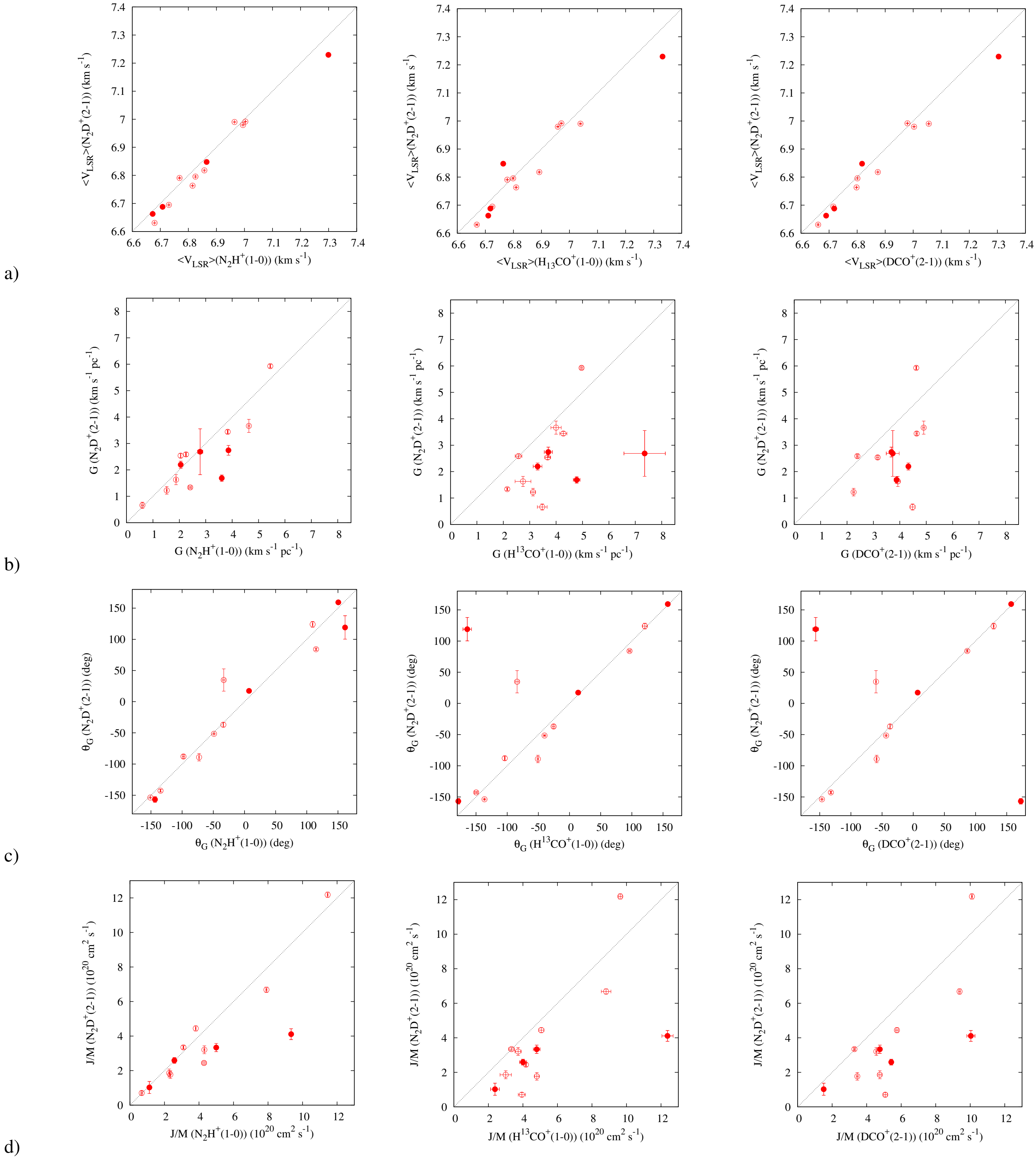

**Fig. B.2.** $N_2D^+$(2–1) compared to $N_2H^+$(1–0), $H^{13}CO^+$(1–0) and $DCO^+$(2–1): a) average velocity across the core; b) total velocity gradient; c) position angle of the total velocity gradient; d) specific angular momentum. Open circles show starless cores and filled circles show protostellar cores. For the comparison, all maps are convolved to the largest beam of 29.9″ with Nyquist spacing and only the area which contains $N_2D^+$(2–1) emission is used.

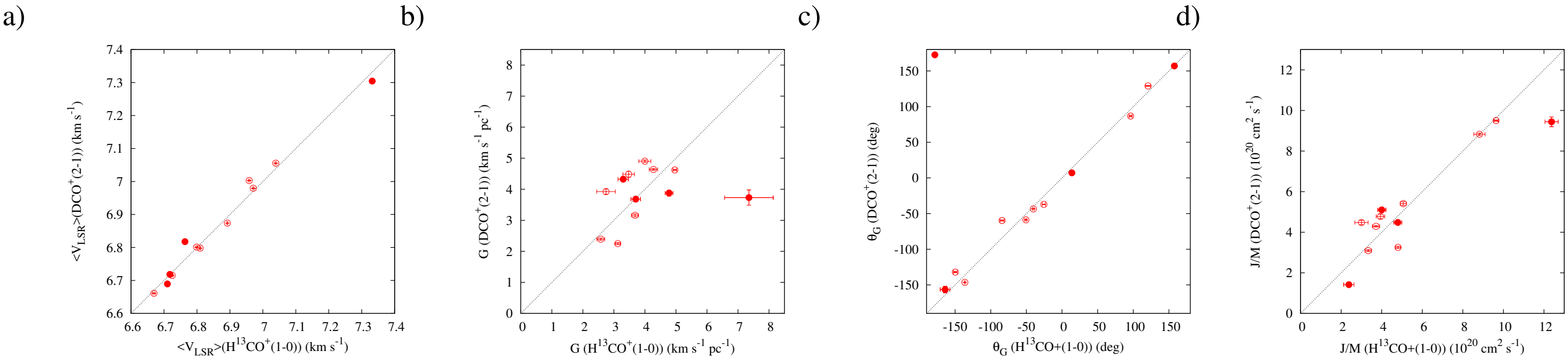

**Fig. B.3.** $DCO^+$(2–1) compared to $H^{13}CO^+$(1–0): a) average velocity across the core; b) total velocity gradient; c) position angle of the total velocity gradient; d) specific angular momentum. Open circles show starless cores and filled circles show protostellar cores. For the comparison, all maps are convolved to the largest beam of 29.9″ with Nyquist spacing and only the area which containes both $H^{13}CO^+$(1–0) and $DCO^+$(2–1) emission is used.

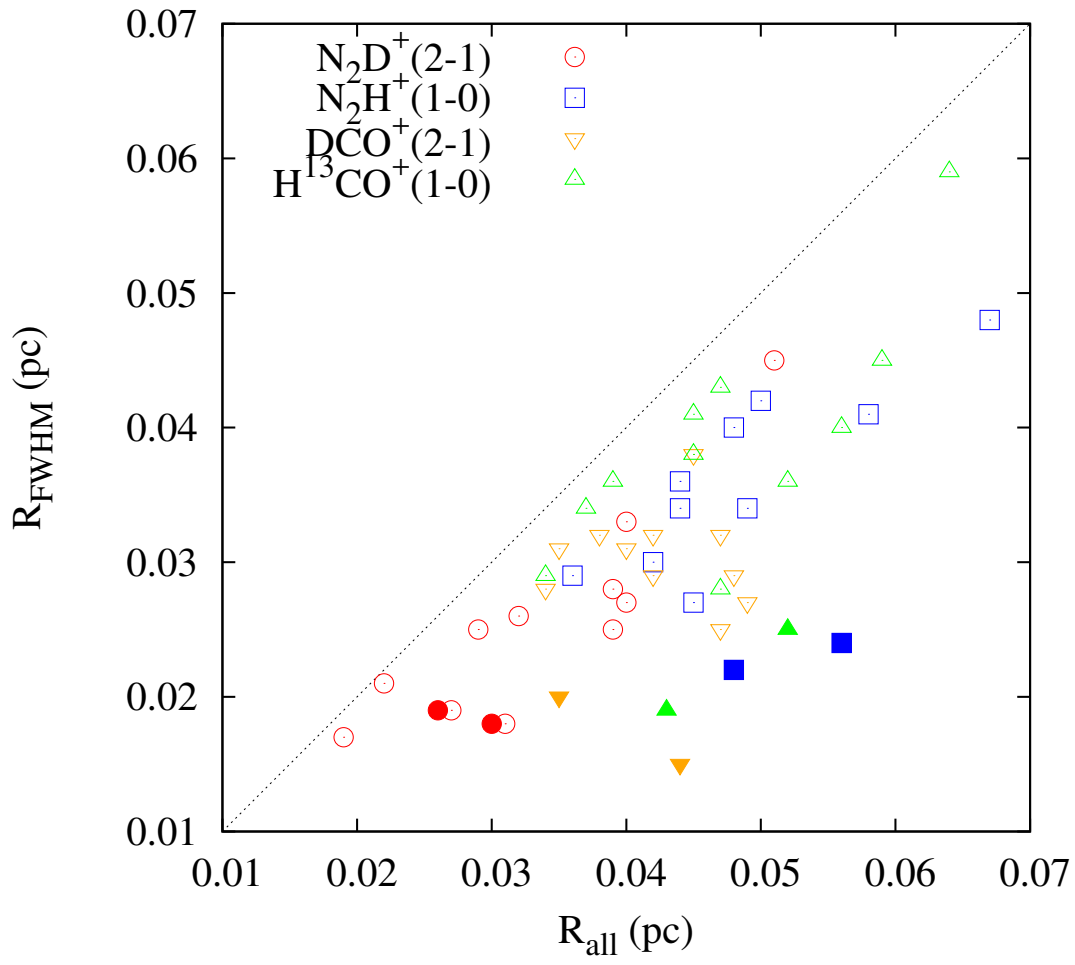

**Fig. B.4.** Core radius at FWHM of maximum emission versus the radius of the total emitting area, measured with different species. Filled symbols show protostellar cores, and open symbols show starless cores. The dashed line shows a one-to-one correlation.

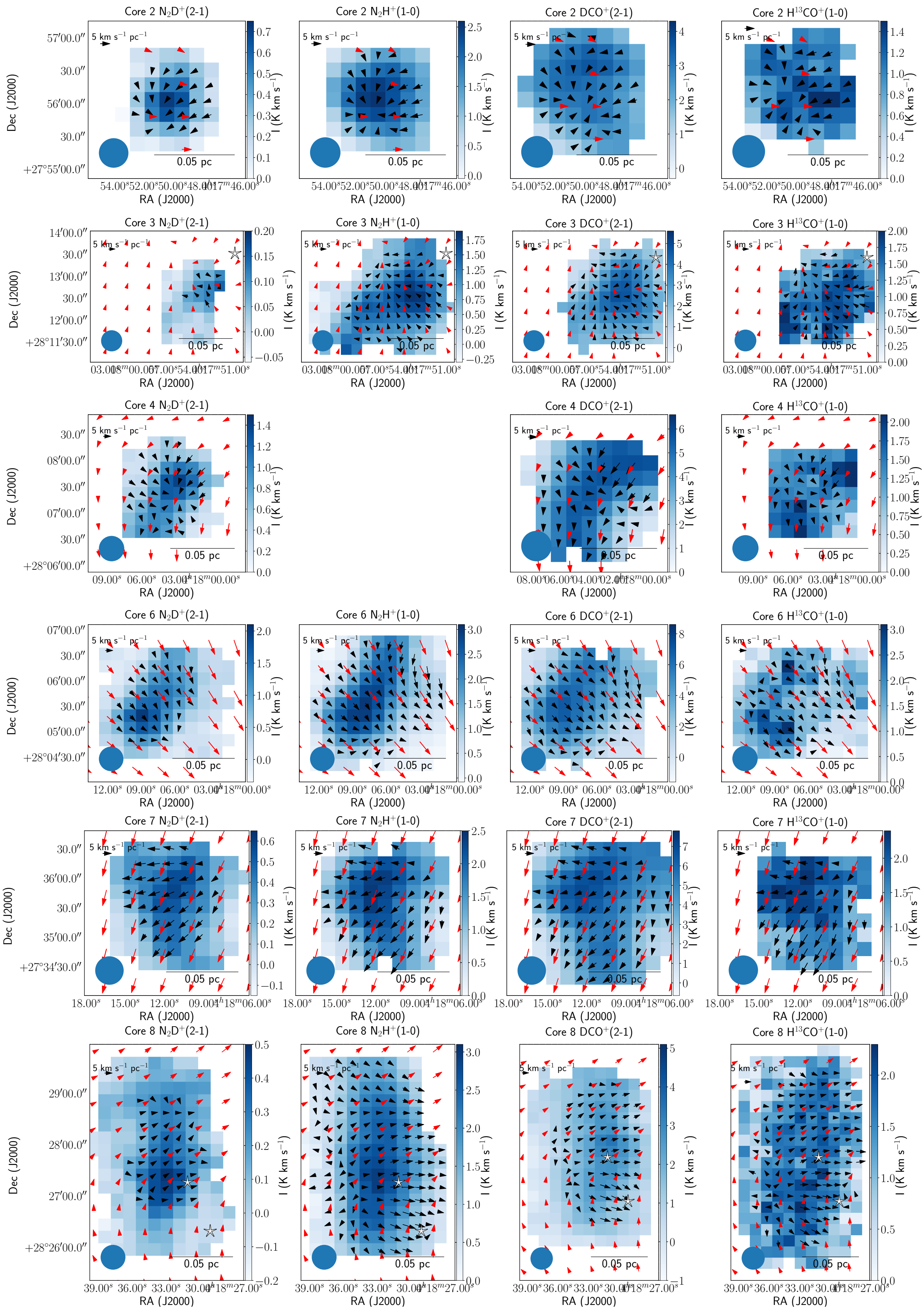

**Fig. B.5.** $V_{LSR}$ local gradients of the $N_2D^+$(2–1), $N_2H^+$(1–0), $DCO^+$(2–1), and $H^{13}CO^+$(1–0) lines across the cores (black arrows) and $V_{LSR}$ local gradients of the $C^{18}O$(1–0) line from Hacar et al. (2013) (red arrows). The scale of the $C^{18}O$(1–0) gradients is 2.5 times larger then the scale indicated in the figures. Colorscale shows integrated intensities of corresponding species across the cores. Stars show the positions of young stellar objects (YSOs) from Rebull et al. (2010): black stars are young, flat and class I objects, white stars are more evolved, class II and III objects.

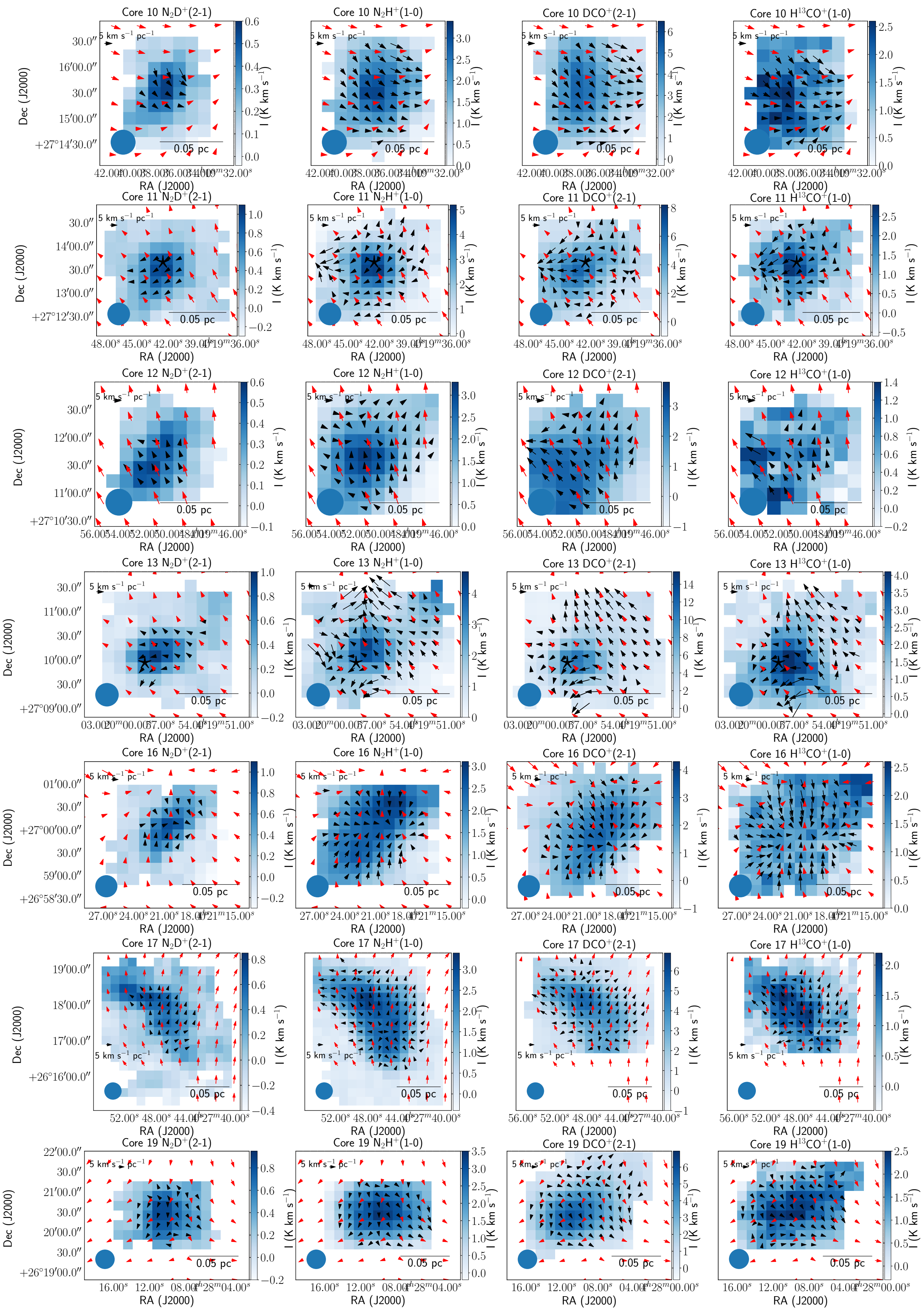

**Fig. B.5.** continued.